\documentclass[a4paper,leqno,11pt]{article}
\begin{document}
\begin{titlepage}
%
%
\title{\bf
Irreducible Green Functions Method and\\
Many-Particle Interacting Systems on a Lattice \thanks{"Rivista
del Nuovo Cimento" vol.25, N 1 (2002) pp.1-91 }}
%
\author{
A.L.Kuzemsky
\thanks{E-mail:kuzemsky@thsun1.jinr.ru; http://thsun1.jinr.ru/~ kuzemsky}
\\
{\it Bogoliubov Laboratory of Theoretical Physics,} \\
{\it  Joint Institute for Nuclear Research,}\\
{\it 141980 Dubna, Moscow Region, Russia.}}
\date{}
\maketitle
%
\begin{abstract}
The Green-function technique, termed the irreducible Green
functions (IGF) method, that is a certain reformulation of the
equation-of motion method for double-time temperature dependent
Green functions (GFs) is presented. This method was developed to
overcome  some ambiguities in terminating the hierarchy of the
equations of motion of double-time Green functions and to give a
workable technique to systematic way of decoupling. The approach
provides a practical method for description of the many-body
quasi-particle dynamics of correlated systems on a lattice with
complex spectra. Moreover, it provides a very compact and
self-consistent way of taking into account the damping effects
and finite lifetimes of   quasi-particles due to   inelastic
collisions. In addition, it  correctly  defines the Generalized
Mean Fields (GMF), that determine   elastic scattering
renormalizations and  , in general, are not   functionals of the
mean particle densities only. The purpose of this article is to
present the foundations of the  IGF method. The technical details
and examples are given as well. Although some space is devoted to
the formal structure of the method, the emphasis is on its
utility. Applications to the lattice fermion models such as
Hubbard/Anderson models and to the Heisenberg ferro- and
antiferromagnet, which manifest the operational ability of the
method are given. It is shown that the IGF method provides a
powerful tool  for the construction of  essentially new dynamical
solutions for  strongly interacting  many-particle systems with
complex spectra.\\
\end{abstract}
\thispagestyle{empty}
\end{titlepage}
\newpage
\tableofcontents
\newpage
\section{Introduction}
 The basic problems of field theory and
statistical mechanics are much similar in many aspects,
especially, when we use  the method of second quantization and
Green functions\cite{bogo}. In  both the cases, we are dealing
with systems possessing a large number of degrees of freedom (the
energy spectrum is practically a continuous one) and with
averages of quantum mechanical operators~\cite{law}.  In quantum
field theory,  we mostly consider averages over the ground state,
while in statistical mechanics, we consider finite temperatures
(ensemble averages) as well as ground-state averages. Great
advances have been made during the last decades in statistical
physics and condensed matter theory through the use of methods of
quantum field theory ~\cite{tsve1} - \cite{naga1}. It was widely
recognized that a successful approximation for  determining
exited states is based on the quasi-particle concept and the
Green function method.  For example, the study of highly
correlated electron systems has attracted much attention
recently~\cite{bic} - \cite{naga2}, especially after discovery of
copper oxide superconductors, a new class of heavy
fermions~\cite{kuz1}, and low-dimensional compounds~\cite{tsve1},
~\cite{oki}. Although much work for strongly correlated systems
has been performed during the last years, it is worthy to remind
that the investigation of  excitations in many-body systems has
been one of the most important and interesting subjects for the
last few decades. \\ The quantum field theoretical techniques
have been widely applied to statistical treatment of a large
number of interacting particles. Many-body calculations are often
done for model many-particle systems by using a perturbation
expansion. The basic procedure in many-body theory~\cite{bog1} is
to find a suitable unperturbed Hamiltonian and then to take into
account a small perturbation operator. This procedure that works
well for  weakly interacting systems needs a special
reformulation for  many-body systems with complex spectra and
strong interaction. For many practically interesting cases ({\it
e.g.} in quantum chemistry problems ), the standard schemes of
perturbation expansion must be reformulated greatly~\cite{hur8} -
\cite{tar12}. Moreover,  many-body systems on a lattice have
their own specific features and in some
important aspects differ greatly from  continuous systems.\\
In this review that is largely pedagogical we are primarily
dealing with the spectra of elementary excitations to learn about
quasi-particle many-body dynamics of interacting systems on a
lattice. Our analysis is based on the equation-of-motion approach,
the derivation of the exact representation of the Dyson equation
and construction of an approximate scheme of calculations in a
self-consistent way. In this review only  some topics in the
field are discussed. The emphasis is on the methods rather than
on a detailed comparison with the experimental results. We
attempt to prove that the approach we suggest produces a more
advanced physical picture of the
problem of the quasi-particle many-body dynamics.\\
The most characteristic feature of the recent advancement in the
basic research on electronic properties of solids is the
development of variety of the new classes of materials with
unusual properties: high-$T_c$ superconductors, heavy fermion
compounds, complex oxides, diluted magnetic semiconductors,
perovskite manganites, {\it etc}. Contrary to  simple metals,
where the fundamentals are very well known and the electrons can
be represented  so that they weakly interact with each other, in
these materials, the electrons interact strongly, and moreover
their spectra are complicated, {\it i.e.} have many branches.
This gives rise to interesting phenomena~\cite{tokur} such as
magnetism, metal-insulator transition in oxides, heavy fermions,
colossal negative magnetoresistance in manganites, {\it etc.},
but the understanding of
what is going on is in many cases only partial.\\
The subject of the present paper is a microscopic many-body
theory of strongly correlated electron models. A principle
importance of these studies is concerned with a fundamental
problem of electronic solid state theory, namely with the tendency
of 3d(4d) electrons in transition metal compounds and 4f(5f)
electrons in rare-earth metal compounds and alloys to exhibit
both the localized and delocalized (itinerant) behaviour.
Interesting electronic and magnetic properties of these
substances are intimately related to this dual behaviour of electrons\cite{ku1}-\cite{akwa}.\\
The problem of  adequate description of  strongly correlated
electron systems has been studied intensively during the last
decade\cite{acqua1},\cite{acqua2}, especially in context of the
physics of magnetism, heavy fermions and high-$T_c$
superconductivity~\cite{kuz1}. The understanding of the true
nature of  electronic states and their quasi-particle dynamics is
one of the central topics of the current experimental and
theoretical studies in the field. A plenty of experimental and
theoretical results show that this many-body quasi-particle
dynamics is quite nontrivial. A vast amount of theoretical
searches for a suitable  description of  strongly correlated
fermion systems deal with  simplified model Hamiltonians. These
include, as workable patterns, the single-impurity Anderson model
(SIAM) and Hubbard model. In spite of certain drawbacks, these
models exhibit the key physical feature: the competition and
interplay between kinetic energy (itinerant) and potential energy
(localized) effects.  A fully consistent theory of quasi-particle
dynamics of both the models is believed to be crucially important
for a deeper understanding of the true nature of  electronic
states in the above-mentioned class of materials. In spite of
experimental and theoretical achievements,   it remains still much
to be understood concerning such systems~\cite{ku2},\cite{acqua3}. \\
Recent theoretical investigations of strongly correlated systems
have brought forth a significant variety of the approaches
 to solve these controversial problems. There is an
important aspect of the problem under consideration, namely, how
to take  adequately into account the lattice (quasi-localized)
character of  charge carriers, contrary to  simplified theories
of the type of a weakly interacting electron gas.  To match such a
trend, we need to develop a systematic theory of  correlated
systems, to describe, from the first principles of the condensed
matter theory and statistical mechanics, the physical properties
of this class of  materials.\\ In  previous papers, we set up the
practical technique of the method of the irreducible Green
functions (IGF) ~\cite{kuz2} -\cite{kuz12}. This IGF method
allows one to describe  quasi-particle spectra with damping for
systems with complex spectra and strong correlation in a very
general and natural way. This scheme differs from the traditional
methods of decoupling or terminating an infinite chain of the
equations and permits one to construct the relevant dynamic
solutions in a self-consistent way on the level of the Dyson
equation without decoupling the chain of the equations of motion
for the double-time temperature Green functions. The essence of
our consideration of  dynamic properties of many-body system with
strong interaction is related closely with the field theoretical
approach, and we use the advantage of the Green- function
language and the Dyson equation. It is possible to say that our
method  emphasizes the fundamental and central role of the Dyson
equation for the single-particle dynamics of many-body systems at
finite temperature. This approach has been suggested as essential
for various many-body systems, and we believe that it bears the
real physics of interacting many-particle interacting
systems~\cite{kuz3}, \cite{kuz4}.\\ It is the purpose of the
present paper to introduce the concepts of irreducible Green
functions (or irreducible operators) and Generalized Mean Fields
(GMF ) in a simple and coherent fashion to assess the validity of
quasi-particle description and mean field theory. The irreducible
Green function method is a reformulation of the
equation-of-motion approach for the double-time thermal GFs, aimed
of operating with the correct functional structure of the required
solutions. In this sense, it has all advantages and shortcomings
of the Green-function method in comparison, say, with the
functional integration technique, that, in turn, has also its own
advantages and shortcomings. The usefulness of one or another
method depends on the problem we are trying to solve. For the
calculation of  quasi-particle spectra, the Green-function method
is the best. The irreducible-Green-function method adds to this
statement: "for the calculation of the quasi-particle spectra
with damping" and gives a workable
recipe how to do this in a self-consistent way.\\
The distinction between elastic and inelastic scattering effects
is a fundamental one in the physics of many-body systems, and it
is also reflected in a number of other ways than in the
mean-field and finite lifetimes.  The present review attempts to
offer a balanced view of quasi-particle interaction effects in
terms of division into elastic- and inelastic-scattering
characteristics. For this aim, in the present paper, we  discuss
the background of the IGF approach more thoroughly. To
demonstrate  the general analysis, we consider here the
calculation of  quasi-particle spectra and their damping within
various types of correlated electron models to extend the
applicability of the general formalism and show flexibility and
practical usage of the IGF method.
\section{ Varieties of Green Functions}
It is appropriate to remind the ideas underlying the Green-
function method, and to discuss briefly why they are particularly
useful in the study of
interacting many-particle systems.\\
The Green functions of potential theory~\cite{green} were
introduced to find the field which is produced by a source
distribution ({\it e.g.} the electromagnetic field which is
produced by current and charge distribution). The Green functions
in field theory are the so-called propagators which describe the
temporal development of quantized fields, in its particle aspect,
as was shown by Schwinger in his seminal works~\cite{schw1} -
~\cite{schw3}.  The idea of the Green function method is
contained in the observation that it is not necessary to attempt
to calculate all the wave functions and energy levels of a system.
Instead, it is more instructive to study the way in which it
responds to simple perturbations, for example, by adding or
removing particles, or by applying external fields.\\
There is a variety of Green functions~\cite{mah} and there are
Green functions for one particle, two particles..., n particles.
A considerable progress in studying the spectra of elementary
excitations and thermodynamic properties of many-body systems has
been for most part due to the development of the
temperature-dependent Green-functions methods.
\subsection{Temperature Green Functions }
The temperature dependent Green functions were introduced by
Matsubara~\cite{matsu}. He  considered a many-particle system with
the Hamiltonian
\begin{equation}
\label{eq.1} H = H_{0} + V
\end{equation}
and  observed a remarkable similarity that exists between the
evaluation of the grand partition function of the system and the
vacuum expectation of the so-called S-matrix in quantum field
theory
\begin{equation}
\label{eq.2} Z = Tr \exp[(\mu N - H_{0})\beta] S(\beta);\\
\nonumber S(\beta) = 1 - \int_{0}^{\beta} V(\tau) S(\tau) d \tau
\end{equation}
where $\beta = (kT)^{-1}$. In essence, Matsubara observed and
exploited, to great advantage,  formal similarities between the
statistical operator $\exp (-\beta H) $ and the
quantum-mechanical time-evolution operator $ \exp (i H t)$. As a
result, he introduced  thermal ( temperature-dependent ) Green
functions which
we call now the Matsubara Green functions.\\
We note that the thermodynamic perturbation theory has been
invented by Peierls~\cite{pei}.  For the free energy of a weakly
interacting system he derived the expansion up to second order in
perturbation:
\begin{equation}
\label{eq.3} F = F_{0} + \sum_{n} V_{nn} \rho_{n} + \sum_{m,n}
\frac{|V_{nm}|^{2} \rho_{n}}{E^{0}_{n} - E^{0}_{m}} - \\ \nonumber
\frac{\beta}{2} \sum_{n} V^{2}_{nn} \rho_{n} + \frac{\beta}{2}
\Bigl ( \sum_{n} V_{nn} \rho_{n} \Bigr )^{2}
\end{equation}
where $\rho_{n} = \exp[\beta (F_{0} - E^{0}_{n})]$ and $\exp
(-\beta F_{0}) = \sum_{n} \exp (-\beta E_{n}^{0})$.  By using the
expansion of $S(\beta)$ up to second order
\begin{equation}
\label{eq.4} S(\beta) = 1 - \int_{0}^{\beta} V(\tau) d\tau +
\int_{0}^{\beta} d\tau_{1} \int_{0}^{\tau_1} d\tau_{2} V(\tau_1)
V(\tau_2) +...
\end{equation}
and rearranging the terms in the  expression for $Z$, it can be
shown that the Peierls result for the thermodynamic potential
$\Omega$ can be reproduced by the Matsubara technique (for a
canonical ensemble).\\
Though the use of Green functions is related traditionally with
the perturbation theory through the use of diagram techniques, in
paper~\cite{schw1}  a prophetic remark has been made:
\begin{quote}
"... it is desirable to avoid founding the formal theory of the
Green functions on the restricted basis provided by the assumption
of expandability in powers of coupling constants".
\end{quote}
Since the most important aspect of the many-body  theory is the
necessity of taking properly into account  the interaction
between particles, that changes ( sometimes drastically ) the
behaviour of  non-interacting particles, this remark of Schwinger
is still extremely actual and important.\\ Since that time, a
great deal of work has been done, and  many different variants of
the Green functions have been proposed for  studies  of
equilibrium and non-equilibrium properties of many-particle
systems. We can mention, in particular, the methods of Martin and
Schwinger~\cite{schw2} and of Kadanoff and Baym~\cite{baym2}.
 Martin and Schwinger formulated  the GF
theory not in terms of conventional diagrammatic techniques, but
 in terms of functional-derivative techniques that reduces the many-body problem directly
to the solution of a coupled set of nonlinear integral equations
(see also\cite{strin}). The approach of Kadanoff and Baym
establishes  general rules for obtaining approximations which
preserve the conservation laws ( sometimes called conserving
approximations~\cite{bic} ). As many transport coefficients are
related to conservation laws, one should take care of it when
calculating the two-particle and one-particle Green
functions\cite{strin}. The random-phase approximation, that is an
essential point of the whole Kadanoff-Baym method, does this and
so preserves the appropriate conservation laws. It should be
noted, however, that the Martin-Schwinger and Kadanoff-Baym
methods in their initial form were formulated for treating the
continuum models
and should be adapted to study lattice models, as well.\\
However,  as was claimed by Matsubara in his subsequent
paper~\cite{mat2}, the most convenient way to describe the
equilibrium average of any observable or time-dependent response
of a system to external disturbances is to express them in terms
of a set of the double-time, or
Bogoliubov-Tyablikov, Green functions.\\
The aim of the present paper is to suggest and justify  that an
approach , the irreducible Green functions (IGF)
method~\cite{kuzem1},~\cite{kuz3}, that is in essence a suitable
reformulation of an equation-of-motion approach for the
double-time temperature-dependent Green functions provides an
effective and self-consistent scheme for description the
many-body quasi-particle dynamics of  strongly interacting
many-particle systems with complex spectra. This IGF method
provides some systematization of approximations  and removes (at
least partially) the difficulties usually encountered in the
termination of the hierarchy of  equations of motion for the GF.
\subsection{Double-time Green Functions}
In this Section, we  briefly review the double-time
temperature-dependent Green functions .\\ The double-time
temperature-dependent Green functions were introduced by
Bogoliubov and Tyablikov~\cite{bog2} and reviewed by
Zubarev~\cite{zub} and Tyablikov~\cite{tyab}.\\ Consider a
many-particle system with the time-independent Hamiltonian ${\bf
H} = H - \mu N$; $\mu$ is the chemical potential, $N$ is the
operator of the total number of particles, and we have chosen our
units so that $\hbar = 1$. Let $A(t)$ and $B(t')$ be some
operators . The time development of these operators in the
Heisenberg representation is given by:
\begin{equation}
\label{eq.5} A(t) =\exp (i{\bf H}t) A(0) \exp (-i{\bf H}t)
\end{equation}
We define three types of Green functions, the retarded, advanced,
and  causal Green functions:
\begin{eqnarray}
\label{eq.6} G^{r} = <<A(t), B(t')>>^{r} = -i\theta(t -
t')<[A(t)B(t')]_{\eta}>, \eta = \pm 1.
 \\ \label{eq.7}
G^{a} = <<A(t), B(t')>>^{a} =
i\theta(t' - t)<[A(t)B(t')]_{\eta}>, \eta = \pm 1.  \\
G^{c} = <<A(t), B(t')>>^{c} = i T <A(t)B(t')> =   \\ \nonumber
i\theta(t - t')<A(t)B(t')> + \eta i\theta (t'- t) <B(t')A(t)> ,
\eta = \pm 1. \label{eq.8}
\end{eqnarray}
where $<...>$ is the average over a grand canonical ensemble,
$\theta(t)$ is a step function, and square brackets represent the
commutator or anticommutator
\begin{equation} \label{eq.9}
[A,B]_{\eta} = AB - \eta BA
\end{equation}
Differentiating a Green function with respect to one of the
arguments, for example, the first argument, we can obtain the
equation (equation-of-motion) describing the development of this
function with time
\begin{equation}
\label{eq.10}
id/dt G^{\alpha}(t,t') = \delta (t - t') <[A,B]_{\eta}> + << [A,H](t), B(t')>>^{\alpha};\\
\alpha = r,a,c
\end{equation}
Since this differential equation contains an inhomogeneous term
with $\delta$-type factors, we are dealing formally with the
equation similar to the usual one for the Green
function~\cite{green} and for this  reason, we use the term the
{\it Green function}. We note that the equation of motion is of
the same functional form for all the three types of Green
functions  ( {\it i.e.} retarded, advanced, and causal ). However,
the boundary conditions for $t$ are different for the
retarded, advanced, and causal functions~\cite{bog2}.\\
 The next differentiation gives an infinite chain of
coupled equations of motion
\begin{eqnarray}
\label{eq.11} (i)^{n}d^{n}/dt^{n} G(t,t') = \\ \nonumber
\sum_{k=1}^{n} (i)^{n-k} d^{n-k}/dt^{n-k} \delta (t - t')
<[[...[A, \underbrace{H]...H]}_{k-1}, B]_{\eta}> \\  \nonumber +
<<[[... [A, \underbrace{H]...H]}_{n}(t), B(t')>> \nonumber
\end{eqnarray}
To solve the differential equation-of-motion, we should consider
the Fourier time transforms of the Green functions:
\begin{equation}
\label{eq.12}
 G_{AB}(t-t') =
(2\pi)^{-1}\int_{\infty}^{\infty} d\omega G_{AB}( \omega )
 \exp [-i{\omega}( t - t')],
\end{equation}
\begin{equation}
 \label{eq.13}
 G_{AB}(\omega) = << A | B >>_{\omega} =
\int_{\infty}^{\infty} dt G_{AB}( t )
 \exp (i \omega t) ,
\end{equation}
By inserting (\ref{eq.12}) into (\ref{eq.10}) and (\ref{eq.11}),
we obtain
\begin{eqnarray}
\label{eq.14}
\omega G_{AB}(\omega) = <[A,B]_{\eta}> + << [A,H] | B >>_{\omega} ;  \\
\omega^{n} G_{AB} (\omega) = \sum_{k=1}^{n} \omega^{n-k}
<[[...[A, \underbrace{H]...H]}_{k-1}, B]_{\eta}> \\  \nonumber +
<<[[... [A, \underbrace{H]...H}_{n}]| B >>_{\omega} \nonumber
\label{eq.15}
\end{eqnarray}
It is often convenient to  differentiate of the Green function
with respect to the second time $t'$ . In terms of Fourier time
transforms, the corresponding equations of motion read
\begin{eqnarray}
\label{eq.16}
- \omega G_{AB}(\omega) = - <[A,B]_{\eta}> + << A | [ B,H ] >>_{\omega} ;  \\
\nonumber
(-1)^{n} \omega ^{n} G_{AB} (\omega) = - \sum_{k=1}^{n} ( - 1 )^{n-k}\omega^{n-k}
< [A, [...[B, \underbrace{H]...H]}_{k-1}]_{\eta}> \\
+ << A |[...[ B, \underbrace{H]...H}_{n}] >>_{\omega}
\label{eq.17}
\end{eqnarray}

It is rather difficult problem to solve the infinite chain of
coupled equations of motion (\ref{eq.15}) and (\ref{eq.17}). It
is well established now that the  usefulness of the retarded and
advanced Green functions is  deeply related with the dispersion
relations~\cite{bog2}, that provide the boundary conditions in
the form of  spectral representations of the Green functions.\\
\subsection{Spectral Representations}
The GFs are linear combinations of the time correlation functions:
\begin{eqnarray}
\label{eq.18} F_{AB} (t-t') = < A (t) B(t') > =  \frac {1}{2\pi}
\int^{ + \infty}_{ - \infty} d\omega
\exp [i\omega (t - t')] A_{AB} (\omega)  \\
F_{BA} (t'-t) = < B(t') A(t) > =  \frac {1}{2\pi} \int^{ +
\infty}_{ - \infty} d\omega \exp [i\omega (t'-t)] A_{BA} (\omega)
\label{eq.19}
\end{eqnarray}
Here, the Fourier transforms  $A_{AB}(\omega)$  and
$A_{BA}(\omega)$ are of the form
\begin{eqnarray}
\label{eq.20}
A_{BA} (\omega) = \\
Q^{-1} 2 \pi \sum_{m,n} \exp(-\beta E_{n}) (\psi^{\dagger}_{n} B
\psi_{m})
(\psi^{\dagger}_{m} A \psi_{n}) \delta ( E_{n} - E_{m} - \omega )\nonumber   \\
A_{AB} = \exp( -\beta \omega) A_{BA} ( - \omega) \label{eq.21}
\end{eqnarray}
The expressions (\ref{eq.20}) and (\ref{eq.21}) are  spectral
representations of the corresponding time correlation functions.
The quantities $A_{AB}$ and $A_{BA}$ are  spectral
densities or spectral weight functions. \\
It is convenient to define
\begin{eqnarray}
\label{eq.22} F_{BA} (0) = < B(t) A(t) > =  \frac {1}{2\pi}
\int^{ + \infty}_{ - \infty} d\omega
A (\omega)  \\
F_{AB} (0) = < A (t) B(t) > =  \frac {1}{2\pi} \int^{ + \infty}_{
- \infty} d\omega \exp (\beta \omega ) A (\omega) \label{eq.23}
\end{eqnarray}
Then, the spectral representations of the Green functions can be
expressed in the form
\begin{eqnarray}
\label{eq.24}
G^{r} (\omega) = << A | B >>^{r}_{\omega} = \\
 \frac {1}{2\pi} \int^{ + \infty}_{ - \infty}
\frac {d\omega'}{ \omega - \omega' + i\epsilon}
[\exp(  \beta \omega') - \eta ] A (\omega')   \nonumber \\
G^{a} (\omega) = << A | B >>^{a}_{\omega} =  \\
\frac {1}{2\pi} \int^{ + \infty}_{ - \infty} \frac {d\omega'}{
\omega - \omega' - i\epsilon} [\exp(  \beta \omega')  - \eta ] A
(\omega')  \label{eq.25} \nonumber
\end{eqnarray}
The most important practical consequence of  spectral
representations for the retarded and advanced GFs is the so-called
{\it spectral theorem}. The spectral theorem can be written as
\begin{eqnarray}
\label{eq.26}
< B(t') A(t) > = \\
\nonumber
- \frac {1}{\pi} \int^{ + \infty}_{ - \infty} d\omega
\exp [i\omega (t-t')] [\exp ( \beta \omega) - \eta ]^{-1} Im G_{AB} (\omega + i\epsilon) \\
\label{eq.27}
 < A(t) B(t') > = \\ \nonumber - \frac {1}{\pi} \int^{ +
\infty}_{ - \infty} d\omega \exp  (\beta \omega)  \exp [i\omega
(t-t')] [\exp ( \beta \omega) - \eta ]^{-1} Im G_{AB} (\omega +
i\epsilon)
\end{eqnarray}
Expressions (\ref{eq.26}) and (\ref{eq.27}) are of fundamental
importance. They directly relate the statistical averages with
the Fourier transforms of the corresponding GFs. The problem of
evaluating the latter is thus reduced to finding their Fourier
transforms , providing the practical usefulness of the Green
functions technique~\cite{zub},~\cite{tyab}.
\section{Irreducible Green Functions Method}
In this Section, we  discuss the main ideas of the IGF approach
which allows one to describe completely  quasi-particle spectra
with damping in a very natural way.\\
We  reformulated  the two-time GF method~\cite{kuzem1},
\cite{kuz3} to the form, which is especially adjusted~\cite{kuz2},
\cite{kuzem1} for  correlated fermion systems on a lattice and
systems with complex spectra~\cite{kuz5},\cite{kuz6}. A similar
method was proposed in paper~\cite{plak} for Bose systems (
anharmonic phonons and spin dynamics of  Heisenberg ferromagnet ).
The very important concept of the whole method is the {\it
Generalized Mean Field} (GMF), as it was formulated in
ref.~\cite{kuz3}.  These GMFs have a complicated structure for
the strongly correlated case and complex spectra and are not
reduced to the functional of  mean densities of the electrons or
spins when
one calculates excitation spectra at finite temperatures. \\
\subsection{ Outline of IGF Method}
To clarify the foregoing, let us consider a retarded GF of the
form~\cite{tyab}
\begin{equation}
\label{eq.28} G^{r} = <<A(t), A^{\dagger}(t')>> = -i\theta(t -
t')<[A(t) A^{\dagger}(t')]_{\eta}>, \eta = \pm 1
\end{equation}
As an introduction to the concept of IGFs, let us describe the
main ideas of this approach in a symbolic and simplified form. To
calculate the retarded GF $G(t - t')$,  let us write down the
equation of motion for it:
\begin{equation}
 \label{eq.29}
\omega G(\omega) = <[A, A^{\dagger}]_{\eta}> + <<[A, H]_{-}\mid
A^{\dagger}>>_{\omega}
\end{equation} The essence of the method
is as follows~\cite{kuz3}: \\ It is based on the notion of the
{\it "IRREDUCIBLE"} parts of GFs (or the irreducible parts of the
operators, $A$ and $A^{\dagger}$, out of which the GF is
constructed) in terms of which it is possible, without recourse
to a truncation of the hierarchy of equations for the GFs, to
write down the exact Dyson equation and to obtain an exact
analytic representation for the self-energy operator. By
definition, we introduce the irreducible part {\bf (ir)} of the GF
\begin{equation}
\label{eq.30} ^{(ir)}<<[A, H]_{-}\vert A^{\dagger}>> = <<[A,
H]_{-} - zA\vert A^{\dagger}>>
\end{equation}
The unknown constant z is defined by the condition (or constraint)
\begin{equation}
\label{eq.31} <[[A, H]^{(ir)}_{-}, A^{\dagger}]_{\eta}> = 0
\end{equation}
which is an analogue of the orthogonality condition in the Mori
formalism (  see ref.\cite{lee}). From the condition
(\ref{eq.31}) one can find:
\begin{equation}
\label{eq.32} z = \frac{<[[A, H]_{-}, A^{\dagger}]_{\eta}>}{<[A,
A^{\dagger}]_{\eta}>} =
 \frac{M_{1}}{M_{0}}
\end{equation}
Here $M_{0}$ and $M_{1}$ are the zeroth and first order moments
of the spectral density. Therefore, the irreducible GFs  are
defined so that they cannot be reduced to the lower-order ones by
any kind of decoupling. It is worth  noting that the term {\it
"irreducible"} in a group theory means a representation of a
symmetry operation that cannot be expressed in terms of lower
dimensional representations. Irreducible (or connected )
correlation functions are known in statistical mechanics ({\it
cf.}\cite{strin}). In the diagrammatic approach, the irreducible
vertices are defined as  graphs that do not contain inner parts
connected by the $G^{0}$-line. With the aid of the definition
(30) these concepts are translated into the language of retarded
and advanced GFs. This procedure extracts all relevant (for the
problem under consideration) mean-field contributions and puts
them into the generalized mean-field GF which  is defined here as
\begin{equation}
\label{eq.33} G^{0}(\omega) = \frac{<[A,
A^{\dagger}]_{\eta}>}{(\omega - z)}
\end{equation}
To calculate the IGF $\quad  ^{(ir)}<<[A, H]_{-}(t),
A^{\dagger}(t')>>$ in (\ref{eq.29}), we have to write the
equation of motion for it after differentiation with respect to
the second time variable $t'$. It should be noted that the trick
of two-time differentiation with respect to the first time $t$
and second time $t'$ (in one equation of motion) was
introduced for the first time by Tserkovnikov~\cite{tser1}.\\
The condition of orthogonality (31) removes the inhomogeneous
term from this equation and is a very crucial point of the whole
approach. If one introduces the irreducible part for the
right-hand side operator as discussed above for the ``left"
operator, the equation of motion (\ref{eq.29}) can be exactly
rewritten in the following form \begin{equation} \label{eq.34} G
= G^{0} + G^{0}PG^{0}
\end{equation} The scattering operator $P$ is given by
\begin{equation}
\label{eq.35} P = (M_{0})^{-1}(\quad ^{(ir)}<<[A,
H]_{-}\vert[A^{\dagger}, H]_{-}>>^{(ir)}) (M_{0})^{-1}
\end{equation}
The structure of  equation (34) enables us to determine the
self-energy operator $M$, by  analogy with the diagram technique
\begin{equation} \label{eq.36} P = M + MG^{0}P
\end{equation} From
the definition (\ref{eq.36}) it follows that  the self-energy
operator $M$ is defined as a proper (in the diagrammatic language,
``connected") part of the scattering operator $M = (P)^{p}$. As a
result, we obtain the exact Dyson equation for the thermodynamic
double-time Green functions:  \begin{equation} \label{eq.37} G =
G^{0} + G^{0}MG
\end{equation}
The difference between $P$ and $M$ can be regarded as two
different solutions of two integral equations (\ref{eq.34}) and
(\ref{eq.37}). But from  the Dyson equation (\ref{eq.37}) only the
full GF  is seen to be expressed as a  formal solution of the form
\begin{equation}
\label{eq.38}
G = [ (G^{0})^{-1} - M ]^{-1}
\end{equation}
Equation (38) can be regarded as an alternative form of the Dyson
equation (\ref{eq.37}) and the {\it definition} of $M$ provided
that the generalized mean-field GF $G^{0}$ is specified. On the
contrary , for the scattering operator $P$, instead of property
$G^{0}G^{-1} + G^{0}M = 1$, one has the property
$$(G^{0})^{-1} - G^{-1} = P G^{0}G^{-1}$$  Thus, the { \it very functional
form} of the formal solution (38) determines the difference
between $P$ and $M$ precisely. \\ Thus, by introducing
irreducible parts of GF (or  irreducible parts of the operators,
out of which the GF is constructed) the equation of motion
(\ref{eq.29}) for the GF can exactly be  ( but using orthogonality
constraint (\ref{eq.31})) transformed into the Dyson equation for
the double-time thermal GF (\ref{eq.37}). This result is very
remarkable , because  the traditional form of the GF method does
not include this point. Notice that all quantities thus considered
are  exact. Approximations can be generated not by truncating the
set of coupled equations of motions but by a specific
approximation of the functional form of the mass operator $M$
within a self-consistent scheme, expressing $M$ in terms of
initial GF
$$ M \approx F[G]$$
Different approximations are relevant to different physical
situations.\\The projection operator technique~\cite{forst} has
essentially the same philosophy, but with using the constraint
(\ref{eq.31}) in our approach we emphasize the fundamental and
central role of the Dyson equation for the calculation of
single-particle properties of  many-body systems. The problem of
reducing the whole hierarchy of equations involving higher-order
GFs by a coupled nonlinear set of integro-differential equations
connecting the single-particle GF to the self-energy operator is
rather nontrivial ( {\it cf.}\cite{strin}). A characteristic
feature of these equations is that, besides the single-particle
GF, they involve also  higher-order GF. The irreducible
counterparts of the GFs, vertex functions, {\it etc},  serve to
identify correctly the self-energy as
$$  M = G^{-1}_{0}  - G^{-1}$$
The integral form of Dyson equation (\ref{eq.37}) gives  $M$ the
physical meaning of a nonlocal and energy-dependent effective
single-particle potential. This meaning can be verified for the
exact self-energy through the diagrammatic expansion for the
causal GF.\\
It is important to note that for the retarded and advanced GFs,
the notion of the proper part $M = (P)^{p}$ is symbolic in
nature~\cite{kuz3}. In a certain sense, it is possible to say that
it is defined here by analogy with the irreducible many-particle
$T$-matrix\cite{strin}. Furthermore, by analogy with the
diagrammatic technique, we can also introduce the proper part
defined as a solution to the integral equation (\ref{eq.36}).
These analogues allow us to understand better the formal structure
of the Dyson equation for the double-time thermal GF but only in
a symbolic form . However, because of the identical form of the
equations for  GFs for all three types ( advanced, retarded, and
causal ), we can convert in each stage of calculations to causal
GF and, thereby, confirm the substantiated nature of definition
(\ref{eq.36})! We therefore should speak of an analogy of the
Dyson equation. Hereafter, we  drop this stipulating, since it
does not cause any misunderstanding. In a sense, the IGF method
is a variant of the Gram-Schmidt orthogonalization procedure (see
Appendix A ).\\It should be emphasized that the scheme presented
above gives just a general idea of the IGF method. A more exact
explanation why one should not introduce the approximation
already in $P$, instead of having to work out $M$, is given below
when working out the application
of the method to  specific problems.\\
The general philosophy of the IGF method is in the separation and
identification of elastic scattering effects and inelastic ones.
This latter point is quite often underestimated, and both effects
are mixed. However, as far as the right definition of
quasi-particle damping is concerned, the separation of elastic
and inelastic scattering processes is believed to be crucially
important for  many-body systems with complicated spectra and
strong interaction.   \\ From a technical point of view, the
elastic GMF renormalizations can exhibit  quite a nontrivial
structure. To obtain this structure correctly, one should
construct the full GF from the complete algebra of  relevant
operators and develop a special projection procedure for
higher-order GFs in accordance with a given algebra. Then the
natural question arises how to select the relevant set of
operators $\{ A_{1}, A_{2}, ... A_{n} \}$ , describing the
"relevant degrees of freedom". The above consideration suggests
an intuitive and heuristic way to the suitable procedure as
arising from an infinite chain of equations of motion
(\ref{eq.14}). Let us consider the column
$$ \pmatrix{ A_{1}\cr  A_{2}\cr \vdots \cr  A_{n}\cr}$$
where
$$ A_{1} = A,\quad A_{2} = [A,H],\quad A_{3} = [[A,H],H], \ldots
A_{n} = [[... [A, \underbrace{H]...H}_{n}]$$ Then the most
general possible Green function can be expressed as a matrix
$$ \hat G = <<\pmatrix{
A_{1}\cr  A_{2}\cr \vdots \cr A_{n}\cr} \vert \pmatrix{
A^{\dagger}_{1}& A^{\dagger} _{2}& \ldots &  A^{\dagger}
_{n}\cr}>>$$

This generalized Green function describes the one-, two- and
$n$-particle dynamics. The equation of motion for it includes, as
a particular case, the Dyson equation for single-particle Green
function, the Bethe-Salpeter equation, which is the equation of
motion for the two-particle Green function and which is an
analogue of the Dyson equation, {\it etc}. The corresponding
reduced equations should be extracted from the equation of motion
for the generalized GF with the aid of the special techniques
such as the projection method and similar techniques. This must
be a final goal towards a real understanding of the true
many-body dynamics. At this point, it is worthwhile to underline
that the above discussion is a heuristic scheme only but not a
straightforward recipe. The specific method of introducing  the
IGFs depends on the form of operators $A_{n}$, the type of the
Hamiltonian, and
 conditions of the problem. The irreducible parts in
higher-order equations and connection with Mori formalism was
 considered by Tserkovnikov~\cite{tser2}. The incorporation
of  irreducible parts in  higher-order equations and connection
with the moment expansion was studied in ref.~\cite{kuz4} ( see
Appendix B ).\\ Here a sketchy form of the IGF method is
presented. The aim  to introduce the general scheme and to lay
the groundwork for generalizations and specific applications is
expounded in the next Sections. We  demonstrate below that the
IGF method is a powerful tool for describing the quasi-particle
excitation spectra, allowing a deeper understanding of elastic and
inelastic quasi-particle scattering effects and corresponding
aspects of damping and finite lifetimes. In the present context,
it provides a clear link between the equation-of-motion approach
and the diagrammatic methods due to derivation of the Dyson
equation (37). Moreover, due to the fact that it allows the
approximate treatment of the self-energy effects on a final stage,
it yields a systematic way of the construction
of approximate solutions.\\
It is necessary to emphasize that there is an intimate connection
between an adequate introduction of  mean fields and internal
symmetries of the Hamiltonian.  To test these ideas further, in
the following Sections,  we  analyze the mean field and
generalized mean field concepts for various many-body systems on
a lattice.
\section{Many-Particle Interacting Systems on a Lattice}
\subsection{Spin Systems on a Lattice}
There exists a big variety of magnetic materials. The group of
magnetic insulators is of a special importance. For the group of
systems considered in this Section, the physical picture can be
represented by a model in which the localized magnetic moments
originating from ions with incomplete shells interact through a
short-range interaction.  Individual spin moments form a regular
lattice.  The first model of a lattice spin system was
constructed to describe  a linear chain of projected electron
spins with nearest-neighbor coupling. This was the famous
Lenz-Izing model which was thought to yield a more sophisticated
description of ferromagnetism than the Weiss uniform molecular
field picture. However, in this model, only one spin component is
significant.  As a result, the system has no collective dynamics.
The quantum states that are eigenstates of the relevant spin
components are stationary states. The collective dynamics of
magnetic systems is of great importance since it is related to the
study of low-lying excitations and their interactions. This is
the main aim of the present consideration. Although the Izing
model was an intuitively right step forward from the uniform
Weiss molecular field picture, the physical meaning of the model
coupling constant remained completely unclear. The concept of the
exchange coupling of  spins of two or more nonsinglet atoms
appeared as a result of the Heitler-London consideration of
chemical bond. This theory and the Dirac analysis of the
singlet-triplet splitting in the helium spectrum  stimulated
Heisenberg to make a next essential step. Heisenberg suggested
that the exchange interaction could be the relevant mechanism
responsible for ferromagnetism.
\subsubsection{ Heisenberg Ferromagnet}
The Heisenberg model of a system of spins on various lattices (
which was actually  written down explicitly by van Vleck ) is
termed the Heisenberg ferromagnet and establishes the origin of
the coupling constant as the exchange energy. The Heisenberg
ferromagnet in  a magnetic field $H$ is described by the
Hamiltonian
\begin{equation}
\label{eq.39} H = -  \sum_{ij}  J(i-j) \vec S_{i} \vec S_{j}
-g\mu_{B}H\sum_{i}S_{i}^{z}
\end{equation}
The coupling coefficient $J(i-j)$ is the measure of the  exchange
interaction between  spins at the lattice sites $i$ and $j$ and
is defined usually to have the property J(i - j = 0) = 0. This
constraint means that only the inter-exchange interactions are
taken into account. However, in some complicated magnetic salts,
it is necessary to consider an "effective" intra-site
(see\cite{maxkuz}) interaction (Hund-rule-type terms). The
coupling, in principle, can be of a more general type
(non-Heisenberg terms). These aspects of  construction of a more
general Hamiltonian are
very interesting, but we do not pause here to give the details.\\
For crystal lattices in which every ion is at the centre of
symmetry, the exchange parameter has the property $$ J(i-j) =
J(j-i)$$  We can rewrite then the Hamiltonian (\ref{eq.39}) as
\begin{equation}
\label{eq.40} H = -  \sum_{ij} J(i-j) ( S^z_{i}S^z_{j} +
 S^+_{i}S^-_{j})
\end{equation}
Here $S^{\pm} = S^x \pm iS^y$ are the raising and lowering spin
angular momentum operators. The complete set of spin commutation
relations is
\begin{eqnarray} \nonumber
[S^{+}_{i},S^{-}_{j}]_{-} = 2S^{z}_{i} \delta_{ij}; \quad
[S^{+}_{i},S^{-}_{i}]_{+} = 2S(S + 1) - 2(S^{z}_{i})^{2};  \\
\nonumber [S^{\mp}_{i},S^{z}_{j}]_{-} = \pm
S^{\mp}_{i}\delta_{ij}; \quad S^{z}_{i} = S(S + 1) -
(S^{z}_{i})^{2} - S^{-}_{i}S^{+}_{i}; \\ \nonumber
(S^{+}_{i})^{2S+1} = 0, \quad (S^{-}_{i})^{2S+1} = 0 \nonumber
\end{eqnarray}
We  omit the term of interaction of the spin with an external
magnetic field for the brevity of notation. The statistical
mechanical problem involving this Hamiltonian was not  exactly
solved, but many approximate
solutions were obtained.\\
To proceed further, it is important to note that for the isotropic
Heisenberg model, the total $z$-component of spin $S^z_{tot} =
\sum_{i}S^z_{i}$ is a constant of motion, i.e. $$ [H,S^z_{tot}] =
0
$$ There are cases when the total spin is not a constant of motion,
as, for instance, for the Heisenberg model with the dipole terms
added.\\Let us define the eigenstate $|\psi_{0}>$  so that
$S^+_{i}|\psi_{0}> = 0$ for all lattice sites $R_{i}$. It is
clear that $|\psi_{0}>$ is a state in which all the spins are
fully aligned and for which $S^z_{i}|\psi_{0}> = S|\psi_{0}>$. We
also have $$ J_{\vec k} = \sum_{i}e^{(i\vec k \vec R_i)} J(i) =
J_{-\vec k}$$, where the reciprocal vectors $\vec k$ are defined
by cyclic boundary conditions. Then we obtain $$ H |\psi_{0}> = -
\sum_{ij} J(i-j)S^2 = - NS^2 J_{0}$$  Here $N$ is the total number
of ions in the crystal. So, for the isotropic Heisenberg
ferromagnet, the ground state $|\psi_{0}>$ has an energy $-NS^2
J_{0}$.\\ The state $|\psi_{0}>$ corresponds to a total spin
$NS$. \\ Let us consider now the first excited state. This state
can be constructed by creating one unit of spin deviation in the
system. As a result, the total spin is $NS - 1$. The state
$$|\psi_{k}> = \frac {1}{\sqrt {(2SN)}}\sum_{j}e^{(i\vec k \vec R_j)}S^-_{j}|\psi_{0}>$$ is
an eigenstate of $H$ which corresponds to a single magnon of the
energy
\begin{equation}
\label{eq.41} \omega^{(fm)}_{0}(k) = 2S (J_{0} - J_{k})
\end{equation}
Note that the role of translational symmetry, i.e. the regular
lattice of spins, is essential, since the state $|\psi_{k}>$ is
constructed from the fully aligned state by decreasing the spin
at each site and summing over all spins with the phase factor
$e^{i\vec k \vec R_j}$. It is easy to verify that
$$<\psi_{k}|S^z_{tot} |\psi_{k}> = NS - 1$$ \\ The above
consideration was possible because we knew the exact ground state
of the Hamiltonian . There are many models where this is not the
case. For example, we do not know the exact ground state of a
Heisenberg ferromagnet with dipolar forces and the ground state
of the Heisenberg antiferromagnet.
\subsubsection{Heisenberg Antiferromagnet}
We now discuss the Heisenberg model of the antiferromagnet which
is more complicated to analyse. The fundamental problem here is
that the exact ground state is unknown. We  consider, for
simplicity, a  two-sublattice structure in which nearest neighbour
ions on opposite sublattices interact through the Heisenberg
exchange. For a system of ions on  two sublattices, the
Hamiltonian is
\begin{equation}
\label{eq.42} H =  J\sum_{m,\delta}  \vec S_{m} \vec S_{m +
\delta} + J\sum_{n,\delta} \vec S_{n} \vec S_{n + \delta}
\end{equation}
Here the notation $m =\vec R_m$ means the position vectors of ions
on one sublattice ($a$) and $n$ for the ions on the other ($b$).
Nearest neighbor ions on different sublattices are a distance $|
\vec \delta|$ apart. ( The anisotropy field $\mu H_{A} ( \sum_m
S^{z}_{m} - \sum_n S^{z}_{n} )$, which is not written down
explicitly, is taken to be parallel to the z axis. ) The simplest
crystal structures that can be constructed from two
interpenetrating identical sublattices are the body-centered and
simple cubic.\\ The exact ground state of this Hamiltonian is not
known. One can use the approximation of taking the ground state
to be a classical ground state,  usually called the Neel state, in
which the spins of the ions on each sublattice are oppositely
aligned along the z axis. However, this state is not even an
eigenstate of the Hamiltonian (\ref{eq.42}). Let us remark that
the total $z$-component of the spin commutes with the Hamiltonian
(\ref{eq.42}). It would be instructive to consider here the
construction of a spin wave theory for the low-lying excitations
of the Heisenberg
antiferromagnet in a sketchy form to clarify the foregoing.\\
To demonstrate  the specifics of Heisenberg antiferromagnet  more
explicitly, it is convenient to rotate the axes of one sublattice
through $\pi$ about the x-axis. This transformation preserves the
spin operator commutation relations and therefore is canonical.
Let us perform the transformation on the $\vec R_{n}$, or
$b$-sublattice
$$ S^z_n \rightarrow - \tilde S^z_n; \quad S^{\pm}_n \rightarrow
\tilde S^{\mp}_n$$ The operators $ S^{\alpha}_m$  and $ \tilde
S^{\beta}_n$  commute, because they refer to different
sublattices.\\ The transformation to the momentum representation
is modified in comparison with the ferromagnet case
$$ S^{\pm}_m = \frac {1}{N}\sum_{\vec q}e^{(\pm i\vec q \vec
R_m)}S^{\pm}_q;\quad \tilde S^{\pm}_m = \frac {1}{N}\sum_{\vec
q}e^{(\mp i\vec q \vec R_m)} \tilde S^{\pm}_q$$  Here $\vec q$ is
the reciprocal lattice vectors for one sublattice, each
sublattice containing $N$ ions. After these transformations,  the
Hamiltonian (\ref{eq.42}) can be rewritten as
\begin{equation}
\label{eq.43} H = \frac {1}{2SN} \sum_{q} 2zJS [(S^-_{q}S^+_{q} +
\tilde S^-_{q} \tilde  S^+_{q}) + \gamma_{q} ( S^+_{q} \tilde
S^+_{q} + S^-_{q} \tilde S^-_{q})]
\end{equation} In
(\ref{eq.43}),  $\gamma_{q} $ is defined as $z\gamma_{q} =
\sum_{m=n.n.}exp({i\vec q \vec R_m})$,  and $z$ is the number of
nearest neighbors; the constant terms and the products of four
operators are omitted. Thus the Hamiltonian of the Heisenberg
antifferomagnet is more complicated than that for the
ferromagnet. Because it contains two types of spin operators that
are coupled together, the diagonalization of (\ref{eq.43}) has its
own specificity. \\ To diagonalize (\ref{eq.43}), let us make a
linear transformation to new operators ( Bogoliubov transformation
)
\begin{equation}
\label{eq.44} S^{+}_{q} = u_q a_q + v_q b^{\dagger}_{q}; \quad
\tilde S^{-}_{q} = u_q b^{\dagger}_{q} + v_q a_q
\end{equation}
with $$[a_q , a^{\dagger}_{q'}] = \delta_{q,q'}; \quad [b_q ,
b^{\dagger}_{q'}] = \delta_{q,q'}$$ The transformation
coefficients $u_{k}$ and $v_{k}$ are purely real. To preserve the
commutation rules for the spin operators
$$ [S^+_k, S^-_{k'}] = 2SN \delta_{k,k'}$$,
they should satisfy $u^{2}(k) - v^{2}(k) = 2SN$. The
transformations from the operators $(S^{+}_{q},\tilde S^{-}_{q})$
to the operators $( a_q, b^{\dagger}_{q})$ give
\begin{eqnarray}
\label{eq.45} [(S^-_{q}S^+_{q} + \tilde S^-_{q} \tilde  S^+_{q}) +
\gamma_{q} ( S^+_{q} \tilde S^+_{q} + S^-_{q} \tilde S^-_{q})] =
\nonumber \\( a^{\dagger}_{q}a_q + b^{\dagger}_{q}b_q )[(
u^{2}(q) + v^{2}(q) ) + 2u_q v_q \gamma_{q}] \nonumber \\ + (
a_{q}b_q + a^{\dagger}_{q}b^{\dagger}_{q} ) [( u^{2}(q) + v^{2}(q)
)\gamma_{q} + 2u_q v_q ] \nonumber \\ + 2u_q v_q \gamma_{q} +
2v^{2}(q)
\end{eqnarray} We  represented Hamiltonian (\ref{eq.43}) as a
form quadratic in the Bose operators $( a_q, b^{\dagger}_{q})$ .
We shall now consider the problem of  diagonalization of this
form\cite{tyab}. To diagonalize (\ref{eq.43}), we should require
that
$$  2u_q v_q + ( u^{2}(q) + v^{2}(q) )\gamma_{q} = 0$$  Then we
obtain
\begin{equation}
\label{eq.46}
 2u^{2}(q) = 2SN {(1 + \kappa_q ) \over \kappa_q }; \quad
2v^{2}(q) = 2SN {(1 - \kappa_q ) \over \kappa_q }
\end{equation}
Here the following notation was introduced: $ \kappa_{q} = \sqrt
{(1 - \gamma^{2}_{q} )}$ and $ 2u_q v_q = - 2SN
\gamma_{q}/\kappa_{q}$ After   the transformation (\ref{eq.44}),
we get, instead  of (\ref{eq.43}),
\begin{equation}
\label{eq.47} H = \sum_{k} \omega^{(afm)}_{0}(k)(
a^{\dagger}_{q}a_q + b^{\dagger}_{q}b_q )
\end{equation}
with
\begin{equation}
\label{eq.48} \omega^{(afm)}_{0}(k) = 2zJS \sqrt {1 -
\gamma^{2}_{k}}
\end{equation}
 Expression (\ref{eq.47}) contains two terms, each with the
same energy spectrum. Thus, there are two degenerate spin wave
modes,  because there can be two kinds of precession of the spin
about the anisotropy direction. The degeneracy is lifted by the
application of an external magnetic field in the $z$ direction,
because in this case the two sublattices become nonequivalent.
These results should be kept in mind when discussing the
quasi-particle many-body dynamics of the spin lattice models.
\subsection{Correlated Electrons on a Lattice}
The importance of intra-atomic correlation effects in determining
the magnetic properties of transition metals and their compounds
and oxides was  recognized many years ago. The essential basis of
 studies of metallic magnetism, namely, that the dominant
physical mechanism responsible for the observed magnetic
properties of the transition metals and their compounds and
alloys is the strong intra-atomic correlation in an otherwise
tight-binding picture, is generally accepted as being most
suitable. The problem of the adequate description of  strongly
correlated electron systems on a lattice was  studied intensively
during the last decade, especially in the context of metallic
magnetism, heavy fermions, and high-$T_c$
superconductivity~\cite{kuz1}. The understanding of the true
nature of  electronic states and their quasi-particle dynamics is
one of the central topics of the current experimental and
theoretical efforts in the field. The source of spin magnetism in
solids is, of course, the Pauli exclusion principle as manifested
in the exchange interaction and higher order mechanism. Of
particular interest is the fact that the Hartree-Fock or mean
field theory, i.e. the theory including exchange but not
correlation effects, invariably overestimates the tendency to
magnetism. This fact obviously complicated the already
complicated problem of magnetism in a metal with the $d$ band
electrons which, as was mentioned above, are really neither
"local" nor "itinerant" in a full sense.\\
The strongly correlated electron systems are systems in which
electron correlations dominate. The theoretical studies of
strongly correlated systems  had as a consequence  the
formulation of two model Hamiltonians which play a central role
in our attempts to get an insight into this complicated problem.
These are the Anderson single-impurity model (SIAM)~\cite{and} and
Hubbard model~\cite{hub}. It was only relatively recently
recognized that  both the models have a very complicated many-body
dynamics, and their "simplicity" manifests itself in the dynamics
of two-particle scattering, as was shown via elegant Bethe-anzatz
solutions. \subsubsection{ Hubbard Model} The model Hamiltonian
 usually referred to as the Hubbard
Hamiltonian\cite{hub},\cite{acqua3}
\begin{equation}
\label{eq.49} H = \sum_{ij\sigma}t_{ij}a^{\dagger
}_{i\sigma}a_{j\sigma} + U/2\sum_{i\sigma}n_{i\sigma}n_{i-\sigma}
\end{equation}
includes the intra-atomic Coulomb repulsion $U$ and the
one-electron hopping energy $t_{ij}$. The electron correlation
forces electrons to localize in the atomic orbitals which are
modelled here by a complete and orthogonal set of the Wannier wave
functions $[\phi({\vec r} -{\vec R_{j}})]$. On the other hand,
the kinetic energy is reduced when electrons are delocalized. The
main difficulty in solving  the Hubbard model correctly is the
necessity of taking into account both these effects
simultaneously. Thus, the Hamiltonian (\ref{eq.49}) is specified
by two parameters: $U$ and the effective electron bandwidth
$$\Delta = (N^{-1}\sum_{ij}\vert t_{ij}\vert^{2})^{1/2}.$$ The
band energy of Bloch electrons $\epsilon(\vec k)$ is defined as
follows $$t_{ij} = N^{-1}\sum_{\vec k}\epsilon(\vec k) \exp[i{\vec
k}({\vec R_{i}} -{\vec R_{j}}],$$ where  $N$ is the number of
lattice sites. It is convenient to count the energy from the
center of gravity of the band, i.e. $t_{ii} = t_{0} =
\sum_{k}\epsilon(k) = 0$ ( sometimes it is useful to retain
$t_{0}$ explicitly ).
\\This conceptually simple model is mathematically very
complicated. The effective electron bandwidth $\Delta$ and
Coulomb intra-site integral $U$ determine   different regimes in
3 dimensions depending on the parameter $\gamma = \Delta/U$. In
addition, the Pauli exclusion principle that does not allow two
electrons of common spin to be at the same site, i.e.
$n^{2}_{i\sigma} = n_{i\sigma}$,  plays a crucial role, and it
should be taking into account properly while  making any
approximations.  It is usually rather a difficult task to find an
interpolating solution for  dynamic properties of the Hubbard
model for various mean particle densities. To solve this problem
with a reasonably accuracy and to describe correctly an
interpolated solution from the ``band" limit ($\gamma \gg 1$) to
the ``atomic" limit ($\gamma \rightarrow 0$), one needs a more
sophisticated approach than usual procedures  developed for
description of the interacting electron gas problem\cite{maha}.
We have evidently  to  improve the early Hubbard theory taking
into account of variety of possible regimes for the model
depending on the electron density, temperature, and values of
$\gamma$. The single-electron GF
\begin{equation}
\label{eq.50} G_{ij\sigma}(\omega) = <<a_{i\sigma} \vert
a^{\dagger}_{j\sigma}>> = N^{-1}\sum_{\vec k}G_{\sigma}({\vec k},
\omega) \exp[-i{\vec k}({\vec R_{i}} -{\vec R_{j}})],
\end{equation}
calculated by Hubbard~\cite{hub}, \cite{hub2}, has the
characteristic two-pole functional structure
\begin{equation}
\label{eq.51} G_{\sigma}(k, \omega) = [ F_{\sigma}(\omega) -
\epsilon(k)]^{-1}
\end{equation}
where
\begin{equation}
\label{eq.52} F^{-1}_{\sigma}(\omega) = \frac{\omega -
(n^{+}_{-\sigma}E_{-} + n^{-}_{-\sigma}E_{+}) - \lambda}{(\omega
- E_{+} - n^{-}_{-\sigma}\lambda) (\omega - E_{-} -
n^{+}_{-\sigma}\lambda) - n^{+}_{-\sigma}n^{-}_{-\sigma}
\lambda^{2}}
\end{equation}
Here  $n^{+} = n$ , $n^{-} = 1 - n$; $E_{+} = U$, $E_{-} = 0$,
and $\lambda$ is a certain function which depends on parameters
of the Hamiltonian. In this approximation, Hubbard took account
of the scattering effect of  electrons with spins $\sigma$ by
electrons with spin $-\sigma$ which are frozen as well as the
"resonance broadening" effect due to the motion of the electrons
with spin $-\sigma$. The "Hubbard III" decoupling procedure
suffered of serious limitations. However, in spite of the
limitations, this solution gave the first clue  to the
qualitative understanding of the   property of narrow-band system
like the metal-insulator transition.
\\ If $ \lambda$ is small ($\lambda \rightarrow 0$), then
expression (\ref{eq.52}) takes the form:
$$ F^{-1}_{\sigma}(\omega) \approx
\frac{n^{-}_{-\sigma}}{\omega - E_{-} - n^{+}_{-\sigma}\lambda} +
\frac{n^{+}_{-\sigma}}{\omega - E_{+} -
n^{-}_{-\sigma}\lambda},$$ which corresponds to  two shifted
subbands with the gap
$$ \omega_{1} - \omega_{2} = (E_{+} - E_{-}) +
(n^{-}_{-\sigma} - n^{+}_{-\sigma})\lambda = U + \lambda
2n^{+}_{-\sigma}. $$  If $\lambda$ is very big, then we obtain
$$F^{-1}_{\sigma}(\omega) \approx \frac{\lambda}
{[(\omega - E_{-})n^{-}_{-\sigma} + (\omega -
E_{+})n^{+}_{-\sigma}] \lambda} = \frac{1}{\omega - (
n^{+}_{-\sigma}E_{+} - n^{-}_{-\sigma}E_{-})}.$$ The latter
solution corresponds to a single band centered at the energy
$\omega \approx n^{+}_{-\sigma}U$.  Thus, this  solution explains
qualitatively the appearance of a gap in the density of states
when the value of the intra-atomic correlation exceeds a certain
critical value, as it was first conjectured
by N. Mott.\\
The two-pole functional structure of the single-particle GF is
easy to understand within the formalism that describes the motion
of electrons in binary alloys~\cite{hub2},\cite{akvar}. If one
introduces the two types of the scattering potentials $t_{\pm}
\approx (\omega - E_{\pm})^{-1}$, then the two kinds of the
t-matrix $ T_{+}$ and $T_{-}$ appears which satisfy the following
system of equations:
$$T_{+} = t_{+} + t_{+}G^{0}_{++}T_{+} + t_{+}G^{0}_{+-}T_{-}$$\\
$$T_{-} = t_{-} + t_{-}G^{0}_{--}T_{-} + t_{-}G^{0}_{-+}T_{+},$$
where $G^{0}_{\alpha\beta}$ is the bare propagator between the
sites with  energies $E_{\pm}$. The solution of this system is of
the following form
\begin{eqnarray}
\label{eq.53} T_{\pm} = \frac{t_{\pm} + t_{\pm}G^{0}_{\pm}t_{\pm}}
{(1 - t_{+}G^{0}_{++})(1 - t_{-}G^{0}_{--}) -
G^{0}_{-+}G^{0}_{+-}t_{+}t_{-}}
 =\nonumber\\
 \frac{t^{-1}_{\mp} + G^{0}_{\pm}}
{( t^{-1}_{+} - G^{0}_{++} )( t^{-1}_{-} - G^{0}_{--}) -
 G^{0}_{-+}G^{0}_{+-}}.
\end{eqnarray}
Thus, by comparing this functional two-pole structure and the
``Hubbard III" solution~\cite{hub2},\cite{akvar}\\
$$\Sigma_{\sigma}(\omega) = \omega - F_{\sigma}(\omega)$$,   it is
possible to identify the ``scattering corrections" and ``resonance
broadening corrections" in the following way:\\
$$F_{\sigma}(\omega) = \frac{\omega(\omega - U) - (\omega -
Un_{-\sigma})A_{\sigma}(\omega)} {\omega - U(1 - n_{-\sigma}) -
A_{\sigma}(\omega)}$$\\ $$A_{\sigma}(\omega) = Y_{\sigma}(\omega)
+ Y_{-\sigma}(\omega) - Y^{*}_{-\sigma}(U - \omega)$$\\
$$Y_{\sigma} = F_{\sigma}(\omega) - G^{-1}_{0\sigma}(\omega);
G_{0\sigma}(\omega) = N^{-1}\sum_{k}G_{k\sigma}(\omega)$$\\ If we
put $A_{\sigma}(\omega) = 0$, we immediately obtain the ``Hubbard
I" solution~\cite{hub}
\begin{equation}
\label{eq.54} G^{(H1)}_{\sigma}(k, \omega) \approx \frac
{n_{-\sigma}}{\omega - U - \epsilon(k)n_{-\sigma}} + \frac {1 -
n_{-\sigma}}{\omega - \epsilon(k) (1 - n_{-\sigma})}
\end{equation}
Despite that this solution is exact in the atomic limit ( $t_{ij}
= 0$), the "Hubbard I" solution has many serious drawbacks. The
corresponding spectral function consists of two $\delta$-function
peaks. The "Hubbard III" solution includes several corrections,
including  scattering corrections which broadens the peaks
and shift them when $U$ is changed. \\
 The ``alloy analogy"
approximation corresponds to $A_{\sigma}(\omega) \approx
Y_{\sigma}(\omega)$. An interesting analysis of the "Hubbard III"
solution was  performed in paper\cite{akvar}.  The Hubbard
sub-band structure was  obtained in an analytic form in the
"Hubbard III" approximation, using the Lorentzian form for the
density of states for non-interacting electrons. This resulted in
an analytical form for the self-energy and the density of states
for  interacting electrons. Note that the ``Hubbard III"
self-energy operator $\Sigma_{\sigma}(\omega)$ is local, i.e.
does not depend on the quasi-momentum.  Another drawback of this
solution is a very inconvenient functional representation of
elastic and inelastic scattering processes. \\ The conceptually
new approach to the theory of very strong but finite electron
correlation for the Hubbard model was proposed by
Roth~\cite{roth}. She clarified microscopically the origination
of the two-pole solution of the single-particle GF in the strongly
correlated limit
\begin{eqnarray}
\label{eq.55} G^{(R)}_{\sigma}(k, \omega) \approx \\ \nonumber
\frac {n_{-\sigma}}{\omega - U - \epsilon(k)n_{-\sigma} -
W_{k-\sigma}( 1 - n_{-\sigma}) } + \frac {1 - n_{-\sigma}}{\omega
- \epsilon(k) (1 - n_{-\sigma}) - n_{-\sigma}W_{k-\sigma}}
\end{eqnarray}
We see that, in addition to a band narrowing effect, there is an
energy shift $W_{k-\sigma}$ given by
\begin{eqnarray}
\label{eq.56} \nonumber n_{\sigma}(1 - n_{\sigma})W_{k\sigma} =
\sum_{ij} t_{ij} <a^{\dagger }_{i\sigma}a_{j\sigma}( 1 -
n_{i-\sigma} - n_{j-\sigma})>
 - \sum_{ij} t_{ij} \exp [ik(j -i)] \\  ( n^{2}_{\sigma} -
<n_{i\sigma}n_{j\sigma}> + <a^{\dagger }_{j-\sigma}a^{\dagger
}_{i\sigma}a_{j\sigma}a_{i-\sigma}> + <a^{\dagger
}_{j-\sigma}a^{\dagger}_{j\sigma}a_{i\sigma}a_{i-\sigma}>)
\end{eqnarray}
 This energy shift
corrects the situation with the "Hubbard I" spectral function and
recovers, in principle, the possibility of describing  the
ferromagnetic solution. Thus, the Roth solution gives an improved
version of "Hubbard I"  two-pole solution and includes the band
shift, that is most important in the case of a nearly-half-filled
band. It is worth noting that this result was a very unusual fact
from the point of view of the standard Fermi-liquid approach,
showing that the naive one-electron approximation of  band
structure calculations is not valid for the description of
electron correlations of lattice
fermions.\\
It is this feature - the strong modification of  single-particle
states by many-body correlation effects - whose importance we
wish to emphasize here.\\ Various attempts were made to describe
the properties of the Hubbard model in both the strong and weak
coupling regimes and to find a better solution ( {\it e.g.}
\cite{harris} - \cite{hein1} ). Different schemes of decoupling of
the equations of motion for the GFs  analysed and compared in
paper\cite{akva0}, when calculating the electron contribution to
the cohesive energy in a narrow band system. These calculations
showed importance of the correlation effects and the right scheme
of approximation.
\\Thus,  a sophisticated many-body technique is to be used
for  calculating  the excitation spectra and other
characteristics at finite temperatures. We shall show here
following  papers~\cite{kuzem1},\cite{kuz2} that  the IGF method
permits us to improve substantially  both the solutions, Hubbard
and Roth, by defining the correct Generalized Mean Fields for the
Hubbard model.

\subsubsection{Single Impurity Anderson Model (SIAM)} The Hamiltonian of SIAM can be
written in the form~\cite{and}
\begin{eqnarray}
\label{eq.57}
 H =\sum_{k\sigma}\epsilon_{k}c^{\dagger}_{k\sigma}c_{k\sigma} +
\sum_{\sigma} E_{0\sigma} f^{\dagger}_{0\sigma}f_{0\sigma} +
U/2\sum_{\sigma}n_{0\sigma}n_{0-\sigma} + \\
\nonumber \sum_{k\sigma}V_{k}(c^{\dagger}_{k\sigma}f_{0\sigma} +
f^{\dagger}_{0\sigma}c_{k\sigma})
\end{eqnarray}
where $c^{\dagger}_{k\sigma}$ and $f^{\dagger}_{0\sigma}$ are,
respectively, the creation operators for conduction and localized
electrons; $\epsilon_{k}$ is the conduction electron energy,
$E_{0\sigma}$ is the localized electron energy level, and $U$  is
the intra-atomic Coulomb interaction at the impurity site;
$V_{k}$ represents the $s-(d)f$ hybridization interaction term
and was written in paper\cite{and} in the following form
\begin{equation}
\label{eq.58} V_{k} = \frac {1}{\sqrt N} \sum_{j} V_{f}(R_{j})
exp(ikR_{j})
\end{equation}
The  hybridization matrix element is $$V_{f}(R_{j}) = \int
\psi^{\dagger}_{k}(\vec r) H^{H-F} \phi(\vec r - \vec R_{j})
dr$$  The use of Hartree-Fock term here is notable, since it
justifies the initial treatment of SIAM in\cite{and} entirely in
the H-F approximation. A number of approaches for  SIAM and other
correlated electronic systems was proposed, aimed at  answering
the  Anderson question: "...whether a real many-body theory would
give answer radically different from the Hartree-Fock
results?"\cite{and}.\\
Our goal is to propose a new combined many-body approach for the
description of  many-body quasi-particle dynamics of SIAM at
finite temperatures. The interplay and competition of the kinetic
energy ($\epsilon_{k}$), potential energy ($U$), and hybridization
($V$)  substantially influence the electronic spectrum. The
renormalized electron energies are temperature-dependent, and
electronic states have finite lifetimes. These effects are
described most suitable by the Green functions method. The
purpose of the present approach is to find the electronic
quasi-particle spectrum renormalized by interactions (U- and
V-terms) in a wide range of temperatures  and model parameters
and to calculate explicitly the  damping of the electronic states.
\subsubsection{Periodic Anderson Model (PAM)}
Let us now consider  a lattice generalization of SIAM,  the
so-called periodic Anderson model (PAM). The basic assumption of
the periodic impurity Anderson model is the presence of two
well-defined subsystems, {\it i.e.} the Fermi sea of nearly free
conduction electrons and the localized impurity orbitals embedded
into the  continuum of conduction electron states ( in rare-earth
compounds, for instance, the continuum is actually a mixture of
$s$, $p$, and $d$ states, and the localized orbitals are $f$
states). The simplest form of PAM
\begin{eqnarray}
\label{eq.59}
 H =\sum_{k\sigma}\epsilon_{k}c^{\dagger}_{k\sigma}c_{k\sigma} +
\sum_{i\sigma} E_{0} f^{\dagger}_{i\sigma}f_{i\sigma} +
U/2\sum_{i\sigma}n_{i\sigma}n_{i-\sigma} + \\
\nonumber \frac {V}{\sqrt N} \sum_{ik\sigma} ( exp(i
kR_{i})c^{\dagger}_{k\sigma}f_{i\sigma} + exp (-ikR_{i})
f^{\dagger}_{i\sigma}c_{k\sigma})
\end{eqnarray}
assumes a one-electron energy level $E_{0}$,  hybridization
interaction $V$, and the Coulomb interaction $U$ at each lattice
site. Using the transformation
$$ c^{\dagger}_{k\sigma} = \frac {1}{\sqrt N} \sum_{j}  exp(-i
kR_{j}) c^{\dagger}_{j\sigma}; \quad c_{k\sigma} = \frac
{1}{\sqrt N} \sum_{j}  exp(ikR_{j})c_{j\sigma}$$ the Hamiltonian
( \ref{eq.59}) can be rewritten in the Wannier representation:
\begin{eqnarray}
\label{eq.60}
 H =\sum_{ij\sigma} t_{ij}c^{\dagger}_{i\sigma}c_{j\sigma} +
\sum_{i\sigma} E_{0} f^{\dagger}_{i\sigma}f_{i\sigma} +
U/2\sum_{i\sigma}n_{i\sigma}n_{i-\sigma} + \\
\nonumber V \sum_{i\sigma} ( c^{\dagger}_{i\sigma}f_{i\sigma} +
f^{\dagger}_{i\sigma}c_{i\sigma})
\end{eqnarray}
If one retains the $k$-dependence of the hybridization matrix
element $V_{k}$ in (\ref{eq.60}), the last term in the PAM
Hamiltonian  describing the hybridization interaction between the
localized impurity states and extended conduction states and
containing the essence of a specificicity  of the Anderson model,
is as follows
$$
\sum_{ij\sigma} V_{ij}( c^{\dagger}_{i\sigma}f_{i\sigma} +
f^{\dagger}_{i\sigma}c_{i\sigma}); \quad V_{ij} = \frac {1}{ N}
\sum_{k} V_{k} exp[ik(R_{j} - R_{i})]$$ The on-site hybridization
$V_{ii}$ is equal to zero for symmetry reasons. A detailed
analysis of the hybridization problem from  a general point of
view and in the context of PAM was made in paper\cite{akva}.  The
Hamiltonian of PAM in the Bloch representation  takes the form
\begin{eqnarray}
\label{eq.61}
 H =\sum_{k\sigma}\epsilon_{k}c^{\dagger}_{k\sigma}c_{k\sigma} +
\sum_{i\sigma} E_{k} f^{\dagger}_{k\sigma}f_{k\sigma} +
U/2\sum_{i\sigma}n_{i\sigma}n_{i-\sigma} + \\
\nonumber  \sum_{k\sigma} V_{k} (
c^{\dagger}_{k\sigma}f_{k\sigma} +
 f^{\dagger}_{k\sigma}c_{k\sigma})
\end{eqnarray}
Note that as compared to the SIAM, the PAM has its own specific
features. This can lead to peculiar magnetic properties for
concentrated rare-earth systems where the criterion for magnetic
ordering depends on the competition between indirect RKKY-type
interaction\cite{akva1} ( not included into SIAM ) and the
Kondo-type singlet-site screening ( contained in SIAM ). The
inclusion of inter-impurity correlations makes the problem  more
difficult. Since these inter-impurity effects play an essential
role in physical behaviour of real
systems\cite{akva1},\cite{akva2}, it is instructive to consider
the two-impurity Anderson model (TIAM) too.
\subsubsection{Two-Impurity Anderson Model ( TIAM )}
The two-impurity Anderson model was  considered by Alexander and
Anderson\cite{and2}.  They  put forward a theory which introduces
the impurity-impurity interaction within a game of parameters.
The Hamiltonian of TIAM reads
\begin{eqnarray}
\label{eq.62}
 H =\sum_{ij\sigma} t_{ij}c^{\dagger}_{i\sigma}c_{j\sigma} +
\sum_{i=1,2\sigma} E_{0i} f^{\dagger}_{i\sigma}f_{i\sigma} +
U/2\sum_{i=1,2\sigma}n_{i\sigma}n_{i-\sigma} + \\
\nonumber  \sum_{i\sigma} (
V_{ki}c^{\dagger}_{i\sigma}f_{i\sigma} +
V_{ik}f^{\dagger}_{i\sigma}c_{i\sigma}) + \sum_{\sigma} (
V_{12}f^{\dagger}_{1\sigma}f_{2\sigma} +
V_{21}f^{\dagger}_{2\sigma}f_{1\sigma})
\end{eqnarray}
where  $E_{0i}$ are the position energies of  localized states (
for simplicity, we consider identical impurities and $s$-type
({\it i.e.} non-degenerate) orbitals: $E_{01} = E_{02} = E_{0}$.
Let us recall that the  hybridization matrix element $V_{ik}$ was
defined in (\ref{eq.58}). As for the TIAM, the situation with the
right definition of the parameters  $V_{12}$ and $V_{ik}$ is not
very clear. The definition of $V_{12}$ in\cite{and2} is the
following
$$ V_{12} = V^{\dagger}_{21} = \int
\phi^{\dagger}_{1}(\vec r) H_{f}\phi_{2}(\vec r)  dr$$ (now
$H_{f}$ without "H-F" mark). The essentially local character of
the Hamiltonian $H_{f}$ clearly shows that $V_{12}$ describes the
direct coupling between nearest neighboring sites ( for a
detailed discussion see\cite{kuz8} where the hierarchy of the
Anderson models was  discussed too ).
\section{Effective
and Generalized Mean Fields}
\subsection{Molecular Field Approximation}
The most common technique for studying the subject of interacting
many-particle systems is to use the mean field theory. This
approximation is especially popular in the theory of
magnetism~\cite{smart}. Nevertheless, it was pointed~\cite{call1}
that
\begin{quote}
"the Weiss molecular field theory plays an enigmatic role in
statistical mechanics of magnetism".
\end{quote}
To calculate the susceptibility and other characteristic
functions of a system of localized magnetic moments, with a given
interaction Hamiltonian, the approximation, termed the "molecular
field approximation" was  used widely. However, it is not an easy
task to give the formal unified definition what the mean field
is. In a sense, the mean field is the umbrella term for a variety
of theoretical methods of reducing  the many-particle problem to
the single-particle one. Mean field theory, that approximates the
behaviour of a system by ignoring the effect of fluctuations and
those spin correlations which dominate the collective properties
of the ferromagnet usually provides a starting and estimating
point only, for studying phase transitions. The mean field
theories miss important features of the dynamics of a system. The
main intention of the mean field theories, starting from the
works of van der Waals and P.Weiss, is to take into account the
cooperative behaviour of a large number of particles. It is well
known that earlier theories of phase transitions based on the
ideas of van der Waals and Weiss lead to predictions which are
qualitatively at variance with results of measurements near the
critical point. Other variants of simplified mean field theories
such as the Hartree-Fock theory for electrons in atoms, {\it etc}
lead to discrepancies of various kinds too. It is therefore
natural to analyze the reasons for such drawbacks of earlier
variants of the mean field theories.
\subsection{Effective Field Theories}
A number of effective field theories which are improved versions
of the "molecular field approximation" were proposed. It is the
purpose of this study to stress a specificity of strongly
correlated many-particle systems on a lattice contrary to
continuum (uniform) systems. Although many important questions
 remain still unresolved, a vision of useful synthesis begins to
emerge. As a workable eye-guide , the set of mean field theories
( most probably incomplete ) is shown in Table 1.
\begin{table}[ht] \label{tab1}
\caption{\it \Large {Evolution of the Mean Field Concept.}}
\begin{tabular*}{\textwidth}{@{\extracolsep{\fill}}lcr}
\hline Type of the mean field&Author&Year\cr \hline Uniform
molecular field\\in dense gases&van der Waals&1873\cr Uniform
internal mean field\\in magnets&P.Weiss&1905\cr Thomas-Fermi
model&L.H.Thomas, E.Fermi&1926-28\cr Uniform mean field\\in
many-electron atoms&D.Hartree , V.Fock&1928-32\cr
Molecular mean field in\\
Heisenberg ferromagnet&F.Bloch&1930\cr
Non-uniform(local) staggered\\
mean field in
antiferromagnet&L.Neel&1932\cr Reaction and cavity field in\\
polar substances&L.Onsager&1936\cr
Stoner mean-field model\\
of band magnetism& E.Stoner&1938 \cr
Slater mean-field model\\
of band antiferromagnetism& J.Slater&1951 \cr BCS-Bogoliubov mean
field
in\\superconductors&N.N.Bogoliubov&1958\cr Tyablikov decoupling\\
for Heisenberg ferromagnet& S.Tyablikov&1959\cr Mean field theory
for SIAM&P.W.Anderson&1961\cr Density Functional Theory\\ for
inhomogeneous electron gas&W.Kohn&1964\cr Callen decoupling\\ for
Heisenberg ferromagnet&H.B.Callen&1964\cr Alloy analogy (mean
field) approximation\\ in strongly correlated
model&J.Hubbard&1964\cr Generalized Mean Fields\\ in Heisenberg
ferromagnet&N.Plakida&1973\cr Spin Glass Mean Field
Model&S.F.Edwards, P.W.Anderson&1975\cr Generalised Mean Fields
in Strongly\\ Correlated Hubbard Model&A.L.Kuzemsky&1975-78\cr
Generalized Mean Fields\\ in Heisenberg
antiferromagnet&A.L.Kuzemsky, D.Marvakov&1990\cr Generalized Mean
Fields\\ for itinerant antiferromagnet&A.L.Kuzemsky&1999\cr
\hline
\end{tabular*}
\end{table}
The meaning of many these entries and terms will become clearer
in the forthcoming discussion and will put them in a clearer
perspective. My main purpose is to elucidate ( at least in the
mathematical structure ) and to give  plausible arguments for the
tendency, which expounded in  Table 1. This tendency shows the
following. The earlier concepts of molecular field were described
in terms of a functional of mean magnetic moments (in magnetic
terminology ) or mean particle densities ( Hartree-Fock field  ).
The corresponding mean-field functional $F[<n>, <S^{z}>]$
describes the {\it uniform} mean field.\\ Actually, the Weiss
model was not based on discrete "spins"  as is well known, but
the uniformity of the mean internal field was the most essential
feature of the model. In the modern language, one should assume
that the interaction between atomic spins $S_{i}$ and its
neighbors is equivalent to a mean ( or molecular ) field, $ M_{i}
= \chi_{0} [ h_{i}^{(ext)} + h_{i}^{(mf)} ] $ and that the
molecular field $ h_{i}^{(mf)} $ is of the form $ h^{(mf)} =
\sum_{i} J(R_{ji})<S_{i}>$ (above $T_{c}$ ). Here $ h^{ext}$ is
an applied conjugate field, $\chi_{0}$ is the response function,
and $ J(R_{ji})$ is an interaction. In  other words, the mean
field approximation reduces the many-particle problem to a
single-site  problem in which a magnetic moment at any site can
be either parallel or antiparallel to the total magnetic field
composed of the applied field and the molecular field. The
average interaction of  $i$ neighbors was taken into account only,
and the fluctuations were neglected. One particular example,
where the mean field theory works relatively well is the
homogeneous structural phase transitions; in this case the
fluctuations are confined in phase space.\\ The next important
step was made by L. Neel~\cite{neel}. He conjectured that the
Weiss internal field might be either positive or negative in
sign. In the latter case, he showed that below a critical
temperature ( Neel temperature ) an ordered arrangement of equal
numbers of oppositely directed atomic moments could be
energetically favorable. This new magnetic structure was termed
antiferromagnetism. It was conjectured that  the two-sublattice
Neel ( classical ) ground
state is formed by  local staggered internal mean fields. \\
There is a number of the "correlated effective field" theories,
that tend to repair the limitations of  simplified mean field
theories. The remarkable and ingenious one is the Onsager
"reaction field approximation"\cite{onsag}. He suggested that the
part of the molecular field on a given dipole moment which comes
from the reaction of  neighboring molecules to the instantaneous
orientation of the moment should not be included into the
effective orienting field. This "reaction field" simply follows
the motion of the moment and thus does not favor one orientation
over another. The meaning of the mean field approximation for the
spin glass problem is  very interesting but specific,  and we
will not discuss it here. A single-site molecular-field model for
randomly dilute ferro- and antiferromagnets in the framework of
the double-time thermal GFs was  presented in paper\cite{tahir}.

\subsection{Generalized Mean Fields}
It was shown~\cite{pei},~\cite{tyab},~\cite{fey} that mean-field
approximations, for example the molecular field approximation for
a spin system, the Hartree-Fock approximation and the
BCS-Bogoliubov approximation for an electron system are
universally formulated by the Peierls-Bogoliubov-Feynman (PBF)
inequality:
\begin{eqnarray}
\nonumber
   - \beta^{-1} ln (Tr e^{(-\beta H)} )\leq \\ -
\beta^{-1} ln (Tr e^{(-\beta H^{mf})}) + \frac {Tr e^{(- \beta
H^{mf})} ( H - H^{mf})}{Tr e^{(- \beta H^{mf})}} \label{eq.63}
\end{eqnarray}
Here $F$  is the free energy, and $H^{mf}$ is a "trial" or a "mean
field" approximating Hamiltonian. This inequality gives the upper
bound of the free energy of a many-body system. It is important to
emphasize that the BCS-Bogoliubov theory of
superconductivity~\cite{bog1},\cite{kuze} was formulated on the
basis of a trial Hamiltonian  which  consists of a quadratic form
of creation and annihilation operators, including "anomalous" (
off-diagonal ) averages~\cite{bog1}. The functional of the mean
field ( for the superconducting single-band Hubbard model ) is of
the following form~\cite{kuze}:
\begin{equation} \label{eq.64}
\Sigma^{c}_{\sigma} =  U \pmatrix{ <a^{\dagger}_{i-\sigma}
a_{i-\sigma}> & -<a_{i\sigma} a_{i-\sigma}> \cr
-<a^{\dagger}_{i-\sigma} a^{\dagger}_{i\sigma}> &
-<a^{\dagger}_{i\sigma} a_{i\sigma}> \cr}
\end{equation}
The "anomalous" off-diagonal terms fix the relevant
BCS-Bogoliubov vacuum and select the appropriate set of
solutions.\\ Another remark about the BCS-Bogolubov mean-field
approach is instructive.  Speaking in physical terms, this theory
involves a condensation correctly, in spite that such a
condensation cannot be obtained by an expansion in the effective
interaction between electrons. Other mean field theories, {\it
e.g.} the Weiss molecular field theory and the van der Waals
theory of the liquid-gas transition are  much less reliable. The
reason why a mean-field theory of the superconductivity in the
BCS-Bogoliubov form is successful would appear to be that the
main correlations in metal are governed by the extreme degeneracy
of the electron gas. The correlations due to the pair
condensation, although they have dramatic effects, are weak (  at
least in the ordinary superconductors ) in comparison with the
typical electron energies, and may be treated in an average way
with a reasonable accuracy. All above remarks have relevance to
ordinary low-temperature superconductors. In high-$T_c$
superconductors, the corresponding degeneracy temperature is much
lower, and the transition temperatures are much higher. In
addition, the relevant interaction  responsible for the pairing
and its strength are unknown. From this point of view, the
high-$T_c$ systems are more complicated. It should be clarified
what governs the scale of temperatures, {\it i.e.} critical
temperature, degeneracy temperature, interaction strength or
their complex combination, {\it etc.} In this way a useful insight
into this extremely complicated problem would be gained.\\
Generalization of the molecular field approximation on the basis
of the PBF inequality is possible when we know a particular
solution of the model ({\it e.g.}, for one-dimensional Ising model
we know the exact solution  in the field). One can use this
solution to get a better approximation than the mean field
theory. There are some other methods of improvement of the
molecular field theory~\cite{marsh}, \cite{kat}. Unfortunately,
these approaches are nonsystematic.\\ From the point of view of
quantum many-body theory, the problem of adequate introduction of
mean fields for system of many interacting particles can be most
consistently investigated in the framework of the IGF method. A
correct calculation of the quasi-particle spectra and their
damping, particularly, for  systems with a complicated spectrum
and strong interaction~\cite{kuz3} reveals, as it will be shown
below, that the generalized mean fields can have very complicated
structure
which  cannot be described by a functional of the mean-particle density.\\
To illustrate the actual distinction of  description of the
generalized mean field in the equation-of-motion method for the
double-time Green functions, let us compare the two approaches,
namely, that of Tyablikov~\cite{tyab} and of Callen~\cite{call}.
We shall consider the Green function $<<S^{+}|S^{-}>> $ for the
isotropic Heisenberg model
\begin{equation}
\label{eq.65} H = - \frac{1}{2} \sum_{ij} J(i-j) \vec S_{i} \vec
S_{j}
\end{equation}
The equation of motion (\ref{eq.14}) for the spin Green function
is of the form
\begin{eqnarray}
\label{eq.66}
\omega <<S_{i}^{+}|S_{j}^{-}>>_{\omega} = \\
\nonumber 2<S^{z}>\delta_{ij} + \sum_{g} J (i-g)
<<S^{+}_{i}S^{z}_{g} - S^{+}_{g}S_{i}^{z}|S_{j}^{-}>>_{\omega}
\end{eqnarray}
The Tyablikov decoupling  expresses the second-order GF in terms
of the first (initial) GF:
\begin{equation}
\label{eq.67} <<S^{+}_{i}S_{g}^{z}|S^{-}_{j}>> =
<S^{z}><<S^{+}_{i}|S^{-}_{j}>>
\end{equation}
This approximation is an RPA-type; it does not lead to the
damping of spin wave excitations ({\it cf.} (\ref{eq.41}) )
\begin{equation} \label{eq.68}
E(q) = \sum_{g} J(i-g)<S^{z}> \exp [i(\vec R_{i} - \vec R_{g})
\vec q ] = 2<S^{z}>(J_{0} - J_{q})
\end{equation}
The reason for this is rather transparent. This decoupling does
not take into account  the {\it inelastic} magnon-magnon
scattering processes. In a sense, the Tyablikov approximation
consists of approximating the commutation relations of spin
operators to the extent of replacing the commutation relation
$[S^{+}_{i},S^{-}_{j}]_{-} = 2S^{z}_{i}\delta_{ij}$
by $ [S^{+}_{i},S^{-}_{j}]_{-} = 2<S^{z}>\delta_{ij}$ .\\
Callen~\cite{call} has proposed an improved decoupling
approximation in the method of Tyablikov in the following form:
\begin{equation} \label{eq.69}
\nonumber \\
<<S^{z}_{g}S_{f}^{+}|B>> \rightarrow <S^{z}><<S^{+}_{f}|B>> -
\alpha <S^{-}_{g}S^{+}_{f}><<S^{+}_{g}|B>>
\end{equation}
Here $ 0 \leq \alpha \leq 1$. To clarify this point, it should be
reminded that for spin $1/2$ ( the procedure was  generalized by
Callen to an arbitrary spin), the spin operator $ S^{z}$   can be
written as $ S^{z}_{g} = S - S^{-}_{g}S^{+}_{g}$ or $S^{z}_{g} =
{1 \over 2} ( S^{+}_{g}S^{-}_{g} - S^{-}_{g}S^{+}_{g})$. It is
easy to show that
$$
S^{z}_{g}  = \alpha S + \frac {1 - \alpha}{2} S^{+}_{g}S^{-}_{g}
- \frac {1 + \alpha}{2} S^{-}_{g}S^{+}_{g}$$  The operator
$S^{-}_{g}S^{+}_{g}$ represents the deviation of  $<S^{z}>$ from
$S$. In the low-temperature region, this deviation is small, and
$\alpha \sim 1$. Similarly, the operator ${1 \over 2} (
S^{+}_{g}S^{-}_{g} - S^{-}_{g}S^{+}_{g})$ represents the
deviation of $<S^{z}>$ from 0. Thus, when $<S^{z}>$ approaches to
zero, one can expect that $\alpha \sim 0$. Thus, in this way, it
is possible to obtain a correction to the Tyablikov decoupling
with either a positive or negative sign, or no correction at all,
or any intermediate value, depending on the choice of  $\alpha$.
The above Callen arguments are not rigorous , for, although the
difference in the operators $S^{+}S^{-}$ and $S^{-}S^{+}$ is
small if $<S^{z}> \sim 0$, each operator makes a contribution of
the order of $S$, and it is each operator which is treated
approximately, not the difference. There are some other drawbacks
of the Callen decoupling scheme. Nevertheless, the Callen
decoupling was the first conceptual attempt to introduce the
interpolation decoupling procedure. Let us note that the choice
of $\alpha = 0$ over the entire temperature range is just the
Tyablikov decoupling (\ref{eq.67}).\\
The energy spectrum for the Callen decoupling is given by
\begin{equation} \label{eq.70}
E(q) = 2<S^{z}> \bigl ((J_{0} - J_{q}) +\frac {<S^{z}>}{NS^2}
\sum_{k} [ J(k) - J(k - q)] N(E(k)) \bigr )
\end{equation}
Here $N(E(k))$ is the Bose distribution function $ N(E(k)) = [
\exp ( E(k)\beta) - 1 ]^{-1}$. This is an implicit equation for
$N(E(k))$, involving the unknown quantity $<S^{z}>$ . For the
latter an additional equation is given~\cite{call}. Thus, both
these equations constitute a set of coupled equations which must
be solved self-consistently for $<S^{z}>$.\\ This formulation of
the Callen decoupling scheme displays explicitly the tendency of
the improved description of the mean field. In a sense, it is
possible to say that the Callen work  dates really the idea of
the generalized mean field within the equation-of-motion method
for double-time GFs, however, in a semi-intuitive form. The next
essential steps were made by Plakida~\cite{plak} for the
Heisenberg ferromagnet and by Kuzemsky~\cite{kuzem1},\cite{kuz2}
for the Hubbard model. As was mentioned above, the correct
definition of Generalized Mean Fields depends on the condition of
the problem, the strength of interaction, the choice of relevant
operators, and on the symmetry requirements.
\subsection{Symmetry Broken Solutions} In many-body
interacting systems, the symmetry is important in classifying
different phases and in understanding  the phase transitions
between them~\cite{bog3}. According to Bogoliubov~\cite{bog3}(
{\it cf.} refs.~\cite{matt}, \cite{bog4},\cite{heis}) in each
condensed phase, in addition to the normal process, there is an
anomalous process (or processes) which can take place because of
the long-range internal field, with a corresponding propagator.
Additionally, the Goldstone theorem\cite{gold} states that, in a
system in which a continuous symmetry is broken ( {\it i.e.} a
system such that the ground state is not invariant under the
operations of a continuous unitary group whose generators commute
with the Hamiltonian ), there exists a collective mode with
frequency vanishing, as the momentum goes to zero. For
many-particle systems on a lattice, this statement needs a proper
adaptation.   In the above form, the Goldstone theorem is true
only if the condensed and  normal phases have the same
translational properties. When translational symmetry is also
broken, the Goldstone mode appears at a zero frequency but at
nonzero momentum, {\it e.g.}, a crystal and a helical
spin-density-wave (SDW)
ordering (see for discussion\cite{mor1}-\cite{mor3}). \\
The anomalous propagators for an interacting many-fermion system
corresponding to the ferromagnetic (FM) , antiferromagnetic
(AFM),  and superconducting (SC) long-range ordering are given by
\begin{eqnarray} \label{eq.71}
FM: G_{fm} \sim <<a_{k\sigma};a^{\dagger}_{k-\sigma}>>\\
\nonumber AFM: G_{afm} \sim <<a_{k+Q\sigma};a^{\dagger}_{k+Q'\sigma'}>>\\
\nonumber SC: G_{sc} \sim <<a_{k\sigma};a_{ - k -\sigma}>>\\
\nonumber
\end{eqnarray} In the SDW case, a particle picks up a momentum $Q - Q'$
from scattering against the periodic structure of the spiral (
nonuniform ) internal field, and has its spin changed from
$\sigma$ to $\sigma'$ by the spin-aligning character of the
internal field.  The Long-Range-Order (LRO) parameters are:
\begin{eqnarray} \label{eq.72}
FM: m = 1/N\sum_{k\sigma} <a^{\dagger}_{k\sigma}a_{k-\sigma}>\\
\nonumber AFM:
M_{Q} = \sum_{k\sigma} <a^{\dagger}_{k\sigma}a_{k+Q-\sigma}> \\
\nonumber SC: \Delta = \sum_{k} <a^{\dagger}_{
-k\downarrow}a^{\dagger}_{k \uparrow}>\\ \nonumber
\end{eqnarray} It is
important to note that the long-range order parameters are
functions of the internal field, which is itself a function of the
order parameter. There is a more mathematical way of formulating
this assertion. According to the paper~\cite{bog3}, the notion
"symmetry breaking" means that the state fails to have the
symmetry that the Hamiltonian has.\\
A true breaking of symmetry can arise only if there are
infinitesimal "source fields".   Indeed, for the rotationally and
translationally invariant Hamiltonian,  suitable source terms
should be added:
\begin{eqnarray} \label{eq.73}
FM:  \varepsilon\mu_{B}
H_{x}\sum_{k\sigma}a^{\dagger}_{k\sigma}a_{k-\sigma}\\ \nonumber
AFM:
\varepsilon \mu_{B} H \sum_{kQ} a^{\dagger}_{k\sigma}a_{k+Q-\sigma}\\
\nonumber SC: \varepsilon v \sum_{k} (a^{\dagger}_{- k \downarrow}
a^{\dagger}_{k \uparrow} + a_{k \uparrow} a_{ - k \downarrow})
\end{eqnarray} where $\varepsilon \rightarrow 0$ is to be taken at
the end of calculations.\\ For example,  broken symmetry
solutions of the SDW type  imply that the vector $Q$ is a measure
of the inhomogeneity or breaking of translational symmetry.  The
Hubbard model is a very interesting tool for  analyzing   the
symmetry broken concept. It is possible to show that
antiferromagnetic state and more complicated states ( {\it e.g.}
ferrimagnetic) can be made eigenfunctions of the self-consistent
field equations within an "extended"  mean-field approach,
assuming that the "anomalous" averages
$<a^{\dagger}_{i\sigma}a_{i-\sigma}>$ determine the behaviour of
the system on the same footing as the "normal" density of
quasi-particles $<a^{\dagger}_{i\sigma}a_{i\sigma}>$.  It is
clear, however, that these "spin-flip" terms break the rotational
symmetry of the Hubbard Hamiltonian. For the single-band Hubbard
Hamiltonian, the averages $<a^{\dagger}_{i-\sigma}a_{i, \sigma}>
= 0$ because of the rotational symmetry of the Hubbard model.  The
inclusion of  "anomalous" averages leads to the so-called
"unresricted" H-F approximation (UHFA).  This type of
approximation was  used sometimes also for the single-band Hubbard
model for  calculating  the density of states. For this aim, the
following definition of UHFA
\begin{equation} \label{eq.74}
n_{i-\sigma}a_{i\sigma} \approx <n_{i-\sigma}>a_{i\sigma} -
<a^{\dagger}_{i-\sigma}a_{i\sigma}>a_{i-\sigma}
\end{equation}
was  used. Thus, in addition to the standard H-F term, the new
 so-called ``spin-flip" terms are retained. This example
clearly shows that the structure of  mean field follows from the
specificity of the problem and should be defined in a proper
way.  So, one needs a properly defined effective Hamiltonian
$H_{\rm eff}$.  In paper~\cite{kuze1} we thoroughly analyzed
 the proper definition of the irreducible GFs which
includes the ``spin-flip" terms for the case of itinerant
antiferromagnetism\cite{raja} of correlated lattice fermions. For
the single-orbital Hubbard model, the definition of the
"irreducible" part should be modified in the following way:
\begin{eqnarray} \label{eq.75}
^{(ir)}<<a_{k+p\sigma}a^{\dagger}_{p+q-\sigma}a_{q-\sigma} \vert
a^{\dagger}_{k\sigma}>>_ {\omega} =
<<a_{k+p\sigma}a^{\dagger}_{p+q-\sigma}a_{q-\sigma}\vert
a^{\dagger}_{k\sigma}>>_{\omega} - \nonumber\\ \delta_{p,
0}<n_{q-\sigma}>G_{k\sigma} -
<a_{k+p\sigma}a^{\dagger}_{p+q-\sigma}> <<a_{q-\sigma} \vert
a^{\dagger}_{k\sigma}>>_{\omega}
\end{eqnarray}
From this definition it follows that this way of introduction of
the IGF broadens the initial algebra of  operators and the
initial set of the GFs.  This means that the ``actual" algebra of
 operators must include the spin-flip terms from the beginning,
namely:  $(a_{i\sigma}$, $a^{\dagger}_{i\sigma}$, $n_{i\sigma}$,
$a^{\dagger}_{i\sigma}a_{i-\sigma})$. The corresponding initial
GF will be of the form
$$\pmatrix{ <<a_{i\sigma}\vert a^{\dagger}_{j\sigma}>> &
<<a_{i\sigma}\vert a^{\dagger}_{j-\sigma}>> \cr
<<a_{i-\sigma}\vert a^{\dagger}_{j\sigma}>> & <<a_{i-\sigma}\vert
a^{\dagger}_{j-\sigma}>> \cr}$$ With this definition, one
introduces the so-called anomalous (off-diagonal) GFs which fix
the relevant vacuum and select the proper symmetry broken
solutions. In fact, this approximation was
 investigated earlier by Kishore and Joshi~\cite{kisore}. They
clearly pointed out that they assumed a system to be magnetised in
the $x$ direction instead of the conventional $z$ axis.  \\ The
problem of finding  the ferromagnetic and antiferromagnetic
"symmetry broken" solutions of the correlated lattice fermion
models within IGF method was investigated in ref.~\cite{kuze1}. A
unified scheme for the construction of Generalized Mean Fields (
elastic scattering corrections ) and self-energy ( inelastic
scattering ) in terms of the Dyson equation was  generalized in
order to include  the "source fields". The "symmetry broken"
dynamic solutions of the Hubbard model  which correspond to
various types of itinerant antiferromagnetism were  discussed.
This approach complements previous studies of microscopic theory
of the Heisenberg antiferromagnet~\cite{kuz9} and clarifies the
 concepts of Neel sublattices for localized and
itinerant antiferromagnetism and "spin-aligning fields" of
correlated lattice fermions.
\section{Quasi-Particle Many Body Dynamics}
In this Section, we  discuss the microscopic view of a dynamic
behaviour of interacting many-body systems on a lattice. It was
 recognized for many years that  the strong
correlation in solids exist between the motions of various
particles ( electrons and ions, {\it i.e.} the fermion and boson
degrees of freedom ) which arise from the Coulomb forces. The most
interesting objects are metals and their compounds. They are
invariant under the translation group of a crystal lattice and
have lattice vibrations as well as electron degrees of freedom.
There are many evidences for the importance of  many-body effects
in these systems. Within the Landau semi-phenomenological theory
it was suggested that the low-lying excited states of an
interacting Fermi gas can be described in terms of a set of {\it
"independent quasi-particles"}. However, this was a
phenomenological approach and did not reveal the nature of
relevant interactions.
\subsection{Green Function Picture of Quasi-Particles}
An alternative way of viewing quasi-particles,  more general and
consistent, is through the Green function scheme of many-body
theory\cite{mah}, which we sketch below for completeness
and for pedagogical reasons.\\
We should mention that there exist a big variety of
quasi-particles in many-body systems. At sufficiently low
temperatures, few quasi-particles are excited, and therefore this
dilute quasi-particle gas is nearly a non-interacting gas in the
sense that the quasi-particles rarely collide. The success of the
quasi-particle concept in an interacting many-body system is
particularly striking because of a great number of various
applications. However, the range of validity of the
quasi-particle approximation, especially for  strongly
interacting lattice systems, was not discussed properly in many
cases. In  systems like  simple metals,   quasi-particles
constitute long-lived, weakly interacting excitations, since
their intrinsic decay rate varies as the square of the dispersion
law, thereby justifying their use as the
building blocks for the low-lying excitation spectrum.\\
Unfortunately, there are many strongly correlated systems on a
lattice for which we do not have at present the truly the
first-principles proof of a similar correspondence of the
low-lying excited states of  noninteracting and interacting
systems, adiabatic switching on of the interaction, a simple
effective mass spectrum, long lifetimes of  quasi-particles, {\it
etc.} These specific features of  strongly correlated systems are
the main reason of why the usual perturbation theory starting from
noninteracting states does not work properly.   Many other subtle
nonanalytic effects which are present even in normal systems have
the similar nature . This lack of a rigorous foundation for the
theory of strongly interacting systems on a lattice is not only a
problem of the mathematical perfectionism, but also that of the
correct physics of interacting systems. \\ As we mentioned
earlier, to describe a quasi-particle correctly, the Green
functions method is a very suitable and useful tool. What concerns
us here are  formal expression for the single-particle GF (38)
  and the corresponding quasi-particle excitation
spectrum. From the equation ( \ref{eq.24}) it is thus seen that
the GF is completely determined by the spectral weight function
$A(\omega)$. The spectral weight function reflects the
microscopic structure of the system under consideration. The
other term in ( \ref{eq.24}) is a separation of the purely
statistical aspects of GF.  From the equation ( \ref{eq.20}) it
follows that the spectral weight function can be written formally
in terms of many-particle eigenstates. Its Fourier transform
origination ( \ref{eq.18}) is then the density of states that can
be reached by adding or
removing a particle of a given momentum and energy.\\
Consider a system of interacting fermions as an example. For a
noninteracting system, the  spectral weight function of the
single-particle GF $G_{k}(\omega) =
<<a_{k\sigma};a^{\dagger}_{k\sigma}>>$ has the simple peaked
structure
$$A_{k}(\omega) \sim \delta (\omega - \epsilon_{k})$$.
For an interacting system, the spectral function $ A_{k}(\omega)$
has no   such a simple peaked structure, but it   obeys the
following conditions
$$ A_{k}(\omega) \ge 0; \quad \int A_{k}(\omega)d\omega =
<[a_{k\sigma},a^{\dagger}_{k\sigma}]_{+}> = 1$$  Thus, we can see
from these expressions that  for a noninteracting system, the sum
rule is exhausted by a single peak. A sharply peaked spectral
function for an interacting system means a long-lived
single-particle-like excitation. Thus, the spectral weight
function was established here as the physically significant
attribute of   GF. The question of how best to extract it from a
microscopic theory is the main aim of the present review. \\The
GF for the non-interacting system is
 $G_{k}(\omega) = (\omega - \epsilon_{k})^{-1}$. For a weakly
interacting Fermi system, we have $G_{k}(\omega) = (\omega -
\epsilon_{k} - M_{k}(\omega))^{-1}$  where $M_{k}(\omega)$ is the
mass operator. Thus, for a weakly interacting system, the
$\delta$-function for $ A_{k}(\omega)$ is spread into a peak of
finite width due to the   mass operator. We have
$$M_{k}(\omega \pm i \epsilon) = Re M_{k}(\omega) \mp Im
M_{k}(\omega) = \Delta_{k}(\omega) \mp \Gamma_{k}(\omega)$$ The
single-particle GF can be written in the form \begin{equation}
\label{eq.76} G_{k}(\omega) = \{\omega - [\epsilon_{k} +
\Delta_{k}(\omega)] \pm \Gamma_{k}(\omega)\}^{-1}
\end{equation}
In the weakly interacting case, we can thus find the energies of
  quasi-particles by looking for the poles of single-particle GF
(\ref{eq.76}) $$ \omega = \epsilon_{k} + \Delta_{k}(\omega) \pm
\Gamma_{k}(\omega)$$. The dispersion relation of a quasi-particle
$$ \epsilon(k) = \epsilon_{k} + \Delta_{k}[ \epsilon(k)] \pm
\Gamma_{k}[ \epsilon(k)]$$ and the lifetime $1/\Gamma_{k}$ then
reflects the inter-particle interaction. It is easy to see the
connection between the width of the spectral weight function and
  decay rate. We can write
\begin{eqnarray}
\label{eq.77} A_{k}(\omega) = (\exp(\beta \omega)  + 1)^{-1} (-i)
[G_{k}(\omega + i \epsilon) - G_{k}(\omega - i \epsilon)] = \\
\nonumber (\exp(\beta \omega ) + 1)^{-1} \frac
{2\Gamma_{k}(\omega)}{ [\omega - (\epsilon_{k} +
\Delta_{k}(\omega))]^{2} + \Gamma^{2}_{k}(\omega)}
\end{eqnarray}
In other words, for this case, the corresponding propagator can be
written in the form
$$ G_{k}(t) \approx \exp(-i\epsilon(k)t ) \exp( -\Gamma_{k}t)$$
This form shows under which conditions, the time-development of
an interacting system can be interpreted as the propagation of a
quasi-particle with a reasonably well-defined energy and a
sufficiently long lifetime. To demonstrate this, we consider the
following conditions
$$\Delta_{k}[\epsilon(k)] \ll \epsilon(k); \quad
\Gamma_{k}[\epsilon(k)] \ll \epsilon(k)$$  Then we can write
\begin{equation}
\label{eq.78} G_{k}(\omega) = \frac {1}{ [\omega -
 \epsilon(k)][1 - \frac {d\Delta_{k}(\omega)}{d\omega}
\vert_{\omega = \epsilon(k)}] + i\Gamma_{k}[\epsilon(k)]}
\end{equation}
where the renormalized energy of excitations is defined by
$$\epsilon(k) = \epsilon_{k} + \Delta_{k}[\epsilon(k)]$$
In this case, we have, instead of ( \ref{eq.77}),
\begin{eqnarray}
\label{eq.79} A_{k}(\omega) = \\ \nonumber [\exp(\beta
\epsilon(k)) + 1]^{-1} [1 - \frac {d\Delta_{k}(\omega)}{d\omega}
\vert_{\epsilon(k)}]^{-1} \frac {2\Gamma(k)}{ (\omega -
 \epsilon(k)   )^{2} + \Gamma^{2}(k)}
\end{eqnarray}
As a result, we find
\begin{eqnarray}
\label{eq.80} G_{k}(t) = <<a_{k\sigma}(t);a^{\dagger}_{k\sigma}>> = \\
\nonumber = - i \theta(t) \exp(-i\epsilon(k)t) \exp( -\Gamma(k)t)
[1 - \frac {d\Delta_{k}(\omega)}{d\omega}
\vert_{\epsilon(k)}]^{-1}
\end{eqnarray}
A widely known strategy to justify this line of reasoning is the
perturbation theory\cite{mah}.  A detailed analysis of various
successful approximations for the determination of   excited
states in the framework of the quasi-particle concept and the
Green functions method in metals, semiconductors, and insulators
was done in review paper\cite{wilk}. \\ There are examples of
weakly interacting systems, {\it i.g.} the superconducting phase,
which are not connected perturbatively with   noninteracting
systems. Moreover, the superconductor is a system in which the
interaction between electrons qualitatively changes the spectrum
of excitations. However, quasi-particles are still of use even in
this case, due to the correct redefinition of the relevant {\it
generalized mean field}  which includes the anomalous averages
(see (\ref{eq.72})). In a strongly interacted system on a lattice
with complex spectra,  the concept of a quasi-particle needs a
suitable adaptation and a careful examination. It is therefore
useful to have the workable and efficient IGF  method which, as
we shall see, permits one to determine and correctly separate the
elastic and inelastic scattering renormalizations through a
correct definition of the generalized mean field and to calculate
real quasi-particle spectra, including the damping and lifetime
effects. A careful analysis and   detailed presentations of the
IGF method   will provide an important step to the formulation of
the consistent theory of strongly interacting systems and   the
justification of  approximate methods presently used within
equation-of-motion approaches. These latter remarks will not be
substantiated until next Sections, but it is important to
emphasize that the development which follows is not a  merely
formal exercise but essential for the proper and   consistent
theory of strongly interacting many-body systems on a lattice.
\subsection{Spin-Wave Scattering Effects in Heisenberg Ferromagnet}
In this Section, we briefly describe , mainly for pedagogical
reasons, how the formulation of the quasi-particle picture depends
in an essential way on an analysis of the sort introduced in
Section 3.1. We consider here the most studied case of a
Heisenberg ferromagnet\cite{plak} with the Hamiltonian
(\ref{eq.65}) and the equation of motion (\ref{eq.66}). In an
earlier discussion in Sections 4.11 and 5.3,  we   described the
Tyablikov decoupling procedure (\ref{eq.67}) based on replacing
$S_{i}^{z}$ by $<S_{i}^{z}>$ in the last term of (\ref{eq.66}).
We also discussed  an alternative method of decoupling proposed by
Callen (\ref{eq.69}). Both these decoupling procedures retain
only the elastic spin-wave scattering effects. But for our
purposes, it is essential to retain  also the inelastic scattering
effects,   and therefore, we must carefully identify and separate
the elastic and inelastic spin-wave scattering. This is directly
related with the correct definition of   {\it generalized mean
fields}. Thus, the purpose of the present consideration is to
justify the use of IGF method
for the self-consistent theory of spin-wave interactions.\\
The irreducible part of   GF is introduced according to the
definition (\ref{eq.30}) as
\begin{equation}
\label{eq.81} ^{(ir)}<< (S^{+}_{i}S^{z}_{g} - S^{+}_{g}S_{i}^{z})
\vert S^{-}_{j}>> = <<(S^{+}_{i}S^{z}_{g} - S^{+}_{g}S_{i}^{z}) -
A_{ig}S^{+}_{i} - A_{gi}S^{+}_{g}|S_{j}^{-}>>
\end{equation}
Here the unknown quantities $A_{ig}$ are defined on the basis of
orthogonality constraint (\ref{eq.31})
$$ <[(S^{+}_{i}S^{z}_{g} - S^{+}_{g}S_{i}^{z})^{(ir)},S_{j}^{-}]>
= 0$$ We have ($i \ne g$)
\begin{equation}
\label{eq.82} A_{ig} = A_{gi} =   \frac { 2<S^{z}_{i}S^{z}_{g}> +
<S^{-}_{i}S_{g}^{+}>}{2<S^{z}>}
\end{equation}
The definition (see eq.(\ref{eq.33}) ) of a generalized mean field
GF $G^{MF}$ is given by the equation
\begin{equation}
\label{eq.83} \omega G^{MF}_{ij} = 2<S^{z}> \delta_{ij} + \sum_{g}
J_{ig} A_{ig} ( G^{MF}_{ij} -  G^{MF}_{gj} )
\end{equation}
From the Dyson equation in the form (\ref{eq.37}) we find
\begin{eqnarray}
\label{eq.84}  M_{ij} = ( P_{ij})^{p} = \\ \nonumber <2S^{z}>^{-2}
\sum_{gl} J_{ig} J_{lj}<<(S^{+}_{i}S^{z}_{g} -
S^{+}_{g}S_{i}^{z})^{(ir)}\vert ((S^{+}_{i}S^{z}_{g} -
S^{+}_{g}S_{i}^{z})^{(ir)})^{\dagger}>>^{(p)}
\end{eqnarray}
where the proper $(p)$ part of the irreducible GF is defined by
the equation (\ref{eq.36})
$$
P_{ij} = M_{ij} + \sum_{gl} M_{ig}G^{MF}_{gl}P_{lj}; \quad M_{ij}
= ( P_{ij})^{p}$$ ( in the diagrammatic language, this means that
it has no parts connected by one $G^{MF}$-line). The formal
solution of the Dyson equation is of the form (38):
\begin{eqnarray}
\label{eq.85}
  G_{ij}(\omega) = \\ \nonumber 2<S^{z}>N^{-1} \sum_{k} \exp
[ik(R_{i} - R_{j})] [ \omega - \omega(k) - 2<S^{z}> M_{k}(\omega)
]^{-1}
\end{eqnarray}
The  spectrum of spin excitations in the generalized mean field
approximation is given by
\begin{equation}
\label{eq.86} \omega(k) =  N^{-1}  \sum_{ig} J_{ig} A_{ig} \{1 -
\exp [ik(R_{i} - R_{j})]\}
\end{equation}
Now it is not difficult to see that the result (\ref{eq.86})
includes both the simplest spin-wave dispersion law (\ref{eq.41})
and the result of Tyablikov decoupling (\ref{eq.67}) as the
limiting cases
\begin{eqnarray}
\label{eq.87} \omega(k) = <S^{z}> ( J_{0} - J_{k} ) + \\ \nonumber
(<2S^{z}>N)^{-1} \sum_{q} ( J_{q} -  J_{k-q}) ( \psi^{-+}_{q} + 2
\psi^{zz}_{q})
\end{eqnarray}
where $$ \psi^{-+}_{q} = \sum_{ij} <S^{-}_{i}S_{j}^{+}> \exp
[iq(R_{i} - R_{j})]$$ It is seen that due to the correct
definition of   generalized mean fields (\ref{eq.82}) we get the
spin excitation spectrum in a general way. In the hydrodynamic
limit, it leads to $\omega(k) \sim k^{2}$. The procedure is
straightforward, and the details are left as an exercise. \\ Let
us remind that till now no approximation has been made. The
expressions (\ref{eq.84}), (\ref{eq.85}), and (\ref{eq.86}) are
very useful as the starting point for approximate calculation of
the self-energy, a determination of which can only be
approximate. To do this, it is first necessary to express, using
the spectral theorem (\ref{eq.26}), the mass operator
(\ref{eq.84}) in terms of correlation functions

\begin{eqnarray}
\label{eq.88} <2S^{z}> M_{k}(\omega)  =  \\ \nonumber \frac
{1}{2\pi} \int^{ + \infty}_{ - \infty} \frac {d\omega'}{\omega -
\omega'} (\exp  (\beta \omega')  - 1) \int^{+ \infty}_{- \infty}
dt \exp (i\omega't) \\ \nonumber N^{-1} \sum_{ijgl} J_{ig} J_{lj}
\exp [ik(R_{i} - R_{j})] \\ \nonumber  \frac {1}{<2S^{z}>}
<((S^{+}_{l}(t)S^{z}_{j}(t) - S^{+}_{j}(t)S_{l}^{z}(t)
)^{(ir)})^{\dagger} \vert (S^{+}_{i}S^{z}_{g} -
S^{+}_{g}S_{i}^{z})^{(ir)}>^{(p)}
\end{eqnarray}
This representation is exact, and only the algebraic properties
were used to derive it.  Thus, the expression for the analytic
structure of the single-particle GF ( or the propagator ) can be
deduced without any approximation.  A characteristic feature of
eq.(\ref{eq.84}) is that it involves the higher-order GFs. A whole
hierarchy of equations involving higher-order GFs could thus be
rewritten compactly.   Moreover, it   not only gives a convenient
alternative representation, but avoids some of the algebraic
complexities of   higher-order Green-function theories. Objective
of the present consideration is to give a plausible
self-consistent scheme of the approximate calculation of the
self-energy within the IGF method. To this end, we should express
the higher-order GFs in terms of the initial ones, {\it i.e.}
find the relevant approximate functional form
$$ M \approx F[G]$$
It is clear that this can be done in many ways. As a start, let us
consider how to express higher-order correlation function in
(\ref{eq.88})  in terms of the low-order ones. We use the
following form\cite{plak}
\begin{eqnarray}
\label{eq.89} <((S^{+}_{l}(t)S^{z}_{j}(t) -
S^{+}_{j}(t)S_{l}^{z}(t) )^{(ir)})^{\dagger} \vert
(S^{+}_{i}S^{z}_{g} - S^{+}_{g}S_{i}^{z})^{(ir)}>^{(p)} \approx
\\ \nonumber \psi^{zz}_{jg}(t) \psi^{-+}_{li}(t) -
\psi^{zz}_{lg}(t) \psi^{-+}_{ji}(t) - \psi^{zz}_{ji}(t)
\psi^{-+}_{lg}(t) + \psi^{zz}_{li}(t) \psi^{-+}_{jg}(t)
\end{eqnarray}
We find
\begin{eqnarray}
\label{eq.90} <2S^{z}> M_{k}(\omega)  =  \\ \nonumber \frac
{1}{2\pi} \int^{ + \infty}_{ - \infty} \frac {d\omega'}{\omega -
\omega'} (\exp  (\beta \omega')  - 1) \int^{+ \infty}_{- \infty}
dt \exp (i\omega't) \\ \nonumber N^{-1} \sum_{ijgl} J_{ig} J_{lj}
\exp [ik(R_{i} - R_{j})] \\ \nonumber  \frac {1}{<2S^{z}>} \Bigl (
\psi^{zz}_{jg}(t) \psi^{-+}_{li}(t) - \psi^{zz}_{lg}(t)
\psi^{-+}_{ji}(t) - \psi^{zz}_{ji}(t) \psi^{-+}_{lg}(t) +
\psi^{zz}_{li}(t) \psi^{-+}_{jg}(t) \Bigr )
\end{eqnarray}
It is reasonable to approximate the longitudinal correlation
function by its static value  $ \psi^{zz}_{ji}(t) \approx
\psi^{zz}_{ji}(0)$. The transversal spin correlation functions
are given by the expression
\begin{eqnarray}
\label{eq.91} \psi^{-+}_{ji}(t) = \\ \nonumber \int^{\infty}_{-
\infty} {d\omega
 \over 2\pi}[\exp (\beta \omega ) - 1]^{-1}
\exp (i\omega t )(-2 Im <<S^{+}_{i}|S_{j}^{-}>>_{\omega  +
i\epsilon})
\end{eqnarray}
After the substitution of eq.( \ref{eq.91}) into eq.(
\ref{eq.90})  for the self-energy, we find an approximate
expression in the self-consistent form, which, together with the
exact Dyson equation (\ref{eq.85}), constitute a self-consistent
system of equations for the calculation of the GF. As an example,
we start the calculation procedure ( which can be made iterative
) with the simplest first "trial" expression
$$(-2 Im <<S^{+}_{i}|S_{j}^{-}>>_{\omega + i\epsilon}) \approx
\delta ( \omega - \omega(k))$$  After some algebraic
transformations we find
\begin{equation}
\label{eq.92} <2S^{z}> M_{k}(\omega)  \approx N^{-1} \sum_{q} (
J_{q} -  J_{k-q})^{2} (\omega - \omega(q-k))^{-1} \psi^{zz}_{q}
\end{equation}
This expression gives a  compact representation for the
self-energy of the spin-wave propagator in a Heisenberg
ferromagnet. The above calculations show that the inelastic
spin-wave scattering effects influence  the single-particle
spin-wave excitation energy
$$\omega(k, T) = \omega(k) + Re M_{k}(\omega(k) )$$ and the energy
width
$$\Gamma_{k}( T) =  Im M_{k}(\omega(k) )$$
Both these   quantities are observable, in principle, via the
ferromagnetic resonance or inelastic scattering of neutrons.
There is no  time   to go into details of this aspect of
spin-wave interaction effects. It is worthy to note only that it
is well known that spin-wave interactions in ferromagnetic
insulators have a relatively well-established theoretical
foundation, in contrast to the situation with antiferromagnets.
\section
{Heisenberg Antiferromagnet at Finite Temperatures} As it is
mentioned above, in this article, we describe the foundation of
the IGF method, which is based on the equation-of-motion
approach. The strength of this approach lies in its flexibility
and applicability to systems with complex spectra and strong
interaction. The microscopic theory of the Heisenberg
antiferromagnet is of great interest from the point of view of
application to any novel many-body technique. This is not only
because of the interesting nature of the phenomenon itself but
also because of the intrinsic difficulty of solving the problem
self-consistently in a wide range of temperatures.  In this
Section, we  briefly describe how the generalized mean fields
should be constructed for the case of the Heisenberg
antiferromagnet, which become very complicated  when one uses
other many-body methods, like the diagrammatic
technique~\cite{halper}. Within our IGF scheme, however, the
calculations of quasi-particle spectra seem feasible and very
compact.
\subsection{Hamiltonian of the Model}
The problem to be considered is the many-body quasi-particle
dynamics of the system described by the Hamiltonian~\cite{tyab}
\begin{equation}
\label{eq.93} H = - \frac {1}{2} \sum_{ij}\sum_{\alpha \alpha'}
J^{\alpha \alpha'}(i-j) \vec S_{i \alpha} \vec S_{j \alpha'} = -
\frac {1}{2} \sum_{q}\sum_{\alpha \alpha'} J_{q}^{\alpha \alpha'}
\vec S_{q \alpha} \vec S_{-q \alpha'}
\end{equation}
This is the Heisenberg-Neel model of an isotropic two-sublattice
antiferromagnet (the notation is slightly more general than in
Section 4.1.2 ). Here  $S_{i \alpha}$ is a spin operator situated
on site $i$ of sublattice $\alpha$, and $J^{\alpha \alpha'}(i-j)$
is the exchange energy between atoms on sites $R_{i\alpha}$ and
$R_{j\alpha'}$; $\alpha, \alpha'$ takes two values $(a,b)$ . It is
assumed that all of the atoms on sublattice $\alpha$  are
identical, with spin magnitude $S_{\alpha}$. It should be noted
that, in principle, no restrictions are placed in the Hamiltonian
(\ref{eq.93}) on the number of sublattices, or the number of sites
on a sublattice. What is important is that sublattices are to be
distinguished   on the basis of differences in local magnetic
characteristics rather than merely differences in
geometrical or chemical characteristics.\\
Let us introduce the spin operators $S^{\pm}_{i \alpha} = S^{x}_{i
\alpha} \pm iS^{y}_{i \alpha}$. Then the commutation rules for
spin operators are
$$
[S^{+}_{i\alpha},S^{-}_{j\alpha'}]_{-} = 2(S^{z}_{i\alpha})
\delta_{ij}\delta_{\alpha \alpha'};  \quad
[S^{\mp}_{i\alpha},S^{z}_{j\alpha'}]_{-} = \pm
S^{\mp}_{i\alpha}\delta_{ij}\delta_{\alpha \alpha'} $$ For an
antiferromagnet, an exact ground state is not known.
Neel~\cite{neel} introduced the model concept of two mutually
interpenetrating sublattices to explain the behaviour of the
susceptibility of antiferromagnets. However, the ground state in
the form of two sublattices ( the Neel state ) is only a classical
approximation. In contrast to ferromagnets, in which the mean
molecular field is approximated relatively reasonably by a
function homogeneous and proportional to the magnetisation, in
ferri- and antiferromagnets, the mean molecular field is strongly
inhomogeneous. The local molecular field of Neel~\cite{neel} is a
more general concept. Here, we present the
calculations~\cite{kuz9} of the quasi-particle spectrum and
damping of a Heisenberg antiferromagnet in the framework of the
IGF method.\\ In what follows, it is convenient to rewrite
(\ref{eq.93}) in the form
\begin{equation}
\label{eq.94} H = - \frac {1}{2} \sum_{q}\sum_{\alpha \alpha'}
I_{q}^{\alpha \alpha'} ( S^{+}_{q \alpha} S^{-}_{-q \alpha'} +
S^{z}_{q\alpha} S^{z}_{-q\alpha'})
\end{equation}
where $$ I^{\alpha \alpha'}_{q} = 1/2( J^{\alpha \alpha'}_{q} +
J^{\alpha' \alpha}_{-q})$$  It will be shown that the use of
"anomalous averages" which fix the Neel vacuum  makes it possible
to determine uniquely generalized mean fields and to calculate, in
a very compact manner, the spectrum of spin-wave excitations and
their damping due to inelastic magnon-magnon scattering
processes. A transformation from the spin operators to Bose (or
Pauli ) operators is not required.
\subsection{Quasi-Particle Dynamics of Heisenberg Antiferromagnet}
In this section, to make the discussion more concrete, we
consider the  retarded GF of localized spins  defined as
$G^{AB}(t - t') = <<A(t),B(t')>>$ . Our attention is focused on
the spin dynamics of the model. To describe the spin dynamics of
the model ( \ref{eq.94}) self-consistently,  one should take into
account the full algebra of relevant operators of the suitable
"spin modes" ( "relevant degrees of freedom" ) which are
appropriate for the case. This relevant algebra should be
described by the 'spinor' $A = { S^{+}_{ka}\choose S^{+}_{kb}}$,
$ B = A^{\dagger}$
, according to the IGF strategy of Section 3. \\
Once this has been done, we must introduce the generalized matrix
GF of the form
\begin{equation}
\label{eq.95} \pmatrix{ <<S^{+}_{ka}\vert S^{-}_{-ka}>> &
<<S^{+}_{ka}\vert S^{-}_{-kb}>> \cr <<S^{+}_{kb}\vert
S^{-}_{-ka}>> & <<S^{+}_{kb}\vert S^{-}_{-kb}>> \cr} = \hat
G(k;\omega)
\end{equation} To show the advantages of the IGF
in the most full form, we carry out the calculations in the matrix form.\\
To demonstrate the utility of the IGF method, we consider the
following steps in a more detailed form. Differentiating the GF
$<<S^{+}_{ka} \vert B >> $ with respect to the first time, $t$,
we find
\begin{eqnarray}
\label{eq.96} \omega<<S^{+}_{ka} \vert {S^{-}_{-ka} \choose
S^{-}_{-kb}} >>_{\omega} =
\\
{2<S^{z}_{a}> \brace 0} +  \frac {1}{N^{1/2}}
\sum_{q}I^{ab}_{q}<<S^{ab}_{kq} \vert B_{ab}>>_{\omega} \nonumber \\
 +  \frac {1}{N^{1/2}} \sum_{q}I^{aa}_{q}<<S^{aa}_{kq} \vert
B_{ab}>>_{\omega} \nonumber
\end{eqnarray}
where $S^{ab}_{kq} = (S^{+}_{k-q,a}S^{z}_{qb} - S^{+}_{qb}S^{z}_{k-q,a})$.\\
In (\ref{eq.96}),  we   introduced the notation
$$ B_{ab} = {S^{-}_{-ka} \brace S^{-}_{-kb}}; \quad B_{ba} = {S^{-}_{-kb} \brace
S^{-}_{-ka }}$$\\  Let us define the irreducible $(ir)$ operators
as (equivalently, it is possible to define the irreducible GFs)
\begin{eqnarray}
\label{eq.97}
(S^{ab}_{kq})~^{(ir)} = S^{ab}_{kq} - A^{ab}_{q}S^{+}_{ka} + A^{ba}_{k-q}S^{+}_{kb}  \\
\label{eq.98} (S^{z}_{q\alpha})^{~(ir)} = S^{z}_{q\alpha} -
N^{1/2} <S^{z}_{\alpha}>\delta_{q,0}
\end{eqnarray}
The choice of the irreducible parts is uniquely determined by the
"orthogonality" constraint ( 31)
\begin {equation}
\label{eq.99} < \bigl [(S^{ab}_{kq})~^{(ir)} ,{ S^{-}_{-ka}
\choose S^{-}_{-kb}} \bigr ]> = 0
\end {equation}
From eq.(\ref{eq.99}) we find that
\begin{equation}
\label{eq.100} A^{ab}_{q} = \frac { 2< (S^{z}_{-qa})^{(ir)}
(S^{z}_{qb})^{(ir)}> + < S^{-}_{-qa}S^{+}_{qb} >} {
2N^{1/2}<S^{z}_{a}>}
\end{equation}
By using the definition of the irreducible parts (\ref{eq.97}),
the equation of motion (\ref{eq.96}) can be exactly transformed to
the following form
\begin{eqnarray}
\label{eq.101} (\omega - \omega_{aa} ) <<S^{+}_{ka} \vert B_{ab}
>>_{\omega} + \omega_{ab}<<S^{+}_{kb} \vert
B_{ab}>>_{\omega} = \\ \nonumber
{2<S^{z}_{a}> \brace 0} +
<< \Phi_{a}^{~(ir)}(k) \vert B_{ab}>>_{\omega} \\
\label{eq.102} (\omega - \omega_{bb} ) <<S^{+}_{kb} \vert B_{ba}
>>_{\omega} + \omega_{ba}<<S^{+}_{ka} \vert
B_{ba}>>_{\omega} = \\ \nonumber {2<S^{z}_{b}> \brace 0} + <<
\Phi_{b}^{~(ir)}(k) \vert B_{ba}>>_{\omega}
\end{eqnarray}
The following notation was used:
\begin{eqnarray}
\label{eq.103} \omega_{aa} = \Bigl( (I_{0}^{aa} -
I_{k}^{aa})<S^{z}_{a}> + I^{ab}_{0}<S^{z}_{b}> + \\ \nonumber
\sum_{q} [(I^{aa}_{q} - I^{aa}_{k-q}) A^{aa}_{Nq} + I^{ab}_{q} A_{Nq}^{ab} ] \Bigr)  \\
\label{eq.104} \omega_{ab} = \Bigl(  I_{k}^{ab}<S^{z}_{a}> +
\sum_{q} I^{ab}_{k-q} A^{ba}_{Nq}  \Bigr )   \\
\label{eq.105}
A_{Nq}^{\alpha \beta} = N^{-1/2} A_{q}^{\alpha \beta}  \\
\label{eq.106} \Phi^{(ir)}_{a} ( k ) = \\ \nonumber N^{- 1/2}
\sum_{q} \sum_{ \gamma = a,b} I^{\alpha \gamma}_{q} [
S^{+}_{k-q,a}(S^{z}_{q\gamma})^{~(ir)} - S^{+}_{q\gamma}(
S^{z}_{k-q,a})^{(ir)} ]^{(ir)}
\end{eqnarray}
To calculate the irreducible GFs on the right-hand sides of eqs.
(\ref{eq.101}) and (\ref{eq.102}), we use the device of
differentiating with respect to the second time $t'$. After
introduction of the corresponding irreducible parts into the
resulting equations, the system of equations can be represented
in the matrix form which can be identically transformed to the
standard form (\ref{eq.34})
\begin{equation}
\label{eq.107} \hat G(k,\omega) = \hat G_{0}(k,\omega) + \hat
G_{0}(k,\omega) \hat P(k,\omega) \hat G_{0}(k,\omega)
\end{equation}
Here we  introduced the generalized mean-field (GMF) GF $G_{0}$
and the scattering operator $P$ according to the following
definitions
\begin{equation}
\label{eq.108} \hat G_{0} = \hat \Omega^{-1} \hat I
\end{equation}
\begin{eqnarray}
\label{eq.109} \hat P = \\ \nonumber {1 \over 4<S^{z}_{a}>^{2}}
\pmatrix{ <<\Phi^{(ir)}_{a}(k) \vert \Phi_{a}^{(ir) \dagger}(k)>>
& << \Phi_{a}^{(ir)} (k) \vert \Phi_{b}^{(ir) \dagger}(k)>> \cr
<<\Phi_{b}^{(ir)} (k) \vert \Phi_{a}^{(ir) \dagger} (k) >> &
<<\Phi_{b}^{(ir)} (k) \vert \Phi^{(ir) \dagger}_{b} (k) >> \cr}
\end{eqnarray}
 where
\begin{equation} \label{eq.110}
\hat \Omega = \pmatrix{ (\omega - \omega_{aa})&\omega_{ab} \cr
\omega_{ab} & ( \omega - \omega_{bb}) \cr}
\end{equation}
The Dyson equation can be written exactly in the form
(\ref{eq.37}) where the mass operator $M$ is of the form
\begin{equation}
\label{eq.111} \hat M(k,\omega) = ( \hat P(k,\omega)  )^{(p)}
\end{equation}
It follows from the Dyson equation that
$$ \hat P( k,\omega) = \hat M( k,\omega) +
\hat M(k,\omega)\hat G_{0}(k,\omega) \hat P( k,\omega) $$ Thus,
on the basis of these relations, we can speak of the mass operator
$M$ as the proper part of the operator $P$ by analogy with the
diagram technique, in which the mass operator is the connected
part of the scattering operator.  As it is shown in Section 3, the
formal solution of the Dyson equation is of the form (38). Hence,
the determination of the full GF $\hat G$ was reduced to the
determination of $\hat G_{0}$ and $\hat M$.
\subsection{Generalized Mean-Field GF }
From the definition (\ref{eq.108}), the GF matrix in the
generalized mean-field approximation reads
\begin{eqnarray}
\label{eq.112} \hat G_{0} = \\ \nonumber \pmatrix{
G_{0}^{aa}(k,\omega) & G_{0}^{ab}(k,\omega)\cr
G_{0}^{ba}(k,\omega) &G_{0}^{bb}(k,\omega)\cr} = \frac
{2<S^z_a>}{det \hat \Omega} \pmatrix{ (\omega -
\omega_{aa})&\omega_{ab} \cr \omega_{ab} & ( \omega -
\omega_{bb}) \cr}
\end{eqnarray}
where
$$ det \hat \Omega = (\omega - \omega_{aa})(\omega - \omega_{bb})
- \omega_{aa}\omega_{ab}$$  We find the poles of  GF
(\ref{eq.112}) from the equation
$$ det \hat \Omega = 0$$
from which it follows that
\begin{equation}
\label{eq.113} \omega_{\pm}(k) = \pm \sqrt {(\omega^{2}_{aa}(k) -
\omega^{2}_{ab}(k))}
\end{equation}
It is convenient to adopt here the Bogoliubov
$(u,v)$-transformation notation by analogy with that of Section
4.1.2. The elements of the matrix GF $G_{0}(k,\omega)$ are found
to be
\begin{equation}
\label{eq.114} G^{aa}_{0}(k,\omega) = 2<S^{z}_{a}> \Bigl [
\frac{u^{2}(k)}{\omega -\omega_{+}(k) } - \frac{v^{2}(k)}{\omega
- \omega_{-}(k)} \Bigr ] = G^{bb}_{0}(k,-\omega)
\end{equation}
\begin{equation}
\label{eq.115} G^{ab}_{0}(k,\omega) = 2<S^{z}_{a}> \Bigl [ \frac
{-u(k)v(k)}{\omega - \omega_{+}(k)} + \frac {u(k)v(k)}{\omega -
\omega_{-}(k)} \Bigr ] = G^{ba}_{0}(k,\omega)
\end{equation}
where
\begin{eqnarray}
\label{eq.116} u^{2}(k) = 1/2[(1 - \gamma_{k}^{2})^{-1/2} + 1];
\quad v^{2}(k) =
1/2[(1 - \gamma_{k}^{2})^{-1/2} - 1] \nonumber\\
\gamma_{k} = \frac{1}{z} \sum_{i} \exp (ikR_{i}); \quad
I^{aa}_{q} = I^{bb}_{q} = 0
\end{eqnarray}
The simplest assumption is that each sublattice is s.c. and
$\omega_{ \alpha \alpha}(k) = 0 \quad (\alpha = a,b)$. Although
that we   work  in the GFs formalism, our expressions
(\ref{eq.114}), (\ref{eq.115}) are in accordance with the results
of the Bogoliubov (u,v)-transformation , but, of course, the
present derivation is more general. However, it is possible to
say that we diagonalized the generalized mean-field GF by
introducing a new set of operators. We used the notation
\begin{equation}
\label{eq.117} S^{+}_{1}(k) = u_{k}S^{+}_{ka} + v_{k}S^{+}_{kb};
\quad S^{+}_{2}(k) = v_{k}S^{+}_{ka} + u_{k}S^{+}_{kb}
\end{equation}
This notation permits us to write down the results in a compact
and convenient form,   but all calculations can be done in the
initial notation  too.\\
The spectrum of elementary excitations in the GMF approximation
for an arbitrary spin   $S$ is of the form
\begin{equation}
\label{eq.118} \omega(k) = Iz<S^{z}_{a}> \Bigl [ 1 - \frac
{1}{N^{1/2}<S^{z}_{a}>} \sum_{q} \gamma_{q} A^{ab}_{q} \Bigr ]
\sqrt {( 1 - \gamma^{2}_{k} )}
\end{equation}
where  $I_{q} = zI\gamma_{q}$, and $z$ is the number of nearest
neighbors in the lattice. The first term in (\ref{eq.118})
corresponds to the Tyablikov approximation ( {\it
cf.}(\ref{eq.48})). The second term in (\ref{eq.118}) describes
the elastic scattering of the spin-wave quasi-particles. At low
temperatures, the fluctuations of the longitudinal spin components
are small, and, therefore, for (\ref{eq.118}) we obtain
\begin{equation}
\label{eq.119} \omega (k) \approx I S z [ 1 - C(T) ] \sqrt{ (1 -
\gamma_{k}^{2} )}
\end{equation}
The function $C(T)$ determines the temperature dependence of the spin-wave
spectrum
\begin{equation}
\label{eq.120} C(T) = \frac {1}{2 N S^{2}} \sum_{q} (<S^{-}_{-qa}
S^{+}_{qa}> + \gamma_{q} < S^{-}_{-qa} S^{+}_{qb}>) \nonumber
\end{equation}
In the case when $C(T) \rightarrow 0$, we obtain the result of
the Tyablikov decoupling for the spectrum of the antiferromagnons
\begin{equation}
\label{eq.121} \omega(k) \approx I <S^{z}_{a}> z \sqrt {( 1 -
\gamma_{k}^{2}) }
\end{equation}
In the hydrodynamic  limit, when $ \omega(k) \sim D(T)|\vec k|$,
we can conclude that the stiffness constant $D(T) =zIS (1 - C(T))$
for an antiferromagnet decreases with temperature because of the
elastic magnon-magnon scattering as $T^{4}$. To estimate the
contribution of the inelastic scattering processes, it is
necessary to take into account the corrections due to the mass
operator.
\subsection{Damping of Quasi-Particle Excitations}
An antiferromagnet is a system with a complicated quasi-particle
spectrum. The calculation of the damping due to inelastic
scattering processes in a system of that sort has some important
aspects. When calculating the damping, it is necessary to take
into account the contributions from all matrix elements of the
mass operator $M$
$$M =  G^{-1}_{0}  - G^{-1} $$  It is then convenient to use the
representation in which the generalized mean field GF has a
diagonal form . In terms of the new operators $S_{1}$ and $S_{2}$,
the GF $G$   takes the form $$ \tilde G(k;\omega) =
 \pmatrix{ << \tilde S^{+}_{1}(k)\vert \tilde S^{-}_{1}(-k)>> &
<<\tilde S^{+}_{1}(k)\vert \tilde S^{-}_{2}(-k)>> \cr <<\tilde
S^{+}_{2}(k)\vert \tilde S^{-}_{1}(-k)>> & <<\tilde
S^{+}_{2}(k)\vert \tilde S^{-}_{2}(-k)>> \cr} = \pmatrix{ G_{11}&
G_{12} \cr G_{21}&  G_{22} \cr} $$ In other words, the damping of
the quasi-particle excitations is determined on the basis of a GF
of the form
\begin{equation}
\label{eq.122}  G_{11}(k,\omega) = \frac {2<S^{z}_{a}>}{\omega -
\omega(k) - 2<S^{z}_{a}> \Sigma(k,\omega)}
\end{equation}
Here, the self-energy operator $\Sigma(k,\omega)$ is determined by the
expression
\begin{equation}
\label{eq.123} \Sigma (k,\omega) = M_{11}(k,\omega) - \frac
{2<S^{z}_{a}> M_{12}(k,\omega) M_{21}(k,\omega) } {\omega +
\omega(k) - 2<S^{z}_{a}> M_{22}(k,\omega)}
\end{equation}
In the case when $ k, \omega \rightarrow 0$, one can be restricted
to the approximation
\begin{equation}
\label{eq.124} \Sigma (k,\omega) \approx  M_{11}(k,\omega) =
u^{2}_{k} M_{aa} + v_{k}u_{k}( M_{ab} + M_{ba} ) + v^{2}_{k}
M_{bb}
\end{equation}
It follows from (\ref{eq.111}) that to calculate the damping, it
is necessary to find the GFs $<<\Phi^{(ir)}_{\alpha}(k) \vert
\Phi^{(ir) \dagger}_{\beta}(k)>>$. As an example, we consider the
calculation of one of them. By means of the spectral theorem
(\ref{eq.27}), we can express the GF in terms of the correlation
function $<\Phi^{(ir) \dagger}_{a}(k)  \Phi^{(ir)}_{a}(k, t)>$.
We have
\begin{eqnarray}
\label{eq.125}
<<\Phi^{(ir)}_{a}(k) \vert \Phi^{(ir) \dagger}_{a}(k)>> = \\
\frac {1}{2\pi} \int^{ + \infty}_{ - \infty} \frac
{d\omega'}{\omega - \omega'} (\exp  (\beta \omega')  - 1) \int^{+
\infty}_{- \infty}  dt \exp (i\omega't) <\Phi^{(ir)
\dagger}_{a}(k)
  \Phi^{(ir) }_{a}(k, t)>  \nonumber
\end{eqnarray}
Thus, it is necessary to find a workable "trial" approximation
for the correlation function on the r.h.s. of (\ref{eq.125}). We
consider an approximation of the following form
\begin{eqnarray}
\label{eq.126}
< (S^{z}_{-qb})^{(ir)} S^{-}_{-(k-q')a} S^{+}_{(k-q')a}(t) (S^{z}_{q'b}(t))^{(ir)}> \approx \\
\frac {1}{4 N S^{2}} \sum_{p} \bigl ( \psi^{-+}_{k-p,aa} (t)
\psi^{ -+}_{q+p,bb}(t) \psi^{+-}_{p,bb} (t) +
\psi^{-+}_{k-q,ab}(t) \psi^{-+}_{q+p,ab}(t) \psi^{+-}_{p,bb}(t)
\bigr )\delta_{q,q'} \nonumber
\end{eqnarray}
where $\psi^{-+}_{q,ab} (t) = < S^{-}_{-qa} S^{+}_{ qb} (t)>$. By
analogy with the diagram technique, we can say that the
approximation (\ref{eq.126}) corresponds to the neglect  of the
vertex corrections to the magnon-magnon inelastic collisions.
Using (\ref{eq.126}) in (\ref{eq.125}), we obtain
\begin{eqnarray}
\label{eq.127} <<\Phi^{(ir)}_{a}(k) \vert \Phi^{(ir)
\dagger}_{a}(k)>> \approx
\\   \nonumber \frac {1}{16 N S^{4}} \sum _{qp} \int \frac
{d\omega_{1} d\omega_{2} d\omega_{3}}
{\omega - \omega_{1} -\omega_{2} + \omega_{3}} F(\omega_{1}, \omega_{2}, \omega_{3}) \\
\nonumber
[-\frac {1}{\pi} Im G_{aa} (k-q, \omega_{1})]
[-\frac {1}{\pi} Im G_{bb} (q+p, \omega_{2})]
[-\frac {1}{\pi} Im G_{bb} (p, \omega_{3})]
\end{eqnarray}
where
\begin{equation}
\label{eq.128} F(\omega_{1}, \omega_{2}, \omega_{3}) =
N(\omega_{2}) [ N(\omega_{3}) - N(\omega_{1})] + [ 1 +
N(\omega_{1})] N(\omega_{3})
\end{equation}
Equations (\ref{eq.37}), (\ref{eq.111}), and (\ref{eq.127})
constitute a self-consistent system of equations. To solve this
system of equations, we can, in principle, use any convenient
initial representation for the GF, substituting it into the
right-hand side of eq. (\ref{eq.127}). The system can then be
solved iteratively. To estimate the damping, it is usually
sufficient, as the first iteration, to use the simplest
single-pole approximation
\begin{equation}
\label{eq.129} -\frac {1}{\pi} Im G (k, \omega) \approx \delta
(\omega - \omega(k))
\end{equation}
As a result, for the damping of the spin-wave excitations we obtain
\begin{eqnarray}
\label{eq.130} \Gamma (k,\omega) = - 2S Im \Sigma(k,\omega) = \\
\nonumber
 = \frac {\pi}{N} (z I )^{2} (1 - e^{(- \beta \omega)} ) \\
\nonumber \sum_{qp} N_{p} ( 1 + N_{q+p} ) ( 1 + N_{k-q} ) M_{11}
(k,p ; k-q, p+q ) \delta ( \omega - \omega ( k-q ) + \omega (p))
\nonumber
\end{eqnarray}
The explicit expression for   $ M_{11}$ is given in
ref.~\cite{kuz9}. In our approach, it is possible to take into
account the inelastic scattering of   spin waves due to
scattering by the longitudinal spin fluctuations too~\cite{kuz9}.
In general, the correct estimates of the temperature dependence
of the damping of antiferromagnons depend  strongly on the reduced
temperature and energy scales and are   rather a nontrivial task.
However, under the normal conditions, the damping is weak $
\omega(k) / \Gamma \sim 10^{2} - 10 ^{3}$, and the
antiferromagnons are the well-defined quasi-particle
excitations\cite{wind}.\\ In summary, in this Section, we have
shown that the IGF method permits us to calculate the spectrum and
the damping for a two-sublattice Heisenberg antiferromagnet in a
wide range of temperatures in a compact and self-consistent way.
At the same time, a certain advantage is that all the calculation
can be made in the representation of spin operators for an
arbitrary   spin $S$. The theory we have developed can be
directly extended to the case of a large number of magnetic
sublattices with inequivalent spins, i.e., it can be used to
describe the complex ferrimagnets.\\ In the framework of our IGF
approach, it was shown that the mean fields in an antiferromagnet
must include the "anomalous" averages which represent the local
nature of the Neel molecular fields. Thus, the mean field in an
antiferromagnet, like the mean field in a superconductor, has a
more complicated structure.
\section{Quasi-Particle Dynamics of Lattice Fermion Models}
\subsection{Hubbard Model. Weak Correlation}
The concept of   GMFs and the relevant algebra of operators from
which GFs are constructed are important for our treatment of
electron correlations in solids. It is convenient (and much
shorter) to discuss these concepts for weakly and strongly
correlated cases separately.   First,   we should construct a
suitable state vector space of a many-body system~\cite{bog1}.
The fundamental assumption implies that   states of a system of
interacting particles can be expanded in terms of   states of
non-interacting particles~\cite{bog1}. This approach originates
from perturbation theory and finds support for weakly interacting
many-particle systems. For the strongly correlated case, this
approach needs a  suitable reformulation, and  just at this point,
the right definition of the GMFs is vital. Let us consider the
weakly correlated Hubbard model (\ref{eq.49}). In some respect,
this case is similar to the ordinary interacting electron gas but
with very local singular interaction. The difference is in the
lattice ( Wannier ) character of   electron  states. It is shown
below that the usual creation $a^{\dagger}_{i\sigma}$ and
annihilation $a_{i\sigma}$ second-quantized operators with the
properties
$$a^{\dagger}_{i}\Psi^{(0)} = \Psi^{(1)}_{i} ; \quad a_{i}\Psi^{(1)} =
\Psi^{(0)}$$ $$a_{i}\Psi^{(0)} = 0 ;\quad  a_{j}\Psi^{(1)}_{i} =
0 \quad ( i \not= j)$$ are suitable variables for  description of
a  system under consideration. Here $\Psi^{(0)}$ and $\Psi^{(1)}$
are vacuum and single-particle states, respectively. The question
now is how to describe our system in terms of  quasi-particles.
For a translationally invariant system, to describe the low-lying
excitations of a system in terms of quasi-particles~\cite{mah},
one has to choose eigenstates such that they all correspond to a
definite momentum. For the single-band Hubbard model
(\ref{eq.49}), the exact transformation reads $$a_{{\vec k}\sigma}
= N^{-1/2}\sum_{i}\exp(-i{\vec k}{\vec R_{i}}) a_{i\sigma}$$ Note
that for a degenerate band  model, a more general transformation
is necessary\cite{kuz42}. Then the Hubbard Hamiltonian
(\ref{eq.49}) in the Bloch vector state space is given by
\begin{equation}
\label{eq.131} H =
\sum_{k\sigma}\epsilon(k)a^{\dagger}_{k\sigma}a_{k\sigma} +
U/2N\sum_{pqrs}a^{\dagger}_{p+r-q\sigma}a_{p\sigma}a^{\dagger}_{q-\sigma}a_{r-\sigma}
\end{equation}
If the interaction is weak, the algebra of   relevant operators
is very simple: it is an algebra of a non-interacting fermion
system ($a_{k\sigma}, a^{\dagger}_{k\sigma}, n_{k\sigma} =
a^{\dagger}_{k\sigma}a_{k\sigma}$). To calculate of the electron
quasi-particle spectrum of the Hubbard model in this limit, let us
consider the single-electron GF  defined as
\begin{equation}
\label{eq.132} G_{k\sigma}(t - t') = <<a_{k\sigma},
a^{\dagger}_{k\sigma}>> = -i\theta(t - t')<[a_{k\sigma}(t),
a^{\dagger}_{k\sigma}(t')]_{+}>
\end{equation}
The equation of motion for the Fourier transform of   GF
$G_{k\sigma}(\omega)$ is of the form
\begin{equation}
\label{eq.133} (\omega - \epsilon_{k})G_{k\sigma}(\omega) = 1 +
U/N\sum_{pq}<<a_{k+p\sigma}a^{\dagger}_{p+q-\sigma}a_{q-\sigma}
\vert a^{\dagger}_{k\sigma}>>_{\omega}
\end{equation}
Let us introduce  an ``irreducible" GF in the following way
\begin{eqnarray}
\label{eq.134}
^{(ir)}<<a_{k+p\sigma}a^{\dagger}_{p+q-\sigma}a_{q-\sigma}
\vert a^{\dagger}_{k\sigma}>>_{\omega} =\nonumber\\
<<a_{k+p\sigma}a^{\dagger}_{p+q-\sigma}a_{q-\sigma} \vert
a^{\dagger}_{k\sigma}>>_{\omega} -\delta_{p,
0}<n_{q-\sigma}>G_{k\sigma}
\end{eqnarray}
The irreducible $(ir)$ GF in (\ref{eq.134}) is defined so   that
it cannot be reduced to GF of lower order with respect to the
number of fermion operators by an arbitrary pairing of operators
or, in  other words, by any kind of decoupling. Substituting
(\ref{eq.134}) into (\ref{eq.133}), we obtain
\begin{eqnarray}
\label{eq.135}
G_{k\sigma}(\omega) = G^{MF}_{k\sigma}(\omega) + \\
\nonumber G^{MF}_{k\sigma}(\omega)
U/N\sum_{pq}{}^{(ir)}<<a_{k+p\sigma}a^{\dagger}_{p+q-\sigma}a_{q-\sigma}
\vert a^{\dagger}_{k\sigma}>>_{\omega} \nonumber
\end{eqnarray}
Here we   introduced the notation
\begin{equation}
\label{eq.136} G^{MF}_{k\sigma}(\omega) = (\omega -
\epsilon(k\sigma))^{-1} ; \epsilon(k\sigma) = \epsilon(k)
+U/N\sum_{q}<n_{q-\sigma}>
\end{equation}
In this paper, for brevity, we confine ourselves to considering
the paramagnetic solutions, {\it i.e.}  $<n_{\sigma}> =
<n_{-\sigma}>$. To calculate the higher-order GF on the r.h.s. of
(\ref{eq.135}), we have to write the equation of motion obtained
by means of differentiation with respect to the second variable
$t'$. Constraint (\ref{eq.31}) allows us to remove the
inhomogeneous term from this equation for
$\frac{d}{dt'}^{(ir)}<<A(t), a^{\dagger}_{k\sigma}(t')>>$. \\ For
the Fourier components, we   have
\begin{eqnarray}
\label{eq.137} (\omega - \epsilon(k)){}^{(ir)}<<A\vert
a^{\dagger}_{k\sigma}>>_{\omega} =
<^{(ir)}[A, a^{\dagger}_{k\sigma}]_{+}> +\nonumber\\
U/N\sum_{rs}{}^{(ir)}<<A\vert
a^{\dagger}_{r-\sigma}a_{r+s-\sigma}a^{\dagger}_{k+s\sigma}>>_
{\omega}.
\end{eqnarray}
The anticommutator in (\ref{eq.137}) is calculated on the basis of
the definition of the irreducible part
\begin{eqnarray}
\label{eq.138}
<[^{(ir)}(a_{k+p\sigma}a^{\dagger}_{p+q-\sigma}a_{q-\sigma}),
a^{\dagger}_{k\sigma}]_{+}> =
\nonumber\\
<[a_{k+p\sigma}a^{\dagger}_{p+q-\sigma}a_{q-\sigma} -
<a^{\dagger}_{p+q-\sigma}a_{q-\sigma}>a_{k+p\sigma},
a^{\dagger}_{k\sigma}]_{+}> = 0
\end{eqnarray}
If one introduces the irreducible part for the r.h.s. operators by
analogy with expression (\ref{eq.134}), the equation of motion
(\ref{eq.133}) takes the following exact form ( {\it cf.}
eq.(\ref{eq.34}))
\begin{equation}
\label{eq.139} G_{k\sigma}(\omega) = G^{MF}_{k\sigma}(\omega) +
G^{MF}_{k\sigma}(\omega)
P_{k\sigma}(\omega)G^{MF}_{k\sigma}(\omega)
\end{equation}
where we   introduced the following notation for the operator $P$
(35)
\begin{eqnarray}
\label{eq.140} P_{k\sigma}(\omega) =
\frac{U^{2}}{N^{2}}\sum_{pqrs}D^{(ir)}_{k\sigma} (p, q\vert r, s,
;\omega) =  \\ \nonumber
\frac{U^{2}}{N^{2}}\sum_{pqrs}{}(^{(ir)}<<a_{k+p\sigma}a^{\dagger}_{p+q-\sigma}
a_{q-\sigma}\vert
a^{\dagger}_{r-\sigma}a_{r+s-\sigma}a^{\dagger}_{k+s\sigma}>>^{(ir)}_
{\omega})
\end{eqnarray}
To define the self-energy operator according to   (\ref{eq.36}),
one should separate the "proper" part in the following way
\begin{eqnarray}
\label{eq.141} D^{(ir)}_{k\sigma}(p, q\vert r, s;\omega) =
 L^{(ir)}_{k\sigma}(p, q\vert r, s;\omega)
\nonumber\\
+ \frac{U^{2}}{N^{2}}\sum_{r's'p'q'}L^{(ir)}_{k\sigma}(p, q\vert
r's';\omega) G^{MF}_{k\sigma}(\omega)D^{(ir)}_{k\sigma}(p',
q'\vert r, s;\omega)
\end{eqnarray}
Here $L^{(ir)}_{k\sigma}(p, q\vert r, s;\omega)$ is the ``proper"
part of   GF $D^{(ir)}_{k\sigma}(p, q \vert r, s;\omega)$  which,
in accordance with the definition (36), cannot be reduced to the
lower-order one by any type of decoupling. We find
\begin{equation}
\label{eq.142} G_{k\sigma} = G^{MF}_{k\sigma}(\omega) +
G^{MF}_{k\sigma}(\omega) M_{k\sigma}( \omega)G_{k, \sigma}(\omega)
\end{equation}
Equation (\ref{eq.142}) is the Dyson equation for the
single-particle double-time thermal GF. According to (38),
 it has the formal solution
\begin{equation}
\label{eq.143} G_{k\sigma}(\omega) = [\omega -\epsilon(k\sigma)
-M_{k\sigma}( \omega)]^{-1}
\end{equation}
where the self-energy operator $M$ is given by
\begin{eqnarray}
\label{eq.144} M_{k\sigma}(\omega) =  \frac{U^{2}}{N^{2}}
\sum_{pqrs}L^{(ir)}_{k\sigma}
(p, q \vert r, s;\omega) =  \\
 \frac{U^2}{N^2}{} \sum_{pqrs}{} \bigl (^{(ir)}<<a_{k+p\sigma}a^{\dagger}_{p+q-\sigma}
a_{q-\sigma} \vert
a^{\dagger}_{r-\sigma}a_{r+s-\sigma}a^{\dagger}_{k+s\sigma}>>^{(ir)}
\bigr )^{(p)} \nonumber
\end{eqnarray}
We wrote   explicitly   equation (\ref{eq.140}) for $P$ and
equation (\ref{eq.144}) for $M$ to illustrate the general
arguments of Section 3 and to give   concrete equations for
determining  both the quantities,  $P$ and $M$. \\ The latter
expression (\ref{eq.144}) is an exact representation ( no
decoupling was made till now ) for the self-energy in terms of
higher-order GF up to second order in $U$ ( for the consideration
of   higher-order equations of motion, see ref.~\cite{kuz4}). The
explicit difference between $P$ and $M$ follows from the
functional form (\ref{eq.38}). Thus, in contrast to the standard
equation-of-motion approach, the calculation of   full GF was
substituted by the calculation of the mean-field GF $G^{MF}$ and
the self-energy operator $M$. The main reason for this method of
calculation is that the decoupling is only introduced into the
self-energy operator, as it will be shown in a detailed form
below. The formal solution of the Dyson equation (\ref{eq.38})
determines the right reference frame for the formation of the
quasi-particle spectrum due to  its own correct functional
structure. In the standard equation-of-motion approach, that
structure could be lost by using decoupling approximations {\bf
before} arriving at the correct functional structure of the formal
solution of the Dyson equation. This is a crucial point of the
IGF method.\\ The energies of   electron  states in the
mean-field approximation are given by the poles of $G^{MF}$ . Now
let us consider the damping effects and finite lifetimes. To find
an explicit expression for the self-energy $M$ (\ref{eq.144}), we
have to evaluate approximately the higher-order GF in it. It will
be shown below that the IGF method permits one to derive the
damping in a self-consistent way simply and much more generally
than within other formulations. First, it is convenient to write
down the GF in (\ref{eq.144}) in terms of correlation functions
by using the spectral theorem (26)
\begin{eqnarray}
\label{eq.145}
<<a_{k+p\sigma}a^{\dagger}_{p+q-\sigma}a_{q-\sigma} \vert
a^{\dagger}_{k+s\sigma}a^{\dagger}_{r-\sigma}a_{r+s-\sigma}>>_{\omega} =\nonumber\\
{1 \over 2\pi}\int_{-\infty}^{+\infty}{d\omega' \over \omega - \omega'}
(\exp(\beta \omega') +1)     \int_{-\infty}^{+\infty}\exp(i\omega't)
\nonumber\\
<a^{\dagger}_{k+s\sigma}(t)a^{\dagger}_{r-\sigma}(t)a_{r+s-\sigma}(t)
a_{k+p\sigma}a^{\dagger}_{p+q-\sigma}a_{q-\sigma}>
\end{eqnarray}
Further insight is gained if we select the suitable relevant
``trial" approximation for the correlation function on the r.h.s.
of (\ref{eq.145}). In this paper, we show that the earlier
formulations  based on the decoupling or/and diagrammatic methods
can be obtained from our technique but in a self-consistent way.
It is clear  that a  relevant trial approximation for the
correlation function in (\ref{eq.145}) can be chosen in many ways.
For example, the reasonable and workable one can be the following
``pair approximation" that is especially suitable for a low
density of  quasi-particles:
\begin{eqnarray}
\label{eq.146}
<a^{\dagger}_{k+s\sigma}(t)a^{\dagger}_{r-\sigma}(t)a_{r+s-\sigma}(t)
a_{k+p\sigma}a^{\dagger}_{p+q-\sigma}a_{q-\sigma}>^{(ir)} \approx \nonumber\\
<a^{\dagger}_{k+p\sigma}(t)a_{k+p\sigma}><a^{\dagger}_{q-\sigma}(t)a_{q-\sigma}>
<a_{p+q-\sigma}(t)a^{\dagger}_{p+q-\sigma}>\nonumber\\
\delta_{k+s, k+p} \delta_{r, q} \delta_{r+s, p+q}
\end{eqnarray}
Using (\ref{eq.146}) and (\ref{eq.145}) in (\ref{eq.144}) we
obtain the self-consistent approximate expression for the
self-energy operator  ( the self-consistency means that we
express approximately the self-energy operator in terms of the
initial GF, and, in principle, one can obtain the required
solution by a suitable iteration procedure ):
\begin{eqnarray}
\label{eq.147}
M_{k\sigma}( \omega) = \\
 \frac{U^2}{N^2} \sum_{pq} \int
\frac{d\omega_{1}d\omega_{2}d\omega_{3}}{\omega + \omega_{1} -
\omega_{2} - \omega_{3}} \nonumber \\
~\Bigl [n(\omega_{2})n(\omega_{3}) + n(\omega_{1})\Bigl (1 -
n(\omega_{2}) - n(\omega_{3})\Bigr ) \Bigr]
g_{p+q-\sigma}(\omega_{1})g_{k+p\sigma}(\omega_{2})g_{q-\sigma}(\omega_{3})
\nonumber
\end{eqnarray}
where we   used the notation
$$g_{k\sigma}(\omega) = -{1 \over \pi}Im G_{k\sigma}(\omega + i\varepsilon);
\quad n(\omega) = [\exp(\beta\omega) + 1]^{-1}$$   Equations
(\ref{eq.147}) and (\ref{eq.142}) constitute a closed
self-consistent system of equations for the single-electron GF of
the Hubbard model in the weakly correlated limit. In principle, we
can use, on the   r.h.s.  of (\ref{eq.147}), any workable first
iteration-step form of the GF and find a solution by iteration (
see Appendix D ). It is most convenient to choose, as the first
iteration step, the following simple one-pole approximation:
\begin{equation}
\label{eq.148} g_{k\sigma}(\omega) \approx \delta(\omega -
\epsilon(k\sigma))
\end{equation}
Then, using (\ref{eq.148}) in (\ref{eq.147}), we get, for the
self-energy, the explicit and compact expression
\begin{equation}
\label{eq.149} M_{k\sigma}( \omega) = \frac{U^2}{N^2} \sum_{pq}
\frac{n_{p+q-\sigma}(1 - n_{k+p\sigma} - n_{q-\sigma}) +
n_{k+p\sigma} n_{q-\sigma}}{\omega + \epsilon(p+q\sigma) -
\epsilon(k+p\sigma) - \epsilon(q\sigma)}
\end{equation}
Formula (\ref{eq.149})   for the self-energy operator   shows the
role of correlation effects ( inelastic scattering processes ) in
the formation of quasi-particle spectrum of the Hubbard model.
This formula can be derived by several different methods,
including   perturbation theory. Here we derived it from our IGF
formalism as a known limiting case.  The numerical calculations
of the typical behaviour of real and imaginary parts of the
self-energy (\ref{eq.149}) were
performed~\cite{kuz42},\cite{kuz10} for the model density of
states of the FCC lattice. These calculations and many other (see
{\it e.g.}\cite{wilk},\cite{treg},\cite{jelit}) show clearly that
the conventional one-electron approximation of the band theory is
not always a sufficiently good approximation for transition metals
like nickel. A more concrete discussion of the numerical
calculations and their comparison with experiments deserve  a
separate consideration and will be considered elsewhere
(for a detailed recent discussion, see \cite{wilk}). \\
Although   the solution deduced above is a good evidence for the
efficiency of the IGF formalism, there is one more stringent test
of the method that we can perform. It is instructive to examine
other types of   possible trial solutions for the six-operator
correlation function in the eq.(\ref{eq.145}). The approximation
  we propose now reflects the interference between the
one-particle branch of the spectrum and the collective ones:
\begin{eqnarray}
\label{eq.150}
<a^{\dagger}_{k+s\sigma}(t)a^{\dagger}_{r-\sigma}(t)a_{r+s-\sigma}(t)
a_{k+p\sigma}a^{\dagger}_{p+q-\sigma}a_{q-\sigma}>^{(ir)} \approx \nonumber\\
<a^{\dagger}_{k+s\sigma}(t)a_{k+p\sigma}><a^{\dagger}_{r-\sigma}(t)a_{r+s-\sigma}(t)
a^{\dagger}_{p+q-\sigma}a_{q-\sigma}> + \nonumber\\
<a_{r+s-\sigma}(t)a^{\dagger}_{p+q-\sigma}><a^{\dagger}_{k+s\sigma}(t)a^{\dagger}_{r-\sigma}(t
)
a_{k+p\sigma}a_{q-\sigma}> +\nonumber\\
<a^{\dagger}_{r-\sigma}(t)a_{q-\sigma}><a^{\dagger}_{k+s\sigma}(t)a_{r+s-\sigma}(t)
a_{k+p\sigma}a^{\dagger}_{p+q-\sigma}>
\end{eqnarray}
It is seen   that the three contributions in this trial solution
describe the self-energy corrections that take into account the
collective motions of electron density, the spin density and the
density of ``doubles", respectively. An essential feature of this
approximation is   that a correct calculation of the
single-electron quasi-particle spectra with damping requires a
suitable incorporation of the influence of   collective degrees of
freedom on the single-particle ones. The most interesting
contribution comes from spin degrees of freedom, since the
correlated systems are often  magnetic or have very well
developed magnetic fluctuations. \\ We follow the above steps and
calculate the self-energy operator (\ref{eq.144}) as
\begin{eqnarray}
\label{eq.151} M_{k\sigma}( \omega) = {U^2\over
N}\int_{-\infty}^{+\infty} d\omega_{1}d{\omega}_{2}\frac{1 +
N(\omega_{1}) - n(\omega_{2})}
{\omega - \omega_{1} - \omega_{2}}\nonumber\\
\sum_{i, j}\exp[-i{\vec k}({\vec R_{i}} - {\vec R_{j}})]
(-{1 \over \pi} Im <<S^{\pm}_{i} \vert S^{\mp}_{j}>>_{\omega_{1}})\nonumber\\
(-{1 \over \pi} Im<<a_{i-\sigma}\vert
a^{\dagger}_{j-\sigma}>>_{\omega_{2}})
\end{eqnarray}
where the following notation was used:
$$S^{+}_{i} = a^{\dagger}_{i\uparrow}a_{i\downarrow}; \quad
S^{-}_{i} = a^{\dagger}_{i\downarrow}a_{i\uparrow}$$  It is
possible to rewrite ( \ref{eq.151} ) in a more convenient way
\begin{eqnarray}
\label{eq.152}
M_{k\sigma}( \omega) = \\
{U^2\over N} \sum_{q} \int d\omega'
(\cot \frac{\omega - \omega'}{2T} + \tan \frac{\omega'}{2T})\nonumber\\
\Bigl(- \frac{1}{\pi} Im \chi^{\mp \pm} (k-q, \omega -
\omega')g_{q\sigma} (\omega')\Bigr) \nonumber
\end{eqnarray}
Equations ( \ref{eq.152}  ) and ( \ref{eq.142} ) constitute again
another self-consistent system of equations for the
single-particle GF of the Hubbard model. Note  that both the
expressions for the self-energy depend on the quasi-momentum; in
other words, the approximate procedure does not break  the
momentum conservation law. The fundamental importance of
equations (\ref{eq.152}) and (\ref{eq.147}) can be appreciated by
examining the problem of the definition of the Fermi surface. It
is rather clear, because   the poles $\omega(k, \sigma) =
\epsilon(k, \sigma) -i\Gamma_{k}$ of   GF (\ref{eq.143}) are
determined by the equation
$$\omega - \epsilon(k\sigma) - Re[M_{k\sigma}( \omega)] = 0$$
It can be shown quite generally that the Luttinger's definition of
the true Fermi surface~\cite{mah} is valid in the framework of the
present theory. It is worthy to note that for electrons in a
crystal where there is a band index, and a quasi-momentum, the
definition of the Fermi surface is a little more complicated than
the single-band one. Before the single particle energies and
Fermi surface are known, one should carry out a diagonalization in
the band index.
\subsection{Hubbard Model. Strong Correlation}
Being convinced that the IGF method can be applied successfully
to the weakly correlated Hubbard model, we now show that the IFG
approach can be extended to the case of an arbitrarily strong but
finite interaction. This development incorporates   main
advantages
of the IGF scheme and proves its efficiency and flexibility. \\
When studying the electron  quasi-particle spectrum of strongly
correlated systems, one should take care of at least three facts
of major importance:
\begin{itemize}
\item[(i)] The ground state is reconstructed radically as
compared with the weakly correlated case. This fact makes it
necessary to redefine   single-particle states. Due to the strong
correlation, the initial algebra of   operators is transformed
into the new algebra of complicated operators. In principle, in
terms of the new operators, the initial Hamiltonian can be
rewritten   as a bilinear form, and the generalized Wick theorem
can be formulated. It is very important to stress  that the
transformation to the new algebra of relevant operators reflects
some important internal symmetries of the problem, and nowadays,
this way of thinking is formulating in the elegant and very
powerful technique of the classification of the integrable models
and exactly soluble models ({\it cf.}\cite{kuz52}).\item[(ii)]
The single-electron GF that describes dynamic  properties, should
have the two-pole functional structure, which gives in the atomic
limit, when the hopping integral tends to zero, the exact
two-level atomic solution.
\item[(iii)] The GMFs have, in the general case, a very non-trivial
structure. The GMFs functional, as a rule, cannot be expressed in
terms of the functional of the mean particle densities.
\end{itemize}  In this section, we   consider the case of a large  but
finite  Coulomb repulsion $U$ in the Hubbard Hamiltonian
(\ref{eq.49}) . Let us consider the single-particle GF
(\ref{eq.132}) in the Wannier basis
\begin{equation}
\label{eq.153} G_{ij\sigma}(t - t') = <<a_{i\sigma}(t);
a^{\dagger}_{j\sigma}(t')>>
\end{equation}
It is convenient to introduce the new set of relevant
operators\cite{hub2}
\begin{eqnarray}
\label{eq.154} d_{i\alpha\sigma} =
n^{\alpha}_{i-\sigma}a_{i\sigma}, (\alpha = \pm);\quad
n^{+}_{i\sigma} = n_{i\sigma},\quad n^{-}_{i\sigma} = (1 -
n_{i\sigma});
\nonumber\\
\sum n^{\alpha}_{i\sigma} = 1; \quad
n^{\alpha}_{i\sigma}n^{\beta}_{i\sigma} =
\delta_{\alpha\beta}n^{\alpha}_{i\sigma}; \quad \sum_{\alpha}
d_{i\alpha\sigma} = a_{i\sigma}
\end{eqnarray}
The new operators $d_{i\alpha \sigma}$ and $d^{\dagger}_{j\beta
\sigma}$ have complicated commutation rules, namely,
$$[d_{i\alpha \sigma}, d^{\dagger}_{j\beta \sigma}]_{+} = \delta_{ij}
\delta_{\alpha \beta}n^{\alpha}_{i-\sigma}$$ The convenience of
the new operators follows immediately if one writes down the
equation of motion for them
\begin{eqnarray}
\label{eq.155} [d_{i\alpha \sigma}, H]_{-} = E_{\alpha}d_{i\alpha
\sigma} + \sum_{ij}t_{ij}(n^{\alpha}_{i-\sigma}a_{j\sigma} +
\alpha a_{i\sigma}
b_{ij-\sigma})\nonumber\\
b_{ij\sigma} = (a^{\dagger}_{i\sigma}a_{j\sigma} -
a^{\dagger}_{j\sigma}a_{i\sigma}).
\end{eqnarray}
It is possible to interpret~\cite{hub}, \cite{hub2} both
contributions to this equation as {\it alloy analogy} and {\it
resonance broadening corrections}. Using the new operator algebra,
it is possible  identically rewrite  GF (\ref{eq.153}) in the
following way
\begin{equation}
\label{eq.156} G_{ij \sigma}(\omega) = \sum_{\alpha
\beta}<<d_{i\alpha \sigma} \vert d^{\dagger}_{j\beta
\sigma}>>_{\omega} = \sum_{\alpha \beta} F^{\alpha \beta}_{
ij\sigma}(\omega)
\end{equation}
The equation of motion for the auxiliary matrix GF
\begin{equation}
\label{eq.157} F^{\alpha \beta}_{ij\sigma}(\omega) =
\pmatrix{<<d_{i+\sigma} \vert
d^{\dagger}_{j+\sigma}>>_{\omega}&<<d_{i+\sigma} \vert
d^{\dagger}_{j-\sigma}>>_{\omega}
 \cr
<<d_{i-\sigma} \vert
d^{\dagger}_{j+\sigma}>>_{\omega}&<<d_{i-\sigma} \vert
d^{\dagger}_{j-\sigma}>>_{\omega}\cr}
\end{equation}
is of the following form
\begin{equation}
\label{eq.158} ({\bf E}{\bf F}_{ij\sigma}(\omega) -{\bf I}
\delta_{ij})_{\alpha \beta} = \sum_{l\not=
i}t_{il}<<n^{\alpha}_{i-\sigma}a_{l\sigma} + \alpha a_{i\sigma}
b_{il-\sigma} \vert d^{\dagger}_{j\beta \sigma}>>_{\omega}
\end{equation}
where the following matrix notations was used:
\begin{equation}
\label{eq.159} {\bf E} = \pmatrix{(\omega- E_{+})&0\cr 0&(\omega -
E_{-})\cr} ; {\bf I} = \pmatrix{n^{+}_{-\sigma}&0\cr
0&n^{-}_{-\sigma} \cr}.
\end{equation}
In accordance with the general method of Section 3, we introduce
by definition the matrix IGF:
\begin{eqnarray}
\label{eq.160} {\bf D}^{(ir)}_{il, j}(\omega) =
\pmatrix{<<Z_{11}\vert d^{\dagger}_{j+\sigma}>>_
{\omega}&<<Z_{12}\vert d^{\dagger}_{j-\sigma}>>_{\omega}\cr
<<Z_{21}\vert d^{\dagger}_{j+\sigma}>>_{\omega}&<<Z_{22}\vert
d^{\dagger}_{j-\sigma}>>_
{\omega}\cr} -\nonumber\\
\sum _{\alpha'}({A^{+\alpha'}_{il}\brack A^{-\alpha'}_{il}}[F^{\alpha'+}_{
ij\sigma} \ F^{\alpha'-}_{ij\sigma}] - {B^{+\alpha'}_{li}\brack
B^{-\alpha'}_{
li}}[F^{\alpha'+}_{lj\sigma} \ F^{\alpha'-}_{lj\sigma}])
\end{eqnarray}
Here the notation was   used:
$$Z_{11} = Z_{12} = n^{+}_{i-\sigma}a_{l\sigma} + a_{i\sigma}b_{il-\sigma};
\ Z_{21} = Z_{22} = n^{-}_{i-\sigma}a_{l\sigma} -
a_{i\sigma}b_{il-\sigma}$$ It is to be emphasized that the
definition (\ref{eq.158}) is the most important and crucial point
of the whole our approach to description of the strong
correlation. The coefficients A and B are determined by the
orthogonality constraint (31), namely,
\begin{equation}
\label{eq.161} <[({\bf D}^{(ir)}_{il, j})_{\alpha \beta},
d^{\dagger}_{j\beta \sigma}]_{+}> = 0
\end{equation}
After some algebra, we obtain from (\ref{eq.161}) ($i \not= j$)
\begin{eqnarray}
\label{eq.162} [A_{il}]_{\alpha \beta} =
\alpha(<d^{\dagger}_{i\beta-\sigma}a_{l-\sigma}> +
<d_{i-\beta-\sigma}a^{\dagger}_{l-\sigma}>)(n^{\beta}_{-\sigma})^{-1}\nonumber\\
~[B_{li}]_{\alpha \beta} =
[<n^{\beta}_{l-\sigma}n^{\alpha}_{i-\sigma}> + \alpha
\beta(<a_{i\sigma}a^{\dagger}_{i-\sigma}a_{l-\sigma}a^{\dagger}_{l\sigma}>
-
\nonumber\\
<a_{i\sigma}a_{i-\sigma}a^{\dagger}_{l-\sigma}a^{\dagger}_{l\sigma}>)](n^{\beta}_{
-\sigma})^{-1}
\end{eqnarray}
As previously, we introduce now   GMF GF ${\bf
F}^{0}_{ij\sigma}$; however, as it is clear from (\ref{eq.162}),
the actual definition of the GMF GF is very nontrivial. After the
Fourier transformation, we get
\begin{equation}
\label{eq.163} \pmatrix{ F^{0++}_{k\sigma}&F^{0+-}_{k\sigma}\cr
F^{0-+}_{k\sigma}&F^{0--}_{k\sigma}\cr} =
\frac 1 {ab - cd} %
\pmatrix{
 n^{+}_{-\sigma}b & n^{-}_{-\sigma}d \cr %
 n^{+}_{-\sigma}c & n^{-}_{-\sigma}a }   %
\end{equation}
The coefficients $a$, $b$, $c$, $d$ are equal to
\begin{eqnarray}
\label{eq.164} {a\atop b} = \Bigl(\omega - E_{\pm} -
N^{-1}\sum_{p}\epsilon(p)[A^{\pm \pm}(-p) -
B^{\pm \pm}(p - q)] \Bigr)\nonumber\\
{c\atop d} = N^{-1}\sum_{p}\epsilon(p)[ A^{\mp \pm}(-p) - B^{\mp
\pm}(p - q)]
\end{eqnarray}
Then, using the definition (\ref{eq.157}), we find the final
expression for   GMF GF
\begin{equation}
\label{eq.165} G^{MF}_{k\sigma}( \omega) = \frac{\omega -
(n^{+}_{-\sigma}E_{-} + n^{-}_{-\sigma}E_{+}) - \lambda(k)}{
(\omega -E_{+} - n^{-}_{-\sigma}\lambda_{1}(k))(\omega - E_{-} -
 n^{+}_{-\sigma}
\lambda_{2}(k)) -
n^{-}_{-\sigma}n^{+}_{-\sigma}\lambda_{3}(k)\lambda_{4}(k)}
\end{equation}
Here we   introduced the following notation:
\begin{eqnarray}
\label{eq.166} {\lambda_{1}(k)\atop \lambda_{2}(k)} =
\frac{1}{n^{\mp}_{-\sigma}}\sum_{p}
\epsilon(p)[A^{\pm \pm}(-p) - B^{\pm \pm}(p - k)]\\
\label{eq.167} {\lambda_{3}(k)\atop \lambda_{4}(k)} =
\frac{1}{n^{\mp}_{-\sigma}}\sum_{p}
\epsilon(p)[A^{\pm \mp}(-p) - B^{\pm \mp}(p - k)]\\
\lambda(k) = (n^{-}_{-\sigma})^{2}(\lambda_{1} + \lambda_{3}) +
(n^{+}_{-\sigma})^{2}(\lambda_{2} + \lambda_{4}) \nonumber
\end{eqnarray}
From the equation (\ref{eq.165}) it is obvious that our two-pole
solution is   more general than the``Hubbard III" \cite{hub2}
solution and the Roth\cite{roth} solution. Our solution has the
correct nonlocal structure and, thus, takes into account the
non-diagonal scattering matrix elements more accurately. Those
matrix elements describe the virtual ``recombination" processes
and reflect the extremely complicated structure of
single-particle states  which virtually include a great number of
intermediate scattering processes. \\ The spectrum of mean-field
quasi-particle excitations follows from the poles of the GF
(\ref{eq.165}) and consists of two branches
\begin{eqnarray}
\label{eq.168}
\omega{1\atop 2}(k) = \\
1/2[(E_{+} - E_{-} + a_{1} + b_{1}) \pm \sqrt {(E_{+} +E_{-} -
a_{1} - b_{1})^{2} -4cd}] \nonumber
\end{eqnarray}
where $a_{1}  = \omega - E_{\pm} -a; \quad b_{1}  = \omega -
E_{\pm} -b $. Thus, the spectral weight function
$A_{k\sigma}(\omega)$ of GF (\ref{eq.165}) consists of two peaks
separated by the distance
\begin{equation}
\label{eq.169} \omega_{1} - \omega_{2} =\sqrt{(U - a_{1} -
b_{1})^{2} - cd} \approx U( 1 - \frac{a_{1} - b_{1}}{U}) +
O(\gamma)
\end{equation}
For a deeper insight into the functional structure of the
solution (\ref{eq.165}) and to compare with  other solutions, we
rewrite   (\ref{eq.165}) in the following form
\begin{equation}
\label{eq.170} {\bf F}^{0}_{k\sigma}(\omega) = \pmatrix{ ({a\over
n^{+}_{-\sigma}} - {db^{-1}c\over n^{+}_{-\sigma}})^{-1}&{d \over
a}({b\over n^{-}_{-\sigma}} - {da^{-1}c\over
n^{-}_{-\sigma}})^{-1}\cr {c\over b}({a\over n^{+}_{-\sigma}} -
{db^{-1}c\over n^{+}_{-\sigma}})^{-1}& ({b\over n^{-}_{-\sigma}}
- {db^{-1}c\over n^{-}_{-\sigma}})^{-1}\cr}
\end{equation}
from which we obtain for   $G^{MF}_{\sigma}(k, \omega)$
\begin{eqnarray}
\label{eq.171} G^{MF}_{k\sigma}( \omega) = \frac
{n^{+}_{-\sigma}(1 + cb^{-1})}{a - db^{-1}c} + \frac
{n^{-}_{-\sigma}(1 + da^{-1})}{b - ca^{-1}d} \approx
\nonumber\\
\frac {n^{-}_{-\sigma}}{\omega - E_{-} -
n^{+}_{-\sigma}W^{-}_{k-\sigma}} + \frac {n^{+}_{-\sigma}}{\omega
- E_{+} - n^{-}_{-\sigma}W^{\dagger}_{k-\sigma}}
\end{eqnarray}
where
\begin{eqnarray}
\label{eq.172}
n^{+}_{-\sigma}n^{-}_{-\sigma}W^{\pm}_{k-\sigma} =
N^{-1}\sum_{ij} t_{ij}
\exp[-ik(R_{i} -R_{j})] \\
\left ((<a^{\dagger}_{i-\sigma}n^{\pm}_{i\sigma}a_{j-\sigma}> +
<a_{i-\sigma}
n^{\mp}_{i\sigma}a^{\dagger}_{j-\sigma}>)    \right . + \nonumber \\
 \left . (<n^{\pm}_{j-\sigma}n^{\pm}_{i-\sigma}> +
<a_{i\sigma}a^{\dagger}_{i-\sigma}
a_{j-\sigma}a^{\dagger}_{j\sigma}> -
<a_{i\sigma}a_{i-\sigma}a^{\dagger}_{j-\sigma}
a^{\dagger}_{j\sigma}>) \right) \nonumber
\end{eqnarray}
are the shifts for   upper and lower splitted subbands due to the
elastic scattering of   carriers in the Generalized Mean Field.
The quantities $W^{\pm}$ are   functionals of the GMF. The most
important feature of the present solution of the strongly
correlated Hubbard model is a very nontrivial structure of the
mean-field renormalizations (\ref{eq.171}), which is crucial for
understanding the physics of strongly correlated systems. It is
important to emphasize that just this complicated form of   GMF is
only relevant to the essence of the physics under consideration.
The attempts to reduce the functional of   GMF to a simpler
functional of the average density of electrons are incorrect from
the point of view of   real   physics of strongly correlated
systems. This physics clearly shows that the mean-field
renormalizations cannot be expressed as   functionals of the
electron mean density. To explain this statement, let us derive
the ``Hubbard I" solution~\cite{hub} (\ref{eq.54}) from our GMF
solution (\ref{eq.165}). If we approximate (\ref{eq.171}) as
\begin{equation}
\label{eq.173} n^{+}_{-\sigma}n^{-}_{-\sigma}W^{\pm}(k) \approx
N^{-1}\sum_{ij}t_{ij} {\exp[-ik(R_{i} -
R_{j})]}<n^{\pm}_{j-\sigma}n^{\pm}_{i-\sigma}>
\end{equation}
and make  the additional approximation, namely,
$$<n_{j-\sigma}n_{i-\sigma}> \approx n^{2}_{-\sigma}$$ then
solution (\ref{eq.165}) turns into the ``Hubbard I" solution
(\ref{eq.54}). This solution, as it is well known, is unrealistic
from   many points of view. \\
As  to our solution (\ref{eq.165}) , the second important aspect
is that the parameters $\lambda_{i}(k)$ do not depend on
frequency, since they depend essentially on   elastic scattering
processes. The dependence on frequency arises due to inelastic
scattering processes which are contained in our self-energy
operator.  We proceed now with the derivation of the
explicit expression for the self-energy. \\
To calculate a high-order GF on the r.h.s. of (\ref{eq.158}), we
should use the second time variable ($t'$) differentiation of it
again. If one introduces the irreducible parts for the
right-hand-side operators by analogy with the expression
(\ref{eq.160}), the equation of motion (\ref{eq.158}) can be
rewritten exactly in the following form
\begin{equation}
\label{eq.174} {\bf F}_{k\sigma}(\omega) = {\bf
F}^{0}_{k\sigma}(\omega) + {\bf F}^{0}_{k\sigma}(\omega){\bf
P}_{k\sigma}(\omega){\bf F}^{0}_{k\sigma} (\omega)
\end{equation}
Here the scattering operator $P$ (36) is of the form
\begin{equation}
\label{eq.175} {\bf P}_{q\sigma}(\omega) = {\bf
I}^{-1}[\sum_{lm}t_{il}t_{mj} <<{\bf D}^{(ir)}_{il, j}|{\bf
D}^{(ir) \dagger}_{i, mj}>>_{\omega}]_{q}{\bf I}^{-1}
\end{equation}
In accordance with the definition (37), we write down the Dyson
equation
\begin{equation}
\label{eq.176} {\bf F} = {\bf F}^{0} + {\bf F}^{0}{\bf M}{\bf F}
\end{equation}
The self-energy operator $M$ is defined by eq. (37). Let us note
again that the self-energy corrections, according to (38),
contribute to the full GF as   additional terms. This is an
essential advantage in comparison with the ``Hubbard III"
solution and other two-pole solutions. It is clear from the form
of Roth  solution (\ref{eq.55}) that it includes the elastic
scattering corrections only and does not incorporate the damping
effects and finite lifetimes.\\
For the full GF we find, using the formal solution of Dyson
equation (\ref{eq.38}), that it is equal to
\begin{eqnarray}
\label{eq.177} G_{k\sigma}( \omega) = \left ( {1\over
n^{+}_{-\sigma}}(a - n^{+}_{-\sigma} M^{++}_{k\sigma}( \omega)) +
{1\over n^{-}_{-\sigma}}(b - n^{-}_{-\sigma}
M^{--}_{k\sigma}(\omega)) \right . \nonumber\\
\left . + {1\over n^{+}_{-\sigma}}(d + n^{+}_{-\sigma}
M^{+-}_{k\sigma}( \omega)) + {1\over n^{-}_{-\sigma}}(c +
n^{-}_{-\sigma}
M^{-+}_{k\sigma}( \omega))\right )\nonumber\\
~[{\rm det}\left ( (F^{0}_{k\sigma} (\omega))^{-1} - M_{k\sigma}(
\omega)\right )]^{-1}
\end{eqnarray}
After some algebra, we can rewrite this expression in the
following form which is essentially new and, in a certain sense,
is the central result of the present theory:
\begin{equation}
\label{eq.178} G = \frac {\omega - (n^{+}E_{-} + n^{-}E_{+}) -
L}{(\omega -E_{+} -n^{-}L_{1}) (\omega - E_{-} -n^{+}L_{2}) -
n^{-}n^{+}L_{3}L_{4}}
\end{equation}
where
\begin{eqnarray}
\label{eq.179} L_{1}(k, \omega) = \lambda_{1}(k) -
{n^{+}_{-\sigma}\over n^{-}_{-\sigma}}
M^{++}_{\sigma}(k, \omega);\nonumber\\
L_{2}(k, \omega) = \lambda_{2}(k) - {n^{-}_{-\sigma}\over
n^{+}_{-\sigma}}
M^{--}_{\sigma}(k, \omega);\nonumber\\
L_{3}(k, \omega) = \lambda_{3}(k) + {n^{-}_{-\sigma}\over
n^{+}_{-\sigma}}
M^{+-}_{\sigma}(k, \omega);\nonumber\\
L_{4}(k, \omega) = \lambda_{4}(k) + {n^{+}_{-\sigma}\over
n^{-}_{-\sigma}}
M^{-+}_{\sigma}(k, \omega);\nonumber\\
L(k, \omega) = \lambda(k) + n^{+}_{-\sigma}n^{-}_{-\sigma}(M^{++}
+ M^{--} - M^{-+} - M^{+-})
\end{eqnarray}
Thus, now we have to find   explicit expressions for the elements
of the self-energy matrix M. To this end, we should use the
spectral theorem again to express the GF in terms of correlation
functions
\begin{equation}
\label{eq.180} M^{\alpha, \beta}_{k\sigma}( \omega) \sim <D^{(ir)
\dagger}_{mj, \beta}(t) D^{(ir)}_{il, \alpha}>
\end{equation}
For the approximate calculation of the self-energy, we propose to
use the following trial solution
\begin{eqnarray}
\label{eq.181} <D^{(ir) \dagger}(t)D^{(ir)}> \approx
<a^{\dagger}_{m\sigma}(t)a_{l\sigma}><n^{\beta}_
{j-\sigma}(t)n^{\alpha}_{i-\sigma}>\nonumber\\
 + <a^{\dagger}_{m\sigma}(t)n^{\alpha}_{i-\sigma}>
<n^{\beta}_{j-\sigma}(t)a_{l\sigma}> +\beta
<b^{\dagger}_{mj-\sigma}(t)a_{l\sigma}>
<a^{\dagger}_{j\sigma}(t)n^{\alpha}_{i-\sigma}>\nonumber\\
 + \beta <b^{\dagger}_{mj-\sigma}(t)
n^{\alpha}_{i-\sigma}><a^{\dagger}_{j\sigma}(t)a_{l\sigma}> +
\alpha
 <a^{\dagger}_{m\sigma}
(t)a_{i\sigma}><n^{\beta}_{j-\sigma}(t)b_{il-\sigma}>\nonumber\\
 + \alpha <a^{\dagger}_{m\sigma}
(t)b_{il-\sigma}><n^{\beta}_{j-\sigma}(t)b_{il-\sigma}>\nonumber\\
 + \alpha\beta
 <b^{\dagger}_{mj-\sigma}(t)a_{i\sigma}><a^{\dagger}_{j\sigma}(t)b_{il-\sigma}>
\nonumber\\
 +
\alpha \beta <b^{\dagger}_{mj-\sigma}(t)b_{il-\sigma}>
<a^{\dagger}_{j\sigma}(t)a_{i\sigma}>
\end{eqnarray}
It is quite natural to interpret the contributions into this
expression in terms of scattering, resonance-broadening, and
interference corrections of different types. For example, let us
consider the simplest approximation. For this aim, we retain the
first contribution in (\ref{eq.181})
\begin{eqnarray}
\label{eq.182} [{\bf I}{\bf M}{\bf I}]_{\alpha \beta} =
\int_{-\infty}^{+\infty}
{d\omega'\over \omega - \omega'}(\exp(\beta \omega') + 1)\nonumber\\
\int_{-\infty}^{+\infty}{dt\over 2\pi}\exp(i\omega't)N^{-1}\sum_{ijlm}
exp[-ik(R_{i}-R_{j})]t_{il}t_{mj} \nonumber \\
\int d\omega_{1}n(\omega_{1})
\exp(i\omega_{1}t)g_{ml\sigma}(\omega_{1})\left (-{1\over \pi} Im
K^{\alpha \beta}_{ij}(\omega_{1} - \omega')\right ).
\end{eqnarray}
Here $K^{\alpha \beta}_{ij}( \omega) =
<<n^{\alpha}_{i-\sigma}\vert n^{\beta}_{j-\sigma}>>_{\omega}$ is
the density-density GF. It is worthy to note that the mass
operator (\ref{eq.182}) contains the term $t_{il}t_{mj}$ contrary
to the expression (\ref{eq.147}) that contains the term $U^{2}$.
The pair of equations (\ref{eq.182}) and (\ref{eq.176}) is a
self-consistent system of equations for the single-particle Green
function. For a simple estimation, for the calculation of the
self-energy (\ref{eq.182}), it is possible to use any initial
relevant approximation of the two-pole structure. As an example,
we take the expression (\ref{eq.54}). We then obtain
\begin{eqnarray}
\label{eq.183} [{\bf I}{\bf M}{\bf I}]_{\alpha \beta} \approx
\sum_{q}|\epsilon(k - q)|^{2}
K^{\alpha \beta}_{q}\nonumber\\
~[\frac{n_{-\sigma}}{\omega - U - \epsilon(k-q)n_{-\sigma}} +
\frac{1 - n_{-\sigma}}{\omega - \epsilon(k-q)(1 - n_{-\sigma})}]
\end{eqnarray}
In the same way, one can use, instead (\ref{eq.54}), another
initial
two-pole solution, {\it e.g.} the Roth  solution (\ref{eq.55}), {\it etc.}\\
On the basis of the self-energy operator (\ref{eq.183}) we can
explicitly find the energy shift and damping due to   inelastic
scattering of
quasi-particles. This is a great advantage of the present approach. \\
In summary, in this Section, we   obtained the most complete
solution to the Hubbard model Hamiltonian in the strongly
correlated case. It has correct functional structure, and,
moreover, it represents correctly the effects of elastic and
inelastic scattering in a systematic and convenient way. The mass
operator contains all inelastic scattering terms including
various scattering and resonance broadening terms in a systematic
way. The obtained solution (\ref{eq.178}) is valid for all band
filling and for arbitrarily strong but finite strength of the
Coulomb repulsion. Our solution contains no approximations except
those contained in the final calculation of the mass operator.
Therefore, we conclude that our solutions to the Hubbard model in
the weakly correlated case (\ref{eq.143}) and in the strongly
correlated case (\ref{eq.178}) describe most fully and
self-consistently the correlation effects in the Hubbard model
and give a unified interpolation description of the correlation
problem. This result is to be contrasted with Hubbard ,  Roth and
many other results in which this interpolation solution cannot be
derived within the unified scheme.\\ It is clear from the present
consideration that for the systematic construction of the
advanced approximate solutions we need to calculate the
collective correlation functions of the electron density and spin
density and the density of doubles, but this problem must be
considered separately.
\subsection{Correlations in Random Hubbard Model}
In this Section, we   apply the IGF method to consider    the
electron-electron correlations in the presence of disorder to
demonstrate the advantage of our approach. The treatment of the
electron motion in substitutionally disordered $A_{x}B_{1-x}$
transition metal alloys is based upon a certain generalization of
the Hubbard model, including random diagonal and off-diagonal
elements caused by substitutional disorder in a binary alloy. The
electron-electron interaction plays an important role for various
aspects of behaviour in alloys, {\it e.g.} for the weak
localization~\cite{kir55}. The approximation which is used widely
for treating   disordered alloys is the single-site Coherent
Potential Approximation (CPA)~\cite{sov58}. The CPA has been
refined and developed in many papers ({\it e.g.} \cite{argy59},
\cite{zin60}) and till now is the most popular approximation for
the theoretical study  of alloys. But the simultaneous effect of
disorder and electron-electron inelastic scattering has been
considered for some limited cases only and not within the
self-consistent scheme. \\ Let us consider the Hubbard model
Hamiltonian (\ref{eq.49}) on a given configuration of an alloy
$(\nu)$
\begin{equation}
\label{eq.184} H^{(\nu)} = H^{(\nu)}_{1} + H^{(\nu)}_{2}
\end{equation}
where
\begin{eqnarray}
\label{eq.185} H^{(\nu)}_{1} =
\sum_{i\sigma}\varepsilon^{\nu}_{i}n_{i\sigma} +
\sum_{ij\sigma}t^{\nu\mu}_{ij}a^{\dagger}_{i\sigma}a_{j\sigma}\nonumber\\
H^{(\nu)}_{2} = {1\over 2}\sum_{i\sigma}U^{\nu}_{i}n_{i\sigma}n_{i-\sigma}
\end{eqnarray}
Contrary to the periodic model (\ref{eq.49}), the atomic level
energy $\varepsilon^{\nu}_{i}$, the hopping integrals
$t^{\nu\mu}_{ij}$, as well as the intraatomic Coulomb repulsion
$U^{\nu}_{i}$ are here    random variables  which take the values
$\varepsilon^{\nu}$, $t^{\nu\mu}$, and $U^{\nu}$, respectively;
the superscript $\nu(\mu)$ refers to   atomic species ($\nu, \mu
= A, B$) located on site i(j). The nearest-neighbor hopping
integrals were only included . \\ To unify the IGF method and CPA
into a completely self-consistent scheme let us consider the
single-electron GF (\ref{eq.153}) $G_{ij\sigma}$ in the Wannier
representation for a given configuration $(\nu)$. The
corresponding equation of motion is of the form (for brevity we
  omit the superscript $(\nu)$ where its presence is clear)
\begin{eqnarray}
\label{eq.186} (\omega -
\varepsilon_{i})<<a_{i\sigma}|a^{\dagger}_{j\sigma}>>_{\omega} =
\delta_{ij} +
\sum_{n}t_{in}<<a_{n\sigma}|a^{\dagger}_{j\sigma}>>_{\omega}
\nonumber\\
+ U_{i}<<n_{i-\sigma}a_{i\sigma}|a^{\dagger}_{j\sigma}>>_{\omega}
\end{eqnarray}
In the present paper, for brevity, we   confine ourselves to the
weak correlation  and the diagonal disorder case. The
generalization to the case of strong correlation or off-diagonal
disorder is straightforward,
but its lengthy consideration preclude us from discussing it this time. \\
Using the definition (30), we define the IGF for a given (fixed) configuration
of atoms in an alloy as follows
\begin{eqnarray}
\label{eq.187}
^{(ir)}<<n_{i-\sigma}a_{i\sigma}|a^{\dagger}_{j\sigma}>> = \\
<<n_{i-\sigma}a_{i\sigma}| a^{\dagger}_{j\sigma}>> -
<n_{i-\sigma}><<a_{i\sigma}|a^{\dagger}_{j\sigma}>> \nonumber
\end{eqnarray}
This time, contrary to (\ref{eq.163}), because of lack of
translational invariance we must take into account the site
dependence of $<n_{i-\sigma}>$. Then we rewrite the equation of
motion (\ref{eq.186}) in the following form
\begin{eqnarray}
\label{eq.188} \sum_{n}[(\omega - \varepsilon_{i} -
U_{i}<n_{i-\sigma}>)\delta_{ij} -
t_{in}]<<a_{n\sigma}|a^{\dagger}_{j\sigma}>>_{\omega} = \nonumber\\
\delta_{ij} +U_{i}(
^{(ir)}<<n_{i-\sigma}a_{i\sigma}|a^{\dagger}_{j\sigma}>>_{\omega})
\end{eqnarray}
In accordance with the general method of Section 3, we find then the
Dyson equation for a given configuration $(\nu)$
\begin{equation}
\label{eq.189} G_{ij\sigma}(\omega) = G^{0}_{ij\sigma}(\omega) +
\sum_{mn}G^{0}_{im\sigma}
(\omega)M_{mn\sigma}(\omega)G_{nj\sigma}(\omega)
\end{equation}
The GMF GF $G^{0}_{ij\sigma}$ and the self-energy operator $M$ are defined
as
\begin{eqnarray}
\label{eq.190}
\sum_{m}H_{im\sigma}G^{0}_{mj\sigma}(\omega) = \delta_{ij}\nonumber\\
P_{mn\sigma} = M_{mn\sigma} + \sum_{ij}M_{mi\sigma}G^{0}_{ij\sigma}
P_{jn\sigma}\nonumber\\
H_{im\sigma} = (\omega - \varepsilon_{i} - U_{i}<n_{i-\sigma}>)\delta_{im} -
t_{im}\nonumber\\
P_{mn\sigma}(\omega) = U_{m}(
^{(ir)}<<n_{m-\sigma}a_{m\sigma}|n_{n-\sigma}
a^{\dagger}_{n\sigma}>>^{(ir)}_{\omega}) U_{n}
\end{eqnarray}
In order to calculate the self-energy operator $M$
self-consistently, we have to express it approximately by the
lower-order GFs. Employing the same pair approximation as
(\ref{eq.146}) (now in the Wannier representation) and the same
procedure of calculation, we arrive at the following expression
for $M$ for a given configuration $(\nu)$
\begin{eqnarray}
\label{eq.191} M^{(\nu)}_{mn\sigma}(\omega) = U_{m}U_{n}{1\over
2\pi^{4}}\int R(\omega_{1},
\omega_{2}, \omega_{3})\nonumber\\
ImG^{(\nu)}_{nm-\sigma}(\omega_{1})ImG^{(\nu)}_{
mn-\sigma}(\omega_{2})ImG^{(\nu)}_{mn\sigma}(\omega_{3});\nonumber\\
R = \frac{d\omega_{1}d\omega_{2}d\omega_{3}}{\omega + \omega_{1} -
\omega_{2} - \omega_{3}}\frac{(1 - n(\omega_{1}))n(\omega_{2})n(
\omega_{3})}{n(\omega_{2} + \omega_{3} - \omega_{1})}
\end{eqnarray}
As we   mentioned previously, all the calculations just presented
were made for a given configuration of atoms in an alloy. All the
quantities in our theory (G, $G^{0}$, P, M) depend on the whole
configuration of the alloy. To obtain a theory of a real
macroscopic sample, we have to average over various
configurations of atoms in the sample. The configurational
averaging cannot be exactly made for a macroscopic sample. Hence
we must resort to an additional approximation. It is obvious that
the self-energy $M$ is in turn a functional of $G$, namely $M =
M[G]$. If the process of making configurational averaging is
denoted by $\bar{G}$, then we have
$$\bar{G} = \bar{G}^{0} + \overline{G^{0}MG}$$
A few words are now appropriate for the description of general
possibilities. The calculations of $\bar{G}^{0}$ can be performed
with the help of any relevant available scheme. In the present
work, for the sake of simplicity, we choose the single-site
CPA\cite{sov58}, namely, we take
\begin{equation}
\label{eq.192} \bar{G}^{0}_{mn\sigma}(\omega) =
N^{-1}\sum_{k}\frac{\exp(ik(R_{m}-R_{n}))} {\omega -
\Sigma^{\sigma}(\omega) -\epsilon(k)}
\end{equation}
Here $\epsilon(k) = \sum^{z}_{n=1}t_{n, 0}\exp(ikR_{n})$, $z$ is
the number of nearest neighbors of the site $0$, and the coherent
potential $\Sigma^{ \sigma}(\omega)$ is the solution of the CPA
self-consistency equations. For the $A_{x}B_{1-x}$, we have
\begin{eqnarray}
\label{eq.193} \Sigma^{\sigma}(\omega) = x\varepsilon^{\sigma}_{A}
+ (1-x)\varepsilon^{ \sigma}_{B} - (\varepsilon^{\sigma}_{A} -
\Sigma^{\sigma})F^{\sigma} (\omega,
\Sigma^{\sigma})(\varepsilon^{\sigma}_{B} - \Sigma^{\sigma});
\nonumber\\
F^{\sigma}(\omega, \Sigma^{\sigma}) = \bar{G}^{0}_{mm\sigma}(\omega)
\end{eqnarray}
Now, let us return to the calculation of the configurationally
averaged total GF $\bar{G}$. To perform the remaining averaging in
the Dyson equation, we use the approximation $$\overline{G^{0}MG}
\approx \bar{G}^{0}\bar{M}\bar{G}$$ The calculation of $\bar{M}$
requires further averaging of the product of matrices. We again
use the prescription of the factorizability there, namely
$$\bar{M} \approx \overline{(U_{m}U_{n})}~ \overline{(ImG)}~
 \overline{(ImG)}~ \overline{(ImG)}$$
However, the quantities $\overline{U_{m}U_{n}}$ entering into $\bar{M}$
are averaged here according to
\begin{eqnarray}
\label{eq.194}
 \overline{U_{m}U_{n}} = U_{2} + (U_{1} - U_{2})\delta_{mn}\nonumber\\
 U_{1} = x^{2}U^{2}_{A} + 2x(1-x)U_{A}U_{B} + (1-x)^{2}U^{2}_{B}\nonumber\\
 U_{2} = xU^{2}_{A} + (1-x)U^{2}_{B}
\end{eqnarray}
The averaged value for the self-energy is
\begin{eqnarray}
\label{eq.195}
\bar{M}_{mn\sigma}(\omega) \\
= {U_{2}\over 2\pi^{4}}\int
R(\omega_{1}, \omega_{2},
\omega_{3})Im\bar{G}_{nm-\sigma}(\omega_{1})Im\bar{G}_{mn-\sigma}(
\omega_{2})Im\bar{G}_{mn\sigma}(\omega_{3}) +\nonumber\\
\frac{U_{1}-U_{2}}{2\pi^{4}}\delta_{mn}\int R(\omega_{1},
\omega_{2},
\omega_{3})Im\bar{G}_{nm-\sigma}(\omega_{1})Im\bar{G}_{mn-\sigma}(\omega_{2}
) Im\bar{G}_{mn\sigma}(\omega_{3}) \nonumber
\end{eqnarray}
The averaged quantities are periodic, so we can introduce the Fourier
transform of them, i.e.
$$\bar{M}_{mn\sigma}(\omega) = N^{-1}\sum_{k}\bar{M}_{k\sigma}( \omega)
\exp(ik(R_{m} - R_{n}))$$ and similar formulae for $\bar{G}$ and
$\bar{G}^{0}$. Performing the configurational averaging of the
Dyson equation and Fourier transforming of the resulting
expressions according to the above rules, we obtain
\begin{equation}
\label{eq.196} \bar{G}_{k\sigma}(\omega) = [\omega - \epsilon(k) -
\Sigma^{\sigma}(\omega) - \bar{M}_{k\sigma}( \omega)]^{-1}
\end{equation}
where
\begin{eqnarray}
\label{eq.197} \bar{M}_{k\sigma}( \omega) = {1\over
2\pi^{4}}\sum_{pq} \int R(\omega_{1}, \omega_{2},
\omega_{3})N^{-2}Im\bar{G}_{p-q-\sigma}(\omega_{1})
Im\bar{G}_{q-\sigma}(\omega_{2})\nonumber\\
~[U_{2}Im\bar{G}_{k+p\sigma}(\omega_{3}) + {(U_{1}-U_{2})\over N}
\sum_{g}Im\bar{G}_{k+p-g}(\omega_{3})]
\end{eqnarray}
The simplest way to obtain an explicit solution for the
self-energy$\bar{M}$ is to start with a suitable initial trial
solution as it was done for the periodic case. For a disordered
system, it is reasonable to use, as the first iteration
approximation the so-called Virtual Crystal Approximation(VCA):
$${-1\over \pi}Im\bar{G}^{VCA}_{k\sigma}(\omega +i\epsilon)
\approx \delta(\omega - E^{\sigma}_{k})$$ where for the binary
alloy $A_{x}B_{1-x}$ this approximation reads $$\bar{V} = xV^{A} +
(1-x)V^{B};\quad E^{\sigma}_{k} =\bar{\varepsilon} ^{\sigma}_{i} +
\epsilon(k);$$\\ $$\bar{\varepsilon}^{\sigma}_{i} =
x\varepsilon^{\sigma}_{A} + (1-x) \varepsilon^{\sigma}_{B}$$ Note,
that the use of VCA here is by no means a solution of the
correlation problem in VCA. It is only the use of the VCA for the
parametrization of the problem, to start with VCA input
parameters. After the integration of (\ref{eq.197}) the final
result for the self-energy is
\begin{eqnarray}
\label{eq.198}
\bar{M}_{k\sigma}( \omega) = \\
{U_{2}\over
N^{2}}\sum_{pq}\frac{n(E^{-\sigma}_{ p+q})[1 - n(E^{-\sigma}_{q})
- n(E^{\sigma}_{k+p})] + n(E^{\sigma}_{k+p})
n(E^{-\sigma}_{q})}{\omega + E^{-\sigma}_{p+q} - E^{-\sigma}_{q} -
E^{\sigma}_{k+p}} + \nonumber\\
{(U_{1}-U_{2})\over N^{3}}\sum_{pqg}\frac{n(E^{-\sigma}_{p+q}) [1
- n(E^{-\sigma}_{q}) - n(E^{\sigma}_{k+p-g})] +
n(E^{\sigma}_{k+p-g}) n(E^{-\sigma}_{q})}{\omega +
E^{-\sigma}_{p+q} - E^{-\sigma}_{q} - E^{\sigma}_{k+p-g}}
\nonumber
\end{eqnarray}
It is to be emphasized that the equations (\ref{eq.195}) -
(\ref{eq.198}) give the general microscopic self-consistent
description of inelastic electron-electron scattering in an alloy
in the spirit of the CPA. We  took into account the randomness not
only through the parameters of the Hamiltonian but also in a
self-consistent way through the configurational dependence of the
self-energy operator.
\subsection{Electron-Lattice Interaction and MTBA}
To understand quantitatively the electrical, thermal, and
superconducting properties of metals and their alloys, one needs a
proper description of an electron-lattice interaction
too~\cite{kuz27},~\cite{kuz64},\cite{acqua4}.  A systematic,
self-consistent simultaneous treatment of the electron-electron
and electron-phonon interaction plays an important role in recent
studies of strongly correlated systems. It was argued from
different points of view that   to understand quantitatively the
phenomenon of high-temperature superconductivity one needs a
proper inclusion of electron-phonon interaction, too. A lot of
theoretical searches for the relevant mechanism of high
temperature superconductivity deal with strong electron-phonon
interaction models. The natural approach to the description of
superconductivity in that type of compounds is the modified
tight-binding approximation (MTBA)~\cite{kuz27}, \cite{kuz64}.
The papers ~\cite{kuze},\cite{kuz27}, \cite{wys1},\cite{wys2}
contain a self-consistent microscopic theory of the normal and
superconducting properties of transition metals and strongly
disordered binary alloys in the framework of the Hubbard Model
(\ref{eq.49}) and random Hubbard model (\ref{eq.184}). Here we
  derive a system of equations for the superconductivity for
tight-binding electrons of a transition metal interacting with
phonons within the IGF approach. We write the total Hamiltonian
of the electron-ion system as the sum~\cite{kuze}
\begin{equation}
\label{eq.199}
 H = H_{e} + H_{i} + H_{e-i}
\end{equation}
where $H_{e}$ is the electron part of the Hamiltonian represented
by the Hubbard operator (\ref{eq.49}). The Hamiltonian of an ion
subsystem and the operator of   electron-ion interaction have the
form
\begin{eqnarray}
\label{eq.200} H_{i} = \frac {1}{2} \sum_{n} \frac {P^{2}_{n}}{2M}
+
{1\over 2} \sum_{mn \alpha \beta} \Phi^{\alpha\beta}_{nm}u^{\alpha}_{n}u^{\beta}_{m} \\
\label{eq.201} H_{e-i} = \sum_{\sigma} \sum_{n,i \not = j}
V^{\alpha}_{ij} ( \vec R^{0}_{n})
a^{\dagger}_{i\sigma}a_{j\sigma}u^{\alpha}_{n}
\end{eqnarray}
where
\begin{equation}
\label{eq.202} \sum_{n} V^{\alpha}_{ij} ( \vec
R^{0}_{n})u^{\alpha}_{n} = \frac {\partial t_{ij}(\vec
R_{ij}^{0})}{\partial R_{ij}^{0}} (\vec u_{i} - \vec u_{j})
\end{equation}
Here $P_{n}$ is the momentum operator, $M$ is the mass of an ion,
and $u_{n}$ is the displacement of the ion from the equilibrium
position at the lattice site $R_{n}$.\\ In  a more convenient
notation  the electron-phonon interaction Hamiltonian in the
modified tight-binding approximation reads~\cite{kuz27}
\begin{equation}
\label{eq.203} H_{e-i} = \sum_{\nu\sigma}\sum_{kq} V^{\nu}(\vec
k, \vec k + \vec q)Q_{\vec q\nu}a^{\dagger}_{k+q\sigma}
a_{k\sigma}
\end{equation}
where
\begin{equation}
\label{eq.204} V^{\nu}(\vec k, \vec k + \vec q) = \frac{2iq_{0}}{(
N M )^{1/2}}\sum_{\alpha} t(\vec a_{\alpha})e^{\alpha}_{\nu}(\vec
q)[\sin \vec a_{\alpha} \vec k - \sin \vec a_{\alpha} (\vec k -
\vec q)]
\end{equation}
here $q_{0}$ is the Slater coefficient~\cite{kuz27} having the
origin   in the exponential decrease of the wave functions of
$d$-electrons, N is the number of unit cells in the crystal, and M
is the ion mass. The quantities $\vec e_{\nu}(\vec q)$ are
polarization vectors of the phonon modes.\\ For the ion subsystem,
we have
\begin{equation}
\label{eq.205} H_{i} = \frac{1}{2} \sum_{q\nu}
(P^{\dagger}_{q\nu}P_{q\nu} + \omega^{2}(\vec q
\nu)Q^{\dagger}_{q\nu}Q_{q\nu})
\end{equation}
where $P_{q\nu}$ and $Q_{q\nu}$ are   normal coordinates, and
$\omega(q\nu)$ are   acoustical phonon frequencies. It is
important to note that in spite of the fact that in Hubbard model
(\ref{eq.49}), the $d$- and $s(p)$-bands are replaced by one
effective band , the $s$-electrons give rise to screening effects
and were taken into effects by choosing the proper values of $U$
and the acoustical phonon frequencies.
\subsection{Equations of Superconductivity}
To derive the superconductivity equations, we use the IGF method
of Section 3 in which the decoupling procedure is carried out only
for approximate calculation of the mass operator of the matrix
electron GF. According to the arguments of Section 4.3 ,
eqn.(\ref{eq.64}), the relevant matrix GF is of the form
\begin{eqnarray}
\label{eq.206}
 G_{ij} (\omega) =
\pmatrix{ G_{11}&G_{12}\cr G_{21}&G_{22}\cr} = \\ \pmatrix{
<<a_{i\sigma}\vert a^{\dagger}_{j\sigma}>> & <<a_{i\sigma}\vert
a_{j-\sigma}>> \cr <<a^{\dagger}_{i-\sigma}\vert
a^{\dagger}_{j\sigma}>> & <<a^{\dagger}_{i-\sigma}\vert
a_{j-\sigma}>> \cr} \nonumber
\end{eqnarray}
As was discussed in Section 4.4, with this definition, one
introduces the so-called anomalous (off-diagonal) GFs which fix
the relevant BCS-Bogoliubov vacuum and select   proper symmetry
broken solutions. Differentiation of $G_{ij}(t-t')$ with respect
to the first time gives for the Fourier components of the
equations of motion
\begin{eqnarray}
\label{eq.207} \sum_{j}(\omega \delta_{ij} - t_{ij})<<a_{j\sigma}
\vert a^{\dagger}_{i'\sigma}>> = \delta_{ii'} + \\  \nonumber
U<<a_{i\sigma} n_{i-\sigma} \vert a^{\dagger}_{i'\sigma}>> +
\sum_{nj} V_{ijn} <<a_{j\sigma} u_{n} \vert a^{\dagger}_{i'\sigma}>> \\
\label{eq.208} \sum_{j}(\omega \delta_{ij} +
t_{ij})<<a^{\dagger}_{j-\sigma} \vert a^{\dagger}_{i'\sigma}>> =
\\  \nonumber
-U<<a^{\dagger}_{i-\sigma} n_{i\sigma} \vert
a^{\dagger}_{i'\sigma}>> + \sum_{nj} V_{jin}
<<a^{\dagger}_{j-\sigma} u_{n} \vert a^{\dagger}_{i'\sigma}>>
\end{eqnarray}
Following the general strategy of the IGF method, we separate the
renormalization of the electron energy in the
Hartree-Fock-Bogoliubov generalized mean field approximation
(including anomalous averages) from the renormalization of
higher-order due to inelastic scattering. For this, we introduce
irreducible parts of the GF in accordance with the definition ( as
an example, we take two of the four Green functions)
\begin{eqnarray}
\label{eq.209}
(^{(ir)}<<a_{i\sigma}a^{\dagger}_{i-\sigma}a_{i-\sigma} \vert
a^{\dagger}_{i'\sigma}>>_ {\omega} )   =
<<a_{i\sigma}a^{\dagger}_{i-\sigma}a_{i-\sigma}\vert
a^{\dagger}_{i'\sigma}>>_{\omega} - \\ \nonumber
-<n_{i-\sigma}>G_{11} + <a_{i\sigma}a_{i-\sigma}>
<<a^{\dagger}_{i-\sigma} \vert a^{\dagger}_{i'\sigma}>>_{\omega}\\
\nonumber
(^{(ir)}<<a^{\dagger}_{i\sigma}a_{i\sigma}a^{\dagger}_{i-\sigma}
\vert a^{\dagger}_{i'\sigma}>>_ {\omega} )  =
<<a^{\dagger}_{i\sigma}a_{i\sigma}a^{\dagger}_{i-\sigma}\vert
a^{\dagger}_{i'\sigma}>>_{\omega} - \\ \nonumber
-<n_{i\sigma}>G_{21} +
<a^{\dagger}_{i\sigma}a^{\dagger}_{i-\sigma}> <<a_{i\sigma} \vert
a^{\dagger}_{i'\sigma}>>_{\omega} \nonumber
\end{eqnarray}
From this definition it follows that this way of introducing the
IGF broadens the initial algebra of the operators and the initial
set of the GFs.  This means that ``actual" algebra of the
operators must include the anomalous terms from the beginning,
namely:  $(a_{i\sigma}$, $a^{\dagger}_{i\sigma}$, $n_{i\sigma}$,
$a^{\dagger}_{i\sigma}a^{\dagger}_{i-\sigma}$,
$a_{i-\sigma}a^{\dagger}_{i\sigma}) $. The corresponding initial
GF is the form (\ref{eq.206}). The choice of the irreducible
parts of the GF in (\ref{eq.209}) is specified by the
"orthogonality" constraint (31), which makes it possible to
introduce unambiguously the irreducible parts and make the
inhomogeneous terms in the equations for them vanish. Using
(\ref{eq.209}) , we rewrite eqs.(\ref{eq.207}) and (\ref{eq.208})
in the form
\begin{eqnarray}
\label{eq.210} \sum_{j} \Bigl((\omega  -
U<n_{j-\sigma}>)\delta_{ij} -
t_{ij} \Bigr)<<a_{j\sigma} \vert a^{\dagger}_{i'\sigma}>> = \delta_{ii'} \\
\nonumber - U<a_{i\sigma}a_{i-\sigma}><<a_{i\sigma} \vert
a^{\dagger}_{i'\sigma}>> +
\sum_{j} << (a_{j\sigma} \rho_{ij\sigma} )^{(ir)} \vert a^{\dagger}_{i'\sigma}>> \\
\label{eq.211} \sum_{j} \Bigl((\omega + U<n_{j\sigma}>)
\delta_{ij} + t_{ji} \Bigr)<<a^{\dagger}_{j-\sigma} \vert
a^{\dagger}_{i'\sigma}>> =
\\  \nonumber
+U< a^{\dagger}_{i\sigma}a^{\dagger}_{i-\sigma}><<a_{i\sigma}
\vert a^{\dagger}_{i'\sigma}>> - \sum_{j} << (\rho_{ji-\sigma}
a^{\dagger}_{j-\sigma})^{(ir)}  \vert a^{\dagger}_{i'\sigma}>>
\end{eqnarray}
where
\begin{equation}
\label{eq.212} \rho_{ij\sigma} = Un_{j-\sigma} \delta_{ij} +
\sum_{n}V_{ijn}u_{n}( 1 - \delta_{ij})
\end{equation}
In the representation of the Nambu operators~\cite{kuze}
\begin{equation}
\label{eq.213} \psi_{i,-\sigma} = {a_{i-\sigma} \choose
a^{\dagger}_{i\sigma}} \quad \psi^{\dagger}_{i,-\sigma} = (
a^{\dagger}_{i-\sigma}, a_{i\sigma} )
\end{equation}
the equation of motion for   GF (\ref{eq.210}) can be represented
as
\begin{eqnarray}
\label{eq.214} \sum_{j}(\omega \tau_{0}\delta_{ij} - t_{ij}
\tau_{3} - \Sigma^{c}_{i\sigma}) <<\psi_{j} \vert
\psi^{\dagger}_{i'}>> = \\ \nonumber \delta_{ii'}\tau_{0} +
\sum_{j} << ( \rho_{ij} \tau_{3} \psi_{j} )^{(ir)} \vert
\psi^{\dagger}_{i'}>>
\end{eqnarray}
Here the Hartree-Fock-Bogoliubov elastic Coulomb term
(\ref{eq.64}) is of the form
\begin{equation}
\label{eq.215} \Sigma^{c}_{i\sigma} = - U \tau_{3}
<\psi_{i,-\sigma} \psi^{\dagger}_{i,-\sigma}> \tau_{3} + {U \over
2} ( \tau_{0} + \tau_{3} )
\end{equation}
To calculate the irreducible matrix GF in (\ref{eq.214}), we write
down for it the equation of motion with respect to the second
time $t'$ and then separate the irreducible part with respect to
the operators on the right-hand-side of the corresponding GF. This
gives the Dyson equation in the matrix form
\begin{equation}
\label{eq.216} \hat G_{ii'}(\omega) = \hat G^{0}_{ii'}(\omega) +
\sum_{jj'} \hat G^{0}_{ij} (\omega) \hat M_{jj'}( \omega) \hat
G_{j'i'} (\omega)
\end{equation}
The generalized mean field GF $G^{0}$ and the mass operator are defined by
\begin{equation}
\label{eq.217} \sum_{j}(\omega \tau_{0}\delta_{ij} - t_{ij}
\tau_{3} - \Sigma^{c}_{i\sigma}) G^{0}_{ji'} =
\delta_{ii'}\tau_{0}
\end{equation}
\begin{equation}
\label{eq.218} M_{kk'} = \sum_{jj'} (<< ( \rho_{kj} \tau_{3}
\psi_{j} )^{(ir)} \vert (\psi^{\dagger}_{j'} \tau_{3}
\rho_{j'k'})^{(ir)}>>)^{(p)}_{\omega}
\end{equation}
The explicit expression for the mass operator (\ref{eq.218}) is of
the form
\begin{eqnarray}
\label{eq.219} \hat M_{ii'} (\omega) = \\ \sum_{jj'} \pmatrix{
(^{(ir)}<< a_{j \uparrow} \rho_{ij\uparrow} \vert
\rho_{j'i'\uparrow} a^{\dagger}_{j'\uparrow} >>^{(ir)})^{(p)} &
(^{(ir)}<< a_{j \uparrow} \rho_{ij\uparrow} \vert
\rho_{j'i'\downarrow} a_{j'\downarrow} >>^{(ir)})^{(p)} \cr
(^{(ir)}<< a^{\dagger}_{j \downarrow} \rho_{ji\downarrow} \vert
\rho_{j'i'\uparrow} a^{\dagger}_{j'\uparrow}
>>^{(ir)})^{(p)} & (^{(ir)}<< a^{\dagger}_{j \downarrow} \rho_{ji\downarrow} \vert
\rho_{i'j'\downarrow} a_{j'\downarrow} >>^{(ir)})^{(p)} \cr}
\nonumber
\end{eqnarray}
The mass operator (\ref{eq.219}) describes inelastic scattering of
electrons ( the elastic part is contained in
$\Sigma^{c}_{i\sigma}$ ) on fluctuations of the density of a total
electron-ion charge in the lattice. To find an approximating
expression for the mass operator (\ref{eq.219}), we adopt the
following trial approximation
\begin{equation}
\label{eq.220} <\rho_{j'i'\sigma}(t)a^{\dagger}_{j'\sigma}(t)
a_{j\sigma} \rho_{ij\sigma}>^{(ir)} \approx <\rho_{j'i'\sigma}(t)
\rho_{ij\sigma}><a^{\dagger}_{i'\sigma}(t)a_{j\sigma}>
\end{equation}
This approximation was made in the spirit of the approximation of
"two interacting modes" and means ignoring the renormalization of
the vertex, {\it i.e.}, the correlation in the propagation of an
electron (hole) and the
propagation of charge density fluctuations. \\
Writing down further spectral representation for the correlation
functions in (\ref{eq.220}), we represent the mass operator by the
sum
\begin{equation}
\label{eq.221} \hat M_{ii'} (\omega) = \hat M^{1}_{ii'} (\omega) +
\hat M^{2}_{ii'} (\omega)
\end{equation}
The first contribution $M^{1}$ has a form characteristic of an interacting
electron-phonon system
\begin{eqnarray}
\label{eq.222} M^{1}_{ii'}( \omega) = \sum_{nn'} \sum_{jj'}
V_{ijn} V_{j'i'n'}{1 \over 2} \int_{-\infty}^{+\infty} \frac
{d\omega_{1}d{\omega}_{2}}{\omega - \omega_{1} - \omega_{2}}
(\cot \frac{\beta \omega_{1}}{2} + \tan \frac{\beta \omega_{2}}{2}) \nonumber\\
(-{1 \over \pi}Im <<u_{n} \vert u_{n'}>>_{\omega_{2}}) (-{1 \over
\pi} \tau_{3} Im<<\psi_{j} \vert
\psi^{\dagger}_{j'}>>_{\omega_{1}}\tau_{3})
\end{eqnarray}
The contribution $M^{2}_{ii'}$ has a more complicated structure
\begin{equation}
\label{eq.223} M^{2}_{ii'} = {U^2 \over 2}
\int_{-\infty}^{+\infty} \frac {d\omega_{1}d{\omega}_{2}}{\omega
- \omega_{1} - \omega_{2}} (\cot \frac{\beta \omega_{1}}{2} +
\tan \frac{\beta \omega_{2}}{2}) \pmatrix{ m_{11}&m_{12}\cr
m_{21}&m_{22}\cr}
\end{equation}
where
\begin{eqnarray}
m_{11} = (-{1 \over \pi} Im <<n_{i\downarrow} \vert
n_{i'\downarrow}>>_{\omega_{2}}) (-{1 \over \pi} Im
<<a_{i\uparrow} \vert a^{\dagger}_{i'\uparrow}>>_{\omega_{1}})
\nonumber \\
m_{12} = ({1 \over \pi}Im <<n_{i\downarrow} \vert
n_{i'\uparrow}>>_{\omega_{2}}) (-{1 \over \pi} Im<< a_{i\uparrow}
\vert a^{\dagger}_{i'\downarrow}>>_{\omega_{1}})
\nonumber \\
m_{21} = ({1 \over \pi}Im <<n_{i\uparrow} \vert
n_{i'\downarrow}>>_{\omega_{2}}) (-{1 \over \pi} Im<<
a_{i\downarrow} \vert a^{\dagger}_{i'\uparrow}>>_{\omega_{1}})
\nonumber \\
m_{22} = (-{1 \over \pi}Im <<n_{i\uparrow} \vert
n_{i'\uparrow}>>_{\omega_{2}}) (-{1 \over \pi}
Im<<a_{i\downarrow} \vert
a^{\dagger}_{i'\downarrow}>>_{\omega_{1}}) \nonumber
\end{eqnarray}
The equations (\ref{eq.216}) and (\ref{eq.221}) constitute a
self-consistent system of equations for the single-particle GF of
the Hubbard model on a vibrating lattice. Note  that these
equations of superconductivity can be in an obvious way
transformed to the standard form of the Eliashberg
equations~\cite{kuz27}. The numerical calculations of the
electron-phonon spectral function $\alpha^{2}(\omega)F(\omega)$
for a few transition metals were done in ref.~\cite{kuz64}. It is
worthy to emphasize that in paper~\cite{wys2} a very detailed
microscopic theory of the strong coupling superconductivity in
highly disordered transition metal  alloys was developed on the
basis of the IGF method within the MTBA reformulated
approach~\cite{wys1}. The Eliashberg-type strong coupling
equations for highly disordered alloys were derived. It was shown
that the electron-phonon spectral function in alloys is modified
strongly. Thus, the self-consistent system of superconductivity
equations obtained in the Wannier representation makes it
possible to investigate real transition metals, their alloys, and
compounds from a unified point of view.
\section{Quasi-Particle  Dynamics of Anderson Models}
\subsection{Quasi-Particle Dynamics of SIAM} In this Section, we
  consider the many-body quasi-particle dynamics of the
Anderson impurity model at finite temperatures in the framework
of the equation-of-motion method. In spite of many theoretical
efforts, there is no complete solution of the dynamic problem for
the "simple" Anderson/Hubbard model. One of the main reasons for
this is that it has been recognized relatively recently only that
the simplicity of the Anderson model manifests itself not in the
many-body dynamics ( the right definition of quasi-particles via
the poles of   GF; see Section 6.1 ) but rather at quite a
different level - in the dynamics of   two-particle scattering,
resulting in the elegant Bethe-ansatz solution (for the
relativistic  spectrum linear in $k$ ), which gives the static
characteristics (static susceptibility, specific heat, {\it etc}).
In this sense, as to the true many-body dynamics, the complete
analytic  solution of this problem is still   quite an open
subject. This Section is primarily devoted to the analysis of the
relevant many-body dynamic  solution  of the SIAM and its correct
functional structure. We wish to determine which solution
actually arises  both from the self-consistent many-body approach
and intrinsic nature of the model itself. We believe highly that
before numerical calculations of the spectral intensity of the
Green function at low energy and low temperature it is quite
important to have a consistent and closed analytic representation
for the one-particle GF of the SIAM and Hubbard model. The
paper~\cite{bic2} clearly shows the importance of the calculation
of the GF  and spectral densities for SIAM in a self-consistent
way. An alternative approach to dynamics of the Anderson model
was formulated  within a modified version of the Kadanoff-Baym
method ~\cite{neal1},~\cite{neal2}. Unfortunately, the Neal
approach also have certain drawbacks.\\ A proper many-body
description of dynamic correlations is very actual also for the
investigation of the dynamics of the many-impurity Anderson model,
where standard advanced many-body methods do  not work properly
in  usual formulation. Recently, a lot of efforts were devoted to
a better understanding of the static and dynamic  properties of
the Anderson model  in the context of many-impurity
case~\cite{kuz8}. This field is quite important for the
description of magnetic properties of anomalous rare-earth
compounds\cite{akva1},\cite{akva2}. The problem of an adequate and
consistent description of dynamics of single-impurity and
many-impurity Anderson models ( SIAM and MIAM ) and other models
of correlated lattice electrons   was not yet solved analytically
completely . During the last decades, a lot of theoretical papers
were published, attacking the Anderson model by many refined
many-body  analytic  methods. Nevertheless, a fully consistent
dynamic  analytic  solution in the closed form for a
single-particle propagator of SIAM is still lacking. In this
Section, the problem of consistent analytic  description of the
many-body dynamics of SIAM is discussed in the framework of the
equation-of-motion approach for double-time thermodynamic GFs. In
addition to the IGF approach, we find a new exact identity
relating the one-particle and many-particle GFs. Using this
identity, we present a consistent and general scheme for
construction of generalized solutions of SIAM.  A new approach
for the complex expansion for the single-particle propagator in
terms of Coulomb repulsion $U$ and hybridization $V$ is proposed.
Using the exact identity, an essentially new many-body dynamic
solution of SIAM is derived. This approach offers a new way for
the systematic construction of   approximate interpolation
dynamic  solutions of   strongly correlated electron systems.
\subsection{IGF Approach to SIAM}
After discussing some of the basic facts about the correct
functional structure of the relevant dynamic solution of
correlated electron models we are looking for, described in
previous Sections, we   give a similar consideration for SIAM. It
was shown in ~\cite{kuz7}, using the minimal algebra of relevant
operators, that the construction of the GMFs for SIAM is quite
nontrivial for the strongly correlated case, and it is rather
difficult to get it from an intuitive physical point of view. Let
us consider first the following matrix GF

\begin{equation}
\label{eq.224}
 \hat G (\omega) =
\pmatrix{ <<c_{k\sigma}\vert c^{\dagger}_{k\sigma}>> &
<<c_{k\sigma}\vert f^{\dagger}_{0\sigma}>> \cr <<f_{0\sigma}\vert
c^{\dagger}_{k\sigma}>> & <<f_{0\sigma}\vert
f^{\dagger}_{0\sigma}>> \cr}
\end{equation}
Performing the first-time differentiation and defining the
irreducible GF
\begin{eqnarray}
\label{eq.225}
(^{(ir)}<<f_{0\sigma}f^{\dagger}_{0-\sigma}f_{0-\sigma} \vert
f^{\dagger}_{0\sigma}>>_ {\omega})  =
 <<f_{0\sigma}f^{\dagger}_{0-\sigma}f_{0-\sigma}\vert
f^{\dagger}_{0\sigma}>>_{\omega} - \\ \nonumber -<n_{0-\sigma}>
<<f_{0\sigma} \vert f^{\dagger}_{0\sigma}>>_{\omega}
\end{eqnarray}
we obtain the following equation of motion in the matrix form
\begin{equation}
\label{eq.226} \sum_{p} \hat F_{p}(\omega ) \hat G_{p} (\omega) =
\hat 1 + U \hat D^{(ir)} (\omega)
\end{equation}
where all definitions are rather evident. Proceeding  further
with the IGF technique, the equation of motion (\ref{eq.226}) is
exactly rewritten in the form of the Dyson equation
\begin{equation}
\label{eq.227} \hat G (\omega) = \hat G^{0}(\omega) +  \hat
G^{0}(\omega) \hat M( \omega) \hat G (\omega)
\end{equation}
The generalized mean field GF $G^{0}$  is defined by
\begin{equation}
\label{eq.228} \sum_{p} F_{p}(\omega) G_{p}^{0} (\omega) = \hat I
\end{equation}
The explicit solutions for diagonal elements of $G^{0}$ are
\begin{eqnarray}
\label{eq.229} <<f_{0\sigma} \vert f^{\dagger}_{0\sigma}>>^{0}_
{\omega} = \Bigl ( \omega -E_{0\sigma} - Un_{-\sigma} -
S (\omega)) \Bigr )^{-1} \\
\label{eq.230} <<c_{k\sigma} \vert
c^{\dagger}_{k\sigma}>>^{0}_{\omega} = \Bigl ( \omega
-\epsilon_{k} - \frac { |V_{k}|^{2}}{\omega - E_{0\sigma} -
Un_{-\sigma} } \Bigr )^{-1}
\end{eqnarray}
where
\begin{equation}
\label{eq.231} S(\omega) = \sum_{k} \frac { |V_{k}|^{2}}{\omega -
\epsilon_{k}}
\end{equation}
The mass operator, which describes inelastic scattering processes, has the
following matrix form
\begin{equation}
\label{eq.232}
 \hat M (\omega) =
\pmatrix{ 0&0\cr 0&M_{0\sigma}\cr}
\end{equation}
where
\begin{equation}
\label{eq.233} M_{0\sigma} = U ^{2} ( ^{(ir)}<<
f_{0\sigma}n_{0-\sigma} | f^{\dagger}_{0\sigma}n_{0-\sigma}
>>^{(ir)}_{\omega} )^{(p)}
\end{equation}
From the formal solution  of the Dyson equation (38) one obtains
\begin{eqnarray}
\label{eq.234} <<f_{0\sigma} \vert f^{\dagger}_{0\sigma}>>_
{\omega} =
 \Bigl ( \omega -E_{0\sigma} - Un_{-\sigma} - M_{0\sigma} -
S(\omega)  \Bigr )^{-1} \\
\label{eq.235} <<c_{k\sigma} \vert
c^{\dagger}_{k\sigma}>>_{\omega} = \Bigl ( \omega -\epsilon_{k} -
\frac { |V_{k}|^{2}}{\omega - E_{0\sigma} - Un_{-\sigma}-
M_{0\sigma}} \Bigr )^{-1}
\end{eqnarray}
To calculate the self-energy in a self-consistent way, we have to
  approximate it by lower-order GFs. Let us start by
analogy with the Hubbard model with a pair-type approximation
(\ref{eq.146})
\begin{eqnarray}
\label{eq.236} M_{0\sigma}(\omega) = \\ U^2 \int
\frac{d\omega_{1}d\omega_{2}d\omega_{3}}{\omega + \omega_{1} -
\omega_{2} - \omega_{3}}\nonumber\\
~[n(\omega_{2})n(\omega_{3}) + n(\omega_{1})(1 - n(\omega_{2}) -
n(\omega_{3}))]
g_{0-\sigma}(\omega_{1})g_{0\sigma}(\omega_{2})g_{0-\sigma}(\omega_{3})
\nonumber
\end{eqnarray}
where we   used the notation  $$g_{0\sigma}(\omega) = -{1 \over
\pi} Im << f_{0\sigma} | f^{\dagger}_{0\sigma} >>_{\omega }$$ The
equations (\ref{eq.227}) and (\ref{eq.236}) constitute a closed
self-consistent system of equations for the single-electron GF
for SIAM model, but only for weakly correlated  limit. In
principle, we can use, on the r.h.s. of (\ref{eq.236}), any
workable first iteration-step form of the GF and find a solution
by repeated  iteration. If we take for the first iteration step
the expression
\begin{equation}
\label{eq.237} g_{0\sigma}(\omega) \approx \delta(\omega -
E_{0\sigma} - Un_{-\sigma}),
\end{equation}
we get, for the self-energy, the explicit expression
\begin{equation}
\label{eq.238} M_{0\sigma}(\omega) = U^{2} \frac {n(E_{0\sigma} +
Un_{-\sigma})(1 - n(E_{0\sigma} + Un_{-\sigma}))}{\omega -
E_{0\sigma} - Un_{-\sigma}} = U^{2}N_{-\sigma}(1 -
N_{-\sigma})G^{0}_{\sigma}(\omega)
\end{equation}
where $N_{-\sigma} = n(E_{0\sigma} + Un_{-\sigma})$. This is the
well-known "atomic" limit of the self-energy.\\ Let us try again
another type of the approximation for $M$. The approximation which
we will use reflects the interference between the one-particle
branch and the collective one
\begin{eqnarray}
\label{eq.239}
<f_{0\sigma}(t)f^{\dagger}_{0-\sigma}(t)f_{0-\sigma}(t)
f^{\dagger}_{0-\sigma}f_{0-\sigma}f^{\dagger}_{0\sigma}>^{(ir)} \approx \nonumber\\
<f^{\dagger}_{0\sigma}(t)f_{0\sigma}><n_{0-\sigma}(t)n_{0-\sigma}> + \nonumber\\
<f^{\dagger}_{0-\sigma}(t)f_{0-\sigma}><f^{\dagger}_{0-\sigma}(t)f_{0\sigma}(t)
f^{\dagger}_{0\sigma}f_{0-\sigma}> +\nonumber\\
<f^{\dagger}_{0-\sigma}(t)f_{0-\sigma}><f_{0-\sigma}(t)f_{0\sigma}(t)
f^{\dagger}_{0\sigma}f^{\dagger}_{0-\sigma}>
\end{eqnarray}
If we retain  only the first term  in (\ref{eq.239}) and make use
of the same iteration as in (\ref{eq.237}), we obtain
\begin{equation}
\label{eq.240} M_{0\sigma}(\omega) \approx U^{2} \frac {(1 -
n(E_{0\sigma} + Un_{-\sigma}))}{\omega - E_{0\sigma} -
Un_{-\sigma}} < n_{0-\sigma} n_{0-\sigma}>
\end{equation}
If we retain the second term in (\ref{eq.239}), we obtain
\begin{eqnarray}
\label{eq.241} M_{0\sigma}(\omega) = U^2 \int_{-\infty}^{+\infty}
d\omega_{1}d{\omega}_{2}\frac{1 + N(\omega_{1}) - n(\omega_{2})}
{\omega - \omega_{1} - \omega_{2}}\nonumber\\
(-{1 \over \pi}Im <<S^{\pm}_{0} \vert S^{\mp}_{0}>>_{\omega_{1}})\nonumber\\
\Bigl(-{1 \over \pi}Im<<f_{0\sigma}\vert
f^{\dagger}_{0\sigma}>>_{\omega_{2}} \Bigr)
\end{eqnarray}
where the following notation was been used:
$$S^{+}_{0} = f^{\dagger}_{0\uparrow}f_{0\downarrow}; \quad
S^{-}_{0} = f^{\dagger}_{0\downarrow}f_{0\uparrow}$$ It is
possible to rewrite (\ref{eq.241}) in a more convenient way now
\begin{eqnarray}
\label{eq.242} M_{0\sigma}(\omega) = U^2 \int d\omega'
(\cot \frac{\omega - \omega'}{2T} + \tan \frac{\omega'}{2T})\nonumber\\
\Bigl(- \frac{1}{\pi} Im \chi^{\mp \pm} ( \omega -
\omega')g_{0\sigma} (\omega') \Bigr)
\end{eqnarray}
The equations (\ref{eq.227}) and (\ref{eq.242}) constitute a
self-consistent system of equations for the single-particle GF of
SIAM. Note  that spin-up and spin-down electrons are correlated
when they occupy the impurity level. So, this really improves the
H-F theory in which just these correlations were missed. The role
of electron-electron correlation becomes much more crucial for the
case of strong correlation.
\subsection{SIAM. Strong Correlation}
The simplest relevant algebra of the operators used for the
description of the strong correlation has a similar form as for
that of the Hubbard model (\ref{eq.154}). Let us represent the
matrix GF (\ref{eq.224}) in the following form
\begin{equation}
\label{eq.243} \hat G (\omega) = \sum_{\alpha \beta} \pmatrix{
<<c_{k\sigma}\vert c^{\dagger}_{k\sigma}>> & <<c_{k\sigma}\vert
d^{\dagger}_{0 \beta \sigma}>> \cr <<d_{0 \alpha \sigma}\vert
c^{\dagger}_{k\sigma}>> & <<d_{0\alpha \sigma}\vert
d^{\dagger}_{0 \beta \sigma}>> \cr}
\end{equation}
Then we proceed by analogy with the calculations for the Hubbard
model. The equation of motion for the auxiliary matrix GF
\begin{eqnarray}
\label{eq.244} \hat F_{\sigma}(\omega) = \\ \pmatrix{
<<c_{k\sigma}\vert c^{\dagger}_{k\sigma}>> & <<c_{k\sigma}\vert
d^{\dagger}_{0 + \sigma}>> &<<c_{k\sigma} \vert
d^{\dagger}_{0-\sigma}>>\cr <<d_{0 + \sigma}\vert
c^{\dagger}_{k\sigma}>> & <<d_{0 + \sigma}\vert d^{\dagger}_{0 +
\sigma}>>& <<d_{0+\sigma} \vert d^{\dagger}_{0-\sigma}>> \cr
<<d_{0-\sigma} \vert c^{\dagger}_{k\sigma}>>& <<d_{0-\sigma} \vert
d^{\dagger}_{0+\sigma}>>&<<d_{0-\sigma} \vert
d^{\dagger}_{0-\sigma}>>\cr} \nonumber
\end{eqnarray}
is of the following form
\begin{equation}
\label{eq.245} \hat E \hat F_{\sigma}(\omega) - \hat I = \hat D
\end{equation}
where the following matrix notation  was used
\begin{eqnarray}
\label{eq.246} \hat E = \pmatrix{(\omega- \epsilon_{k})& - V_{k} &
- V_{k}\cr 0&(\omega - E_{0\sigma} - U_{+})&0\cr
0&0&(\omega - E_{0\sigma} -U_{-})\cr}  \\
\hat I = \pmatrix{ 1&0&0\cr 0&n^{+}_{0-\sigma}&0\cr
0&0&n^{-}_{0-\sigma} \cr}. \nonumber
\end{eqnarray}
$$ U_{\alpha} = \cases {U, &
$\alpha = +$ \cr 0,& $\alpha = -$ \cr}$$

Here $\hat D$ is a higher-order GF, with the following structure
\begin{equation}
\label{eq.247}
 \hat D (\omega) =
\pmatrix{ 0&0&0\cr D_{21}&D_{22}&D_{23}\cr
D_{31}&D_{32}&D_{33}\cr}
\end{equation}
In accordance with the general method of Section 3, we by define
the matrix IGF:
\begin{equation}
\label{eq.248} \hat D^{(ir)}(\omega) = \hat D - \sum_{\alpha} {
A^{+ \alpha} \choose A^{- \alpha} } (G^{\alpha +}_{\sigma} \quad
G^{\alpha - }_{\sigma} )
\end{equation}
Here the notation was used:
\begin{eqnarray}
\label{eq.249} A^{++} = \frac {
<(f^{\dagger}_{0-\sigma}c_{p-\sigma} +
c^{\dagger}_{p-\sigma}f_{0-\sigma})
(n_{0\sigma} - n_{0-\sigma})>}{ <n_{0-\sigma}>}\\
\label{eq.250} A^{--} = \frac {-
<(f^{\dagger}_{0-\sigma}c_{p-\sigma} +
c^{\dagger}_{p-\sigma}f_{0-\sigma})
(1 + n_{0\sigma} - n_{0-\sigma})>}{ <1 - n_{0-\sigma}>}  \\
A^{-+} = A^{++}, \quad A^{+-} = - A^{--} \nonumber
\end{eqnarray}
The generalized mean-field GF is defined by
\begin{equation}
\label{eq.251} \hat E \hat F^{0}_{\sigma}(\omega) - \hat I = 0;
\quad G^{0} =\sum_{\alpha \beta} F^{0}_{\alpha \beta}
\end{equation}
From the last definition we find that
\begin{eqnarray}
\label{eq.252} <<f_{0\sigma} \vert f^{\dagger}_{0\sigma}>>^{0}_
{\omega} = \frac { <n_{0-\sigma}>}{ \omega -E_{0\sigma} - U_{+} -
\sum_{p} V_{p}A^{++}} \bigl ( 1 + \frac { \sum_{p}
V_{p}A^{-+}}{\omega -E_{0\sigma} - U_{-}}
\bigr ) \nonumber \\
+ \frac { 1 - <n_{0-\sigma}>}{ \omega -E_{0\sigma} - U_{-} -
\sum_{p} V_{p}A^{--}} \bigl ( 1 + \frac { \sum_{p}
V_{p}A^{+-}}{\omega -E_{0\sigma} - U_{+}}
\bigr ) \\
\label{eq.253}
 <<c_{k\sigma} \vert c^{\dagger}_{k\sigma}>>^{0}_{\omega} = \bigl (
\omega -\epsilon_{k} -  |V_{k}|^{2}F^{at}(\omega ) \bigr )^{-1}
\end{eqnarray}
where
\begin{equation}
\label{eq.254} F^{at} = \frac { <n_{0-\sigma}>}{ \omega
-E_{0\sigma} - U_{+} } + \frac { 1 - <n_{0-\sigma}>}{ \omega
-E_{0\sigma} - U_{-} }
\end{equation}
For $V_{p} = 0$, we obtain, from solution (\ref{eq.252}), the
atomic solution $F^{at}$. The conduction electron GF
(\ref{eq.253}) also gives a correct expression for $V_{k} = 0$.
\subsection{IGF Method and Interpolation Solution}
To show  explicitly the flexibility of the IGF method, we
consider  a more extended  new algebra of operators from which the
relevant matrix GF should be constructed to make the connection
with the interpolation solution of the Anderson model. For this
aim, let us consider the following equation of motion in the
matrix form
\begin{equation}
\label{eq.255} \sum_{p}F(p,k)G_{p\sigma}(\omega) = I + \sum_{p}
V_{p}D_{p}(\omega)
\end{equation}
where $G$ is the initial $4 \times 4$ matrix GF and $D$ is the
higher-order GF:
\begin{equation}
\label{eq.256} G_{\sigma} =\pmatrix{
G_{11}&G_{12}&G_{13}&G_{14}\cr G_{21}&G_{22}&G_{23}&G_{24}\cr
G_{31}&G_{32}&G_{33}&G_{34}\cr G_{41}&G_{42}&G_{43}&G_{44}\cr}
\end{equation}
Here the following notation was used
\begin{eqnarray}
\label{eq.257} G_{11} = <<c_{k\sigma}|c^{\dagger}_{k\sigma}>>;
\quad G_{12} = <<c_{k\sigma}|f^{\dagger}_{0\sigma}>>;
\nonumber \\
G_{13} = <<c_{k\sigma}|f^{\dagger}_{0\sigma}n_{0-\sigma}>>; \quad
G_{14} = <<c_{k\sigma}|c^{\dagger}_{k\sigma}n_{0-\sigma}>>;
\nonumber \\
G_{21} = <<f_{0\sigma}|c^{\dagger}_{k\sigma}>>; \quad G_{22} =
<<f_{0\sigma}|f^{\dagger}_{0\sigma}>>;
\nonumber \\
G_{23} = <<f_{0\sigma}|f^{\dagger}_{0\sigma}n_{0-\sigma}>>; \quad
G_{24} =
<<f_{0\sigma}|c^{\dagger}_{k\sigma}n_{0-\sigma}>>; \\
G_{31} = <<f_{0\sigma}n_{0-\sigma}|c^{\dagger}_{k\sigma}>>; \quad
G_{32} = <<f_{0\sigma}n_{0-\sigma}|f^{\dagger}_{0\sigma}>>;
\nonumber \\
G_{33} =
<<f_{0\sigma}n_{0-\sigma}|f^{\dagger}_{0\sigma}n_{0-\sigma}>>;
\quad G_{34} =
<<f_{0\sigma}n_{0-\sigma}|c^{\dagger}_{k\sigma}n_{0-\sigma}>>;
\nonumber \\
G_{41} = <<c_{k\sigma}n_{0-\sigma}|c^{\dagger}_{k\sigma}>>; \quad
G_{42} = <<c_{k\sigma}n_{0-\sigma}|f^{\dagger}_{0\sigma}>>;
\nonumber \\
G_{43} =
<<c_{k\sigma}n_{0-\sigma}|f^{\dagger}_{0\sigma}n_{0-\sigma}>>;
\quad G_{44} =
<<c_{k\sigma}n_{0-\sigma}|c^{\dagger}_{k\sigma}n_{0-\sigma}>>;
\nonumber
\end{eqnarray}
We avoid to write down explicitly the relevant 16 GFs, of which
the matrix GF $D$ consist,   for the brevity. For our aims, it is
enough  to proceed forth in the following way. \\ The equation
(\ref{eq.255}) results from the first-time differentiation of the
GF $G$ and is a starting point for the IGF approach. Let us
introduce the irreducible part for the higher-order GF $D$
  in the following way
\begin{equation}
\label{eq.258} D^{(ir)}_{\beta} = D_{\beta} -
\sum_{\alpha}L^{\beta \alpha}G_{\alpha \beta}; \quad ({\alpha,
\beta}) = (1,2,3,4)
\end{equation}
and define the GMF GF according to
\begin{equation}
\label{eq.259} \sum_{p}{\tilde F(p,k)}G^{MF}_{p\sigma}(\omega) =
I, \end{equation} Then, we are able to write down explicitly the
Dyson equation (37) and the exact expression for the self-energy
M  in the matrix form:
\begin{equation}
\label{eq.260} M_{k\sigma}(\omega) = I^{-1}\sum_{p,q}
V_{p}V_{q}\pmatrix{0&0&0&0 \cr 0&0&0&0 \cr 0&0&M_{33}&M_{34} \cr
0&0&M_{43}&M_{44} \cr}I^{-1} \end {equation} Here the matrix $I$
is given by $$I = \pmatrix{ 1&0&0&<n_{0-\sigma}> \cr
0&1&<n_{0-\sigma}>&0 \cr 0&<n_{0-\sigma}>&<n_{0-\sigma}>&0 \cr
<n_{0-\sigma}>&0&0&<n_{0-\sigma}> \cr}$$ and the the matrix
elements of M are of the form:
\begin{eqnarray}
\label{eq.261} \nonumber
 M_{33} =
(<<A^{(ir)}_{1}(p)|B^{(ir)}_{1}(q)>>)^{(p)} , M_{34} =
(<<A^{(ir)}_{1}(p)| B^{(ir)}_{2}(k,q)>>)^{(p)}  \\ \nonumber
M_{43} = (<<A^{(ir)}_{2}(k,p)|B^{(ir)}_{1}(q)>>)^{(p)} ,  M_{44} =
(<<A^{(ir)}_{2}(k,p)|B^{(ir)}_{2}(k,q)>>)^{(p)} \\
\end{eqnarray}
where
\begin{eqnarray}
\label{eq.262} A_{1}(p) = (
c^{\dagger}_{p-\sigma}f_{0\sigma}f_{0-\sigma} -
c_{p-\sigma}f^{\dagger}_{0-\sigma}f_{0\sigma} );
\nonumber \\
A_{2}(k,p) = ( c_{k\sigma}f^{\dagger}_{0-\sigma}c_{p-\sigma} -
c_{k\sigma}c^{\dagger}_{p-\sigma}f_{0-\sigma} );
\nonumber \\
B_{1}(p) = (
f^{\dagger}_{0\sigma}c^{\dagger}_{p-\sigma}f_{0-\sigma} -
f^{\dagger}_{0\sigma}f^{\dagger}_{0-\sigma}c_{p-\sigma} ); \\
B_{2}(k,p) = (
c^{\dagger}_{k\sigma}c^{\dagger}_{p-\sigma}f_{0-\sigma} -
c^{\dagger}_{k\sigma}f^{\dagger}_{0-\sigma}c_{p-\sigma} );
\nonumber
\end{eqnarray}
Since the self-energy $M$ describes the processes of inelastic
scattering of electrons (c-c , f-f, and c-f types), its
approximate representation would be defined by the nature of
physical assumptions about this scattering.\\ To get an idea
about the functional structure of our GMF solution (\ref{eq.259}),
let us write down the matrix element $G^{MF}_{33}$:
\begin{eqnarray}
\label{eq.263} G^{MF}_{33} =
<<f_{0\sigma}n_{0-\sigma}|f^{\dagger}_{0\sigma}n_{0-\sigma}>> = \nonumber \\
\frac {<n_{0-\sigma}>}{\omega -\epsilon^{MF}_{f} - U -
S^{MF}(\omega) - Y(\omega)} + \nonumber \\
\frac {<n_{0-\sigma}> Z(\omega)}{(\omega - \epsilon^{MF}_{f} - U -
S^{MF}(\omega) - Y(\omega))(\omega - E_{0\sigma} - S(\omega))} \\
\label{eq.264}
Y(\omega) = \frac {UZ(\omega)}{\omega - E_{0\sigma} - S(\omega)} \\
\label{eq.265} Z(\omega) =
S(\omega)\sum_{p}\frac{V_{P}L^{41}}{\omega -
\epsilon^{MF}_{p}} + \nonumber \\
\sum_{p}\frac{|V_{p}|^{2}L^{42}}{\omega -\epsilon^{MF}_{p}}
+S(\omega)L^{31} + \sum_{p}V_{p}L^{32}
\end{eqnarray}
Here the coefficients $L^{41}, L^{42}, L^{31}$, and $L^{32}$ are
  certain complicated averages (see definition (\ref{eq.258}))
from which the functional of the GMF is build. To clarify the
functional structure of the obtained solution, let us consider our
first equation of motion (\ref{eq.255}) , before introducing the
irreducible GFs (\ref{eq.258}). Let us put simply, in this
equation, the higher-order GF $ D = 0$! To distinguish this
simplest equation from the GMF one (\ref{eq.259}), we write it in
the following form
\begin{equation}
\label{eq.266}
 \sum_{p}F(p,k)G^{0}(p,\omega) = I \end{equation}
The corresponding matrix elements   which we are interested in
here read
\begin{eqnarray}
\label{eq.267} G^{0}_{22} = <<f_{0\sigma}|f^{\dagger}_{0\sigma}>> = \\
\frac{1 - <n_{0-\sigma}>}{\omega - E_{0\sigma} - S(\omega)} +
\frac{<n_{0-\sigma}>}{\omega - E_{0\sigma} -S(\omega) - U} \nonumber \\
\label{eq.268} G^{0}_{33} =
<<f_{0\sigma}n_{0-\sigma}|f^{\dagger}_{0\sigma}n_{0-\sigma}>> =
\frac{<n_{0-\sigma}>}{\omega - E_{0\sigma} - S(\omega) - U}\\
\label{eq.269} G^{0}_{32} =
<<f_{0\sigma}n_{0-\sigma}|f^{\dagger}_{0\sigma}>> = G^{0}_{33}
\end{eqnarray}
The conclusion is rather evident. The simplest interpolation
solution  follows from our matrix  GF (\ref{eq.256}) in the lowest
order in $V$, even before introduction of GMF corrections, not
speaking about   the self-energy corrections. The two GFs
$G^{0}_{32}$ and $G^{0}_{33}$  are equal only in the lowest order
in $V$. It is quite clear  that our  full  solution (38) that
includes the self-energy corrections   is much more
richer.\\
It is worthwhile to stress that our $4\times4$ matrix GMF GF
(\ref{eq.256}) gives only approximate description of   suitable
mean fields. If we   consider more extended algebra, we   get the
more correct structure of the relevant GMF.\\
\subsection{Dynamic  Properties of SIAM}
To demonstrate clearly the advantages of the IGF method for SIAM,
it is worthwhile to emphasize a few important points about the
approach   based on the equations-of-motion for the GFs. To give
a more instructive discussion, let us consider the single-particle
GF of localized electrons $ G_{\sigma} = <<f_{0\sigma}\vert
f^{\dagger}_{0\sigma}>> $. The simplest approximate
"interpolation" solution  of SIAM is of the form:
\begin{eqnarray}
\label{eq.270} G_{\sigma}(\omega) = \frac{1}{\omega - E_{0\sigma}
- S(\omega)} +
 \frac{U<n_{0-\sigma}>}{(\omega - E_{0\sigma} - S(\omega) - U)(\omega -
E_{0\sigma} - S(\omega))} = \nonumber \\ \frac{1 -
<n_{0-\sigma}>}{\omega - E_{0\sigma} - S(\omega) } +
\frac{<n_{0-\sigma}>}{\omega - E_{0\sigma} - S(\omega) - U}
\end{eqnarray}
The values of $n_{\sigma}$  are determined through the
self-consistency equation
\begin{equation}
\label{eq.271} n_{\sigma} = <n_{0\sigma}> = -\frac {1}{\pi} \int
dEn(E)Im G_{\sigma}(E,n_{\sigma})
\end{equation}
The "atomic-like" interpolation solution (\ref{eq.270}) reproduces
correctly the two  limits:
\begin{eqnarray}
\label{eq.272} G_{\sigma}(\omega) = \frac{1 -
<n_{0-\sigma}>}{\omega - E_{0\sigma}} +
\frac{<n_{0-\sigma}>}{\omega - E_{0\sigma} - U},\quad for \quad V
= 0 \nonumber \\ G_{\sigma}(\omega) = \frac{1}{\omega -
E_{0\sigma} - S(\omega)},\quad for \quad U = 0
\end{eqnarray}
The important point about formulas (\ref{eq.272}) is that any
approximate solution of SIAM should be consistent with it. Let us
remind how to get solution (\ref{eq.272}). It follows from the
system of equations for small-$V$ limit:
\begin{eqnarray}
\label{eq.273} (\omega - E_{0\sigma} - S(\omega))
<<f_{0\sigma}|f^{\dagger}_{0\sigma}>>_{\omega} = 1 +
U<<f_{0\sigma}n_{0-\sigma}|f^{\dagger}_{0\sigma}>>_{\omega},  \nonumber \\
(\omega - E_{0\sigma} -
U)<<f_{0\sigma}n_{0-\sigma}|f^{\dagger}_{0\sigma}>> _{\omega}
\approx \nonumber \\ <n_{0-\sigma}> + \sum_{k}
V_{k}<<c_{k\sigma}n_{0\sigma}|f^{\dagger}_{0\sigma}>>_{\omega}, \\
(\omega -
\epsilon_{k})<<c_{k\sigma}n_{0-\sigma}|f^{\dagger}_{0\sigma}>>_{
\omega} = \nonumber \\   \nonumber
V_{k}<<f_{0\sigma}n_{0-\sigma}|f^{\dagger}_{0\sigma}>>_{\omega}
\end{eqnarray}
The equations (\ref{eq.273}) are approximate; they include two
more terms,  treated in the limit of small $V$ in
paper~\cite{lac}. \\ We now proceed further. In paper~\cite{lac}
  the GF $G$  was calculated in the limit
of infinitely strong Coulomb correlation $U$ and for small
hybridization $V$. The functional structure of the Lacroix
solution generalizes the solution (\ref{eq.272}).  The starting
point is the system of equations:
\begin{eqnarray}
\label{eq.274} (\omega - E_{0\sigma} -
S(\omega))<<f_{0\sigma}|f^{\dagger}_{0\sigma}>> =
1 + U<<f_{0\sigma}n_{0-\sigma}|f^{\dagger}_{0\sigma}>> \\
(\omega - E_{0\sigma}
-U)<<f_{0\sigma}n_{0-\sigma}|f^{\dagger}_{0\sigma}>> =
<n_{0-\sigma}> +\sum_{k}V_{k} \Bigl
(<<c_{k\sigma}n_{0-\sigma}|f^{\dagger}_{0\sigma}>> - \nonumber
\\
\label{eq.275}
<<c_{k-\sigma}f^{\dagger}_{0-\sigma}f_{0\sigma}|f^{\dagger}_{0\sigma}>>
+
<<c^{\dagger}_{k-\sigma}f_{0\sigma}f_{0-\sigma}|f^{\dagger}_{0\sigma}>>
\Bigr )
\end{eqnarray}
Using a relatively simple decoupling procedure for a higher-order
equation of motion,  a qualitatively correct low-temperature
spectral intensity was calculated. The final expression for $G$
for finite $U$ is of the form
\begin{eqnarray}
\label{eq.276} <<f_{0\sigma}|f^{\dagger}_{0\sigma}>> =
\frac{1}{\omega - E_{0\sigma} -
S(\omega) +US_{1}(\omega)} + \nonumber \\
\frac{U<n_{0-\sigma}> +UF_{1}(\omega)}{K(\omega)(\omega -
E_{0\sigma} - S(\omega) + US_{1}(\omega))}
\end{eqnarray}
where $F_{1}$, $S_{1}$, and $K$ are certain complicated
expressions. We   write down explicitly the infinite $U$
approximate GF~\cite{lac}:
\begin{equation}
\label{eq.277} <<f_{0\sigma}|f^{\dagger}_{0\sigma}>> = \frac{1 -
<n_{0-\sigma}> - F_{\sigma}( \omega)}{\omega - E_{0\sigma} -
S(\omega) - Z^{1}_{\sigma}(\omega)}
\end{equation}
The following notation was used
\begin{eqnarray}
\label{eq.278} F_{\sigma} = V\sum_{k}
\frac{<f^{\dagger}_{0-\sigma}c_{k-\sigma}>}{\omega - \epsilon_{k}}\\
\label{eq.279} Z^{1}_{\sigma} = V^{2}\sum_{q,k}
\frac{<c^{\dagger}_{q-\sigma}c_{k-\sigma}>}{\omega -
\epsilon_{k}} - S(\omega)V\sum_{k}
\frac{<f^{\dagger}_{0-\sigma}c_{k-\sigma}>}{\omega - \epsilon_{k}}
\end{eqnarray} The functional structure of the single-particle GF
(\ref{eq.276}) is quite transparent. The expression in the
numerator of (\ref{eq.276}) plays the role of "dynamic mean
field",   proportional to $<f^{\dagger}_{0-\sigma}c_{k-\sigma}>$.
In the denominator, instead of bare shift $S(\omega)$
(\ref{eq.231})  we have an "effective shift" $S^{1} = S(\omega) +
Z^{1}_{\sigma}(\omega)$. The  choice of the specific procedure of
decoupling for the higher-order equation of motion specifies the
selected "generalized mean fields" (GMFs) and "effective shifts".
\subsection{Interpolation Solutions of Correlated Models}
It is to the point to discuss briefly the general concepts of
construction  of an interpolation dynamic solution of the strongly
correlated electron models. The very problem of the consistent
interpolation solutions of the many-body electron models was
formulated explicitly by Hubbard in the context of the Hubbard
model. Hubbard clearly pointed out one particular feature of
consistent theory, insisting that it should give  exact results
in the two opposite limits of very wide and very narrow bands.
The functional structure of a required interpolation solution can
be clarified if one considers the atomic (very narrow band)
solution of the Hubbard model (\ref{eq.49}):
\begin{equation}
\label{eq.280} G^{at}(\omega) = \frac {1 - n_{-\sigma}}{\omega -
t_{0}} + \frac {n_{-\sigma}}{\omega - t_{0} - U} = \frac { 1
}{\omega - t_{0} - \Sigma^{at}(\omega)}
\end{equation}
where
\begin{equation}
\label{eq.281} \Sigma^{at}(\omega) = \frac {n_{-\sigma}U}{1 -
\frac {(1 - n_{-\sigma})U}{\omega -t_{0}}}; \quad t_{0} = t_{ii}
\end{equation}
Let us consider the expansion in terms of $U$:
\begin{equation}
\label{eq.282} \Sigma^{at}(\omega) \approx n_{-\sigma}U +
n_{-\sigma}(1 - n_{-\sigma})U^{2} \frac {1}{\omega - t_{0}} + O(U)
\end{equation}
The  "Hubbard I" solution (\ref{eq.54})can be written as
\begin{equation}
\label{eq.283} G_{k} = \frac {1}{\omega - \epsilon(k) -
\Sigma^{at}(\omega)} = \frac{1}{(G^{at})^{-1} + t_{0} -
\epsilon(k)}
\end{equation}
The partial "Hubbard III" solution,   called the "alloy analogy"
approximation is of the form:
\begin{equation}
\label{eq.284} \Sigma(\omega) = \frac{n_{-\sigma}U}{1 - (U -
\Sigma(\omega))G(\omega)}
\end{equation}
Equation (\ref{eq.284}) follows from (\ref{eq.281}) when one takes
into account the following relationship:
\begin{equation}
\label{eq.285} \frac {1}{\omega - t_{0}} \propto \frac {1}{1 -
n_{-\sigma}}G(\omega) - \Sigma(\omega)G(\omega)
\end{equation}
The Coherent Potential Approximation (CPA) provides the basis for
physical interpretation of equation (\ref{eq.284})  which
corresponds to elimination of the dynamics of $-\sigma$ electrons.
By analogy with (\ref{eq.282}), it is possible to expand:
\begin{equation}
\label{eq.286} \frac {n_{-\sigma}U}{1 - (U -
\Sigma(\omega))G(\omega)} \approx n_{-\sigma}U + n_{-\sigma}U(U -
\Sigma)G^{0}(\omega - \Sigma) + O(U)
\end{equation}
The solution (\ref{eq.277}) does not reproduce correctly the
U-perturbation expansion for the self-energy:
\begin{eqnarray}
\label{eq.287}
 M_{\sigma}(\omega) \sim U<n_{0-\sigma}>
\nonumber + \\ U^{2}\int dE_{1}\int dE_{2}\int
dE_{3}\frac{n(E_{1})n(E_{2})(1 - n(E_{3})) + (1 - n(E_{1})(1 -
n(E_{2}))n(E_{3})}{\omega - E_{1} - E_{2} + E_{3}} \nonumber \\
Im G_{\sigma}(E_{1})Im G_{-\sigma}(E_{2})Im G_{-\sigma}(E_{3})
\end{eqnarray}
It can be shown that it is possible , in principle, to find a
certain way to incorporate this $U^{2}$ perturbation theory
expansion into the functional structure of an interpolation
dynamic solution of SIAM in a self-consistent way within the
higher-order GFs~\cite{ckw}. The IGF approach  with the use of
minimal algebra of relevant operators allows one to find an
interpolation solution for weak and strong Coulomb interaction $U$
and to calculate explicitly the quasi-particle spectra and their
damping for  both the limits. The U-perturbation expansion
(\ref{eq.147}) is included into the IGF scheme in a
self-consistent way. The correct second-order contribution to the
local approximation for the Hubbard model is of the form
\begin{equation}
\label{eq.288} {\tilde G_{\sigma}} \propto
\frac{G_{\sigma}<<n_{0-\sigma}|n_{0-\sigma}>>}{n_{-\sigma}(1 -
n_{-\sigma})}
\end{equation}
The same arguments are also valid for SIAM.
\subsection{Complex Expansion for a Propagator}
We now proceed with analytic  many-body consideration.  One can
attempt to consider a suitable solution for the SIAM starting
from the following exact relation   derived in
paper~\cite{kuz7}:  \begin{eqnarray} \label{eq.289}
<<f_{0\sigma}|f^{\dagger}_{0\sigma}>> = g^{0} + g^{0}Pg^{0} \\
\label{eq.290}
g^{0} = (\omega - E_{0\sigma} - S(\omega))^{-1}\\
\label{eq.291} P = U<n_{0-\sigma}> +
U^{2}<<f_{0\sigma}n_{0-\sigma}|f^{\dagger}_{0\sigma}n_{0-\sigma}>>
\end{eqnarray}
The advantage of the equation (\ref{eq.289}) is that it is a
pure  identity and does not include any approximation. If we
insert our GMF solution (\ref{eq.277}) into (\ref{eq.289}), we
  get an essentially new dynamic solution of SIAM
constructed on the basis of the complex (combined) expansion of
the propagator in both $U$ and $V$ parameters and   reproducing
  exact solutions of SIAM for $V = 0$ and $U = 0 $. It
generalizes (even on the mean-field level) the solutions of
papers~\cite{lac}, ~\cite{neal1}.\\
Having emphasized the importance of the role of   equation
(\ref{eq.289}) , let us see now what is the best possible fit for
  higher-order GF in (\ref{eq.291}). We   consider
the equation of motion for it:
\begin{eqnarray}
\label{eq.292}
 (\omega - E_{0\sigma} -
U)<<f_{0\sigma}n_{0-\sigma}|f^{\dagger}_{0\sigma}n_{0-\sigma}>> =
<n_{0-\sigma}> +  \\
\sum_{k}V_{k}(<<c_{k\sigma}n_{0-\sigma}|f^{\dagger}_{0\sigma}n_{0-\sigma}>>
+ \nonumber
\\
<<c^{\dagger}_{k-\sigma}f_{0\sigma}f_{0-\sigma}|f^{\dagger}_{0\sigma}n_{0-\sigma}>>
-
<<c_{k-\sigma}f^{\dagger}_{0-\sigma}f_{0\sigma}|f^{\dagger}_{0\sigma}n_{0-\sigma}>>)
\nonumber
\end{eqnarray}
We can think of it as defining   new kinds of elastic and
inelastic scattering processes that contribute to the formation
of   generalized mean fields and self-energy (damping)
corrections. The construction of   suitable mean fields can be
quite nontrivial, and to describe these contributions
self-consistently, let us consider the equations of motion for
higher-order GFs in the r.h.s. of (\ref{eq.292})
\begin{eqnarray}
\label{eq.293} (\omega -
\epsilon_{k})<<c_{k\sigma}n_{0-\sigma}|f^{\dagger}_{0\sigma}n_{0-\sigma}>>
=  \nonumber \\
V<<f_{0\sigma}n_{0-\sigma}|f^{\dagger}_{0\sigma}n_{0-\sigma}>> +  \\
\sum_{p}V
(<<c_{k\sigma}f^{\dagger}_{0-\sigma}c_{p-\sigma}|f^{\dagger}_{0\sigma}n_{0-\sigma}>>
-
<<c_{k\sigma}c^{\dagger}_{p-\sigma}f_{0-\sigma}|f^{\dagger}_{0\sigma}n_{0-\sigma}>>)
\nonumber
\end{eqnarray}
\begin{eqnarray}
\label{eq.294} (\omega - \epsilon_{k} -E_{0\sigma} + E
_{0-\sigma})<<c_{k-\sigma}f^{\dagger}_{0-\sigma}f_{0\sigma}|f^{\dagger}_{0\sigma}n_{0-\sigma}>>
\nonumber \\
= -<f^{\dagger}_{0-\sigma}c_{k-\sigma}n_{0\sigma}> -  \\
V<<f_{0\sigma}n_{0-\sigma}|f^{\dagger}_{0\sigma}n_{0-\sigma}>> + \nonumber \\
\sum_{p}V(<<c_{k-\sigma}f^{\dagger}_{0-\sigma}c_{p\sigma}|f^{\dagger}_{0\sigma}n_{0-\sigma}>>
-
<<c_{k-\sigma}c^{\dagger}_{p-\sigma}f_{0\sigma}|f^{\dagger}_{0\sigma}n_{0-\sigma}>>)
\nonumber
\end{eqnarray}
\begin{eqnarray}
\label{eq.295} (\omega + \epsilon_{k} - E_{0\sigma} - E
_{0-\sigma} -
U)<<c^{\dagger}_{k-\sigma}f_{0\sigma}f_{0-\sigma}|f^{\dagger}_{0\sigma}n_{0-\sigma}>>
\\ = -<c^{\dagger}_{k-\sigma}f_{0\sigma}f^{\dagger}_{0\sigma}f_{0-\sigma}> +
\nonumber
\\
V<<f_{0\sigma}n_{0-\sigma}|f^{\dagger}_{0\sigma}n_{0-\sigma}>> +
\nonumber
\\
\sum_{p}V(<<c^{\dagger}_{k-\sigma}c_{p\sigma}f_{0-\sigma}|f^{\dagger}_{0\sigma}n_{0-\sigma}>>
+
<<c^{\dagger}_{k-\sigma}f_{0\sigma}c_{p-\sigma}|f^{\dagger}_{0\sigma}n_{0-\sigma}>>)
\nonumber
\end{eqnarray}
Now let us see how to proceed further to get a suitable
functional structure of the relevant solution. The intrinsic
nature of the system of the equations of motion (\ref{eq.293}) -
(\ref{eq.295}) suggests to consider the following approximation:
\begin{eqnarray}
\label{eq.296} (\omega -
\epsilon_{k})<<c_{k\sigma}n_{0-\sigma}|f^{\dagger}_{0\sigma}n_{0-\sigma}>>
\approx
V<<f_{0\sigma}n_{0-\sigma}|f^{\dagger}_{0\sigma}n_{0-\sigma}>> \\
\label{eq.297} (\omega - \epsilon_{k} - E_{0\sigma} +
E_{0-\sigma})<<c_{k-\sigma}f^{\dagger}_{0-\sigma}f_{0\sigma}|
f^{\dagger}_{0\sigma}n_{0-\sigma}>> \approx
-<f^{\dagger}_{0-\sigma}c_{k-\sigma}n_{0\sigma}> \nonumber \\ -
V(<<f_{0\sigma}n_{0-\sigma}|f^{\dagger}_{0\sigma}n_{0-\sigma}>> -
<<c_{k-\sigma}c^{\dagger}_{k-\sigma}f_{0\sigma}|f^{\dagger}_{0\sigma}n_{0-\sigma}>>)\\
\label{eq.298} (\omega + \epsilon_{k} - E_{0\sigma} - E_{0-\sigma}
-
U)<<c^{\dagger}_{k-\sigma}f_{0\sigma}f_{0-\sigma}|f^{\dagger}_{0\sigma}n_{0-\sigma}>>
\approx
-<c^{\dagger}_{k-\sigma}f_{0\sigma}f^{\dagger}_{0\sigma}f_{0-\sigma}>
+\nonumber \\
V(<<f_{0\sigma}n_{0-\sigma}|f^{\dagger}_{0\sigma}n_{0-\sigma}>> +
<<c^{\dagger}_{k-\sigma}f_{0\sigma}c_{k-\sigma}|f^{\dagger}_{0\sigma}n_{0-\sigma}>>)
\end{eqnarray}
It is transparent that the construction of   approximations
(\ref{eq.296}) - (\ref{eq.298}) is related with the small-V
expansion and is not unique, but very natural. As a result, we
find the explicit expression for   GF in (\ref{eq.291})
\begin{equation}
\label{eq.299}
<<f_{0\sigma}n_{0-\sigma}|f^{\dagger}_{0\sigma}n_{0-\sigma}>>
\approx \frac{<n_{0-\sigma}> - F^{1}_{\sigma}(\omega)}{\omega -
E_{0\sigma} - U - S_{1}(\omega)}
\end{equation}
Here the following notation was used
\begin{eqnarray}
\label{eq.300} S_{1}(\omega) = S(\omega) \\ +
\sum_{k}|V|^{2}(\frac{1}{\omega - \epsilon_{k} - E_{0\sigma} +
E_{0-\sigma}} + \frac{1}{\omega + \epsilon_{k}
- E_{0\sigma} - E_{0-\sigma} - U}) \nonumber \\
\label{eq.301}
F^{1}_{\sigma} = \sum_{k}(VF_{2} + V^{2}F_{3})\\
\label{eq.302} F_{2} =
\frac{<c^{\dagger}_{k-\sigma}f_{0\sigma}f^{\dagger}_{0\sigma}f_{0-\sigma}>}{\omega
+ \epsilon_{k} - E_{0\sigma} - E_{0-\sigma} - U } +
\frac{<f^{\dagger}_{0-\sigma}c_{k-\sigma}n_{0\sigma}>}{\omega -
\epsilon_{k} -
E_{0\sigma} + E_{0-\sigma}}\\
\label{eq.303}
F_{3} = \\
\frac{<<c_{k-\sigma}c^{\dagger}_{k-\sigma}f_{0\sigma}|f^{\dagger}_{0\sigma}n_{0-\sigma}>>}{
\omega - \epsilon_{k} - E_{0\sigma} + E_{0-\sigma}} +
\frac{<<c^{\dagger}_{k-\sigma}f_{0\sigma}c_{k-\sigma}|f^{\dagger}_{0\sigma}n_{0-\sigma}>>}{
\omega + \epsilon_{k} - E_{0\sigma} - E_{0-\sigma} - U} \nonumber
\end{eqnarray}
Now one can substitute the GF in (\ref{eq.291}) by the expression
(\ref{eq.299}). This   gives  a new approximate dynamic solution
of SIAM where the complex expansion  both in $U$ and $V$ was
incorporated. The important observation is that this new solution
satisfies  both the limits (\ref{eq.272}).  For example, if we
wish to get a lowest order approximation up to $U^{2}$ and
$V^{2}$, it is very easy to notice that for $V = 0$:
\begin{eqnarray}
\label{eq.304}
<<f_{0\sigma}c^{\dagger}_{k-\sigma}c_{k-\sigma}|f^{\dagger}_{0\sigma}n_{0-\sigma}>>
\approx
\frac{<c^{\dagger}_{k-\sigma}c_{k-\sigma}><n_{0-\sigma}>}{\omega -
E_{0\sigma} - U} \nonumber \\
<<c_{k-\sigma}c^{\dagger}_{k-\sigma}f_{0\sigma}|f^{\dagger}_{0\sigma}n_{0-\sigma}>>
\approx
\frac{<c_{k-\sigma}c^{\dagger}_{k-\sigma}><n_{0-\sigma}>}{\omega -
E_{0\sigma} - U}
\end{eqnarray}
This results in the possibility to find explicitly all necessary
quantities and, thus, to solve the problem in a self-consistent
way. \\
In summary, we   presented here a consistent many-body approach
to analytic  dynamic solution of SIAM  at finite temperatures and
for a broad interval of the values of the model parameters.  We
used the exact result (\ref{eq.289}) to connect the
single-particle GF with   higher-order GF to obtain a  complex
combined expansion in terms of $U$ and $V$ for the propagator. To
summarize, we   reformulated the problem of searches for an
appropriate  many-body dynamic solution for SIAM in a way that
provides us with an effective and workable scheme for
constructing of advanced analytic  approximate solutions for the
single-particle GFs on the level of the higher-order GFs in a
rather systematic   self-consistent way. This procedure has the
advantage that it systematically uses the principle of
interpolation solution within the equation-of-motion approach for
  GFs. The leading principle, which we   used here was to
look more carefully for the intrinsic functional structure of the
required relevant solution and then to formulate approximations
for the higher-order GFs in accordance with this structure. \\
The main results of our IGF study are the exact Dyson equations
for the full  matrix GFs  and a new derivation of the GMF GFs .
The approximate explicit calculations of   inelastic self-energy
corrections are quite straightforward but tedious and too
extended for their description. Here we want to emphasize an
essentially new point of view on the derivation of the
Generalized Mean Fields for SIAM when we are interested in the
interpolation finite temperature solution for the single-particle
propagator.  Our final solutions have the correct functional
structure  and differ essentially from previous solutions.
\\ Of course, there are important
criteria to be met (mainly numerically) , such as the question
left open, whether the present approximation satisfies the Friedel
sum rule (this question left open in ~\cite{neal1} and ~\cite{lac}
 ). A quantitative numerical comparison of self-consistent
results ({\it e.g.} the width and shape of the Kondo resonance in
the near-integer regime of the SIAM) would be crucial too. In the
present consideration, we   concentrated on the problem of
correct functional structure of the single-particle GF itself. In
addition to SIAM, it will be instructive to consider sketchy the
PAM and TIAM too for  completness.
\subsection{Quasi-Particle Dynamics of PAM}
The main drawback of the H-F type solution of PAM (\ref{eq.61}) is
that it ignores the correlations of the ""up" and "down"
electrons. In this Section, we   take into account the latter
correlations in a self-consistent way using the IGF method. We
  consider the relevant matrix GF of the form ( {\it cf.}
(\ref{eq.224}) )
\begin{equation}
\label{eq.305}
 \hat G (\omega) =
\pmatrix{ <<c_{k\sigma}\vert c^{\dagger}_{k\sigma}>> &
<<c_{k\sigma}\vert f^{\dagger}_{k\sigma}>> \cr <<f_{k\sigma}\vert
c^{\dagger}_{k\sigma}>> & <<f_{k\sigma}\vert
f^{\dagger}_{k\sigma}>> \cr}
\end{equation}
The equation of motion for GF (\ref{eq.305}) reads
\begin{eqnarray}
\label{eq.306} \pmatrix{(\omega - \epsilon_{k})& - V_{k} \cr -
V_{k} & (\omega - E_{k}) \cr } \pmatrix{ <<c_{k\sigma} \vert
c^{\dagger}_{k\sigma}>> & <<c_{k\sigma}\vert
f^{\dagger}_{k\sigma}>> \cr <<f_{k\sigma} \vert
c^{\dagger}_{k\sigma}>> & <<f_{k\sigma} \vert
f^{\dagger}_{k\sigma}>> \cr} = \nonumber
\\  \pmatrix{ 1&0 \cr 0&1 \cr} + UN^{-1} \sum_{pq} \pmatrix{
0&0 \cr <<A \vert c^{\dagger}_{k\sigma}>>& <<A \vert
f^{\dagger}_{k\sigma}>> \cr}
\end{eqnarray}
where
 $ A = f_{k+p\sigma}f^{\dagger}_{p+q-\sigma}f_{q-\sigma}$.
According to eq.(30), the definition of the irreducible parts in
the equation of motion (\ref{eq.306})  are defined as follows
\begin{eqnarray}
 ^{(ir)}<<f_{k+p\sigma}f^{\dagger}_{p+q-\sigma}f_{q-\sigma} \vert
c^{\dagger}_{k\sigma}>> =
<<f_{k+p\sigma}f^{\dagger}_{p+q-\sigma}f_{q-\sigma} \vert
c^{\dagger}_{k\sigma}>>  - \nonumber \\ \delta_{p,0}<n_{q-\sigma}>
<<f_{k\sigma}\vert c^{\dagger}_{k\sigma}>>  \nonumber \\
^{(ir)}<<f_{k+p\sigma}f^{\dagger}_{p+q-\sigma}f_{q-\sigma} \vert
f^{\dagger}_{k\sigma}>> =
<<f_{k+p\sigma}f^{\dagger}_{p+q-\sigma}f_{q-\sigma} \vert
f^{\dagger}_{k\sigma}>>  - \nonumber \\ \delta_{p,0}<n_{q-\sigma}>
<<f_{k\sigma}\vert f^{\dagger}_{k\sigma}>> \nonumber
\end{eqnarray}
After substituting   these definitions into   equation
(\ref{eq.306}), we obtain
\begin{eqnarray}
\label{eq.307} \pmatrix{(\omega - \epsilon_{k})& - V_{k} \cr -
V_{k} & (\omega - E_{\sigma}(k)) \cr } \pmatrix{ <<c_{k\sigma}
\vert c^{\dagger}_{k\sigma}>> & <<c_{k\sigma}\vert
f^{\dagger}_{k\sigma}>> \cr <<f_{k\sigma} \vert
c^{\dagger}_{k\sigma}>> & <<f_{k\sigma} \vert
f^{\dagger}_{k\sigma}>> \cr} = \nonumber
\\  \pmatrix{ 1&0 \cr 0&1 \cr} + UN^{-1} \sum_{pq} \pmatrix{
0&0 \cr ^{(ir)}<<A \vert c^{\dagger}_{k\sigma}>>& ^{(ir)}<<A \vert
f^{\dagger}_{k\sigma}>> \cr}
\end{eqnarray}
The following notation  was used
$$ E_{\sigma}(k) = E_{k} - Un_{-\sigma}; \quad n_{-\sigma} =
<f^{\dagger}_{k-\sigma}f_{k-\sigma}>$$ The definition of the
generalized mean field GF ( which, for the weak Coulomb
correlation $U$, coincides with the Hartree-Fock mean field ) is
evident. All inelastic renormalization terms are now related to
the last term in the equation of motion (\ref{eq.307}). All
elastic scattering ( or mean field) renormalization terms are
included into the following  mean-field GF
\begin{equation}
\label{eq.308} \pmatrix{(\omega - \epsilon_{k})& - V_{k} \cr -
V_{k} & (\omega - E_{\sigma}(k)) \cr } \pmatrix{ <<c_{k\sigma}
\vert c^{\dagger}_{k\sigma}>>^{0} & <<c_{k\sigma}\vert
f^{\dagger}_{k\sigma}>>^{0} \cr <<f_{k\sigma} \vert
c^{\dagger}_{k\sigma}>>^{0} & <<f_{k\sigma} \vert
f^{\dagger}_{k\sigma}>>^{0} \cr} = \nonumber
\\  \pmatrix{ 1&0 \cr 0&1 \cr}
\end{equation}
It is easy to find that ({\it cf.} (\ref{eq.229}) and
(\ref{eq.230}))
\begin{eqnarray}
\label{eq.309} <<f_{k\sigma} \vert f^{\dagger}_{k\sigma}>>^{0} =
\Bigl ( \omega - E_{\sigma}(k) - \frac { |V_{k}|^{2}}{\omega -
\epsilon_{k}} \Bigr )^{-1} \\
\label{eq.310}
 <<c_{k\sigma} \vert c^{\dagger}_{k\sigma}>>^{0} =
 \Bigl
( \omega - \epsilon_{k} - \frac { |V_{k}|^{2}}{\omega -
E_{\sigma}(k)} \Bigr )^{-1}
\end{eqnarray}
At this point, it is worthwhile to emphasize a significant
difference between  both the models, PAM and SIAM. The
corresponding SIAM equation for generalized mean field GF
(\ref{eq.228}) reads
\begin{eqnarray}
\label{eq.311} \sum_{p} \pmatrix{(\omega -
\epsilon_{p})\delta_{pk}& - V_{p}\delta_{pk} \cr - V_{p} & {1
\over N} (\omega - E_{0\sigma} - Un_{-\sigma}) \cr } \pmatrix{
<<c_{k\sigma} \vert c^{\dagger}_{k\sigma}>>^{0} &
<<c_{k\sigma}\vert f^{\dagger}_{0\sigma}>>^{0} \cr <<f_{0\sigma}
\vert c^{\dagger}_{k\sigma}>>^{0} & <<f_{0\sigma} \vert
f^{\dagger}_{0\sigma}>>^{0}
\cr} = \nonumber \\
 \pmatrix{ 1&0 \cr 0&1 \cr}
\end{eqnarray}
This   matrix notation for SIAM shows a fundamental distinction
between SIAM and PAM. For SIAM,  we have a different number of
states for a strongly localized level and the conduction electron
subsystem: the conduction band contains $2N$ states, whereas the
localized (s-type) level contains only two. The comparison of
(\ref{eq.311}) and (\ref{eq.308}) shows clearly that this
difficulty does not exist for PAM : the number of states   both
in the localized and itinerant subsystems are the same, {\it i.e.} $2N$.\\
This important difference between SIAM and PAM   appears also
when we calculate   inelastic scattering or self-energy
corrections. By analogy with the Hubbard model, the equation of
motion (\ref{eq.307}) for PAM can be transformed exactly to the
scattering equation of the form ( 36). Then, we are able to write
down explicitly the Dyson equation (37) and the exact expression
for the self-energy M  in the matrix form:
\begin{equation}
\label{eq.312} \hat M_{k\sigma}(\omega) = \pmatrix{0&0\cr 0&M_{22}
\cr}
\end {equation}
Here the matrix element $ M_{22}$ is of the form
\begin{eqnarray}
\label{eq.313} M_{22} = M_{k\sigma}(\omega) =
  \\
 \frac{U^2}{N^2}{} \sum_{pqrs}{} \bigl (^{(ir)}<<f_{k+p\sigma}f^{\dagger}_{p+q-\sigma}
f_{q-\sigma} \vert
f^{\dagger}_{r-\sigma}f_{r+s-\sigma}f^{\dagger}_{k+s\sigma}>>^{(ir)}
\bigr )^{(p)} \nonumber
\end{eqnarray}
To calculate the self-energy operator (\ref{eq.313}) in a
self-consistent way, we proceed by analogy with the Hubbard model
in Section 8.1. Then we find both expressions for the self-energy
operator in   form (\ref{eq.149}) and (\ref{eq.152}).
\subsection{Quasi-Particle Dynamics of TIAM}
Let us see now how to rewrite the results of the preceeding
Sections for the case of TIAM Hamiltonian (\ref{eq.62}). We again
consider the relevant matrix GF of the form (cf.(\ref{eq.224}) )
\begin{equation}
\label{eq.314}
 \hat G (\omega) = \pmatrix{G_{11}&G_{12}&G_{13} \cr
 G_{21}&G_{22}&G_{23} \cr
 G_{31}&G_{32}&G_{33}\cr} =
\pmatrix{ <<c_{k\sigma}\vert c^{\dagger}_{k\sigma}>> &
<<c_{k\sigma}\vert f^{\dagger}_{1\sigma}>>&  <<c_{k\sigma}\vert
f^{\dagger}_{2\sigma}>> \cr <<f_{1\sigma}\vert
c^{\dagger}_{k\sigma}>> &<<f_{1\sigma}\vert
f^{\dagger}_{1\sigma}>>& <<f_{1\sigma}\vert
f^{\dagger}_{2\sigma}>> \cr <<f_{2\sigma}\vert
c^{\dagger}_{k\sigma}>> &<<f_{2\sigma}\vert
f^{\dagger}_{1\sigma}>>& <<f_{2\sigma}\vert
f^{\dagger}_{2\sigma}>> \cr}
\end{equation}
The equation of motion for GF (\ref{eq.314}) reads
\begin{eqnarray}
\label{eq.315} \sum_{p} \pmatrix{(\omega -
\epsilon_{p})\delta_{pk}& - V_{1p}\delta_{pk}& -
V_{1p}\delta_{pk} \cr - V_{1p} & {1 \over N} (\omega - E_{0\sigma}
)& -V_{12} \cr -V_{2p} & - V_{21} & {1 \over N} (\omega -
E_{0\sigma}) \cr } \pmatrix{G_{11}&G_{12}&G_{13} \cr
 G_{21}&G_{22}&G_{23} \cr
 G_{31}&G_{32}&G_{33}\cr} = \nonumber \\
 \pmatrix{ 1&0&0 \cr 0&1&0 \cr
 0&0&1 \cr} + \\ \nonumber U
\pmatrix{0&0&0 \cr <<A_{1} \vert c^{\dagger}_{k\sigma}>> & <<
A_{1}\vert f^{\dagger}_{1\sigma}>> & << A_{1} \vert
f^{\dagger}_{2\sigma}>> \cr <<A_{2} \vert c^{\dagger}_{k\sigma}>>
& << A_{2}\vert f^{\dagger}_{1\sigma}>> & << A_{2} \vert
f^{\dagger}_{2\sigma}>> \cr}
\end{eqnarray}
The notation  is as follows $$ A_{1} =
f_{1\sigma}f^{\dagger}_{1-\sigma}f_{1-\sigma}; \quad  A_{2} =
f_{2\sigma}f^{\dagger}_{2-\sigma}f_{2-\sigma}$$ In a compact
notation, the equation (\ref{eq.315}) has the form ({\it cf.}
(\ref{eq.255}))
\begin{equation}
\label{eq.316} \sum_{p}F(p,k)G_{pk}(\omega) = \hat I + U
D_{p}(\omega)
\end{equation}
We thus have the equatin of motion (\ref{eq.316}) which is a
complete analogue of the corresponding equations for the SIAM and
PAM. After introducing the irreducible parts by analogy with the
equation (\ref{eq.225})
\begin{eqnarray}
\nonumber ^{(ir)}<<f_{1\sigma}f^{\dagger}_{1-\sigma}f_{1-\sigma}
\vert B>>_ {\omega} =
<<f_{1\sigma}f^{\dagger}_{1-\sigma}f_{1-\sigma}\vert B>>_{\omega} - \\
\nonumber -<n_{1-\sigma}> <<f_{1\sigma} \vert B>>_{\omega}\\
\nonumber ^{(ir)}<<f_{2\sigma}f^{\dagger}_{2-\sigma}f_{2-\sigma}
\vert B>>_ {\omega} =
<<f_{2\sigma}f^{\dagger}_{2-\sigma}f_{2-\sigma}\vert B>>_{\omega} - \\
\nonumber -<n_{2-\sigma}> <<f_{2\sigma} \vert B>>_{\omega}\\
\nonumber ,
\end{eqnarray}
and performing the second-time   differentiation of the
higher-order GF, and introducing the relevant irreducible parts,
the equation of motion (\ref{eq.316}) is rewritten  in the form of
Dyson equation (37). The definition of the generalized mean field
GF  is as follows
\begin{eqnarray}
\label{eq.317} \sum_{p} \pmatrix{(\omega -
\epsilon_{p})\delta_{pk}& - V_{1p}\delta_{pk}& -
V_{1p}\delta_{pk} \cr - V_{1p} & {1 \over N} (\omega -
E_{0\sigma} - Un_{-\sigma} )& -V_{12} \cr -V_{2p} & - V_{21} & {1
\over N} (\omega - E_{0\sigma} - Un_{-\sigma}) \cr } \nonumber \\
\pmatrix{G^{0}_{11}&G^{0}_{12}&G^{0}_{13} \cr
 G_{21}&G^{0}_{22}&G^{0}_{23} \cr
 G^{0}_{31}&G^{0}_{32}&G^{0}_{33}\cr} =
 \pmatrix{ 1&0&0 \cr 0&1&0 \cr
 0&0&1 \cr}
\end{eqnarray}
The matrix GF (\ref{eq.317}) describes the mean-field solution of
the TIAM Hamiltonian. The explicit solutions for diagonal
elements of $G^{0}$ are ( {\it cf.} (\ref{eq.229}))
\begin{eqnarray}
\label{eq.318}   <<c_{k\sigma} \vert
c^{\dagger}_{k\sigma}>>^{0}_{\omega} =   \Bigl ( \omega
-\epsilon_{k} - \frac { |V_{1k}|^{2}}{\omega - (E_{0\sigma} -
Un_{-\sigma})} - \Delta_{11}(k,\omega)  \Bigr
)^{-1} \\
 \label{eq.319} <<f_{1\sigma} \vert f^{\dagger}_{1\sigma}>>^{0}_
{\omega} = \Bigl ( \omega - ( E_{0\sigma} - Un_{-\sigma}) -
S (\omega)) - \Delta_{22}(k,\omega) \Bigr )^{-1} \\
\label{eq.320} <<f_{2\sigma} \vert f^{\dagger}_{2\sigma}>>^{0}_
{\omega} = \Bigl ( \omega - ( E_{0\sigma} - Un_{-\sigma}) - S
(\omega)) - \Delta_{33}(k,\omega) \Bigr )^{-1}
\end{eqnarray}
Here we   introduced the notation
\begin{eqnarray}
\label{eq.321} \nonumber \Delta_{11}(k,\omega) = \Bigl( V_{2k} +
\frac { V_{1k}V_{12}}{\omega - (E_{0\sigma} - Un_{-\sigma})}
\Bigr) \Bigl( V_{2k} + \frac { V_{1k}V_{21}}{\omega - (E_{0\sigma}
- Un_{-\sigma})} \Bigr)
\\ \nonumber [ \omega - (E_{0\sigma} - Un_{-\sigma}) - \frac {
V_{21}V_{12}}{\omega - (E_{0\sigma} - Un_{-\sigma})}]^{-1}  \\
\nonumber \Delta_{22}(k,\omega) = ( \lambda_{21} (\omega) +
V_{12} )( \lambda_{21} (\omega) + V_{21} )[\omega -
(E_{0\sigma}-  Un_{-\sigma})
- \frac {\sum_{p} |V_{2p}|^{2}}{\omega - \epsilon_{p}}]^{-1}  \\
\nonumber \Delta_{33}(k,\omega) = ( \lambda_{12} (\omega) +
V_{21} )( \lambda_{12} (\omega) + V_{12} )[\omega - (E_{0\sigma}
-  Un_{-\sigma}) - \frac
{\sum_{p} |V_{1p}|^{2}}{\omega - \epsilon_{p}}]^{-1} \\
\lambda_{12} = \lambda_{21} = \sum_{p} \frac {V_{1p}V_{2p}}{\omega
- \epsilon_{p}}
\end{eqnarray}
The formal solution of the Dyson equation for TIAM contains the
self-energy matrix
\begin{equation}
\label{eq.322} \hat M = \pmatrix{0&0&0\cr
 0&M_{22}&M_{23} \cr
 0&M_{32}&M{33}\cr}
\end{equation}
where
\begin{eqnarray}
\label{eq.323} M_{22} =  U^{2}
(^{(ir)}<<f_{1\sigma}n_{1-\sigma}\vert
f^{\dagger}_{1\sigma}n_{1-\sigma}>>^{(ir)})^{p} \\ \nonumber
M_{32} = U^{2} (^{(ir)}<<f_{2\sigma}n_{2-\sigma}\vert
f^{\dagger}_{1\sigma}n_{1-\sigma}>>^{(ir)})^{p} \\ \nonumber
M_{23} = U^{2} (^{(ir)}<<f_{1\sigma}n_{1-\sigma}\vert
f^{\dagger}_{2\sigma}n_{2-\sigma}>>^{(ir)})^{p} \\ \nonumber
M_{33} = U^{2} (^{(ir)}<<f_{2\sigma}n_{2-\sigma}\vert
f^{\dagger}_{2\sigma}n_{2-\sigma}>>^{(ir)})^{p}
\end{eqnarray}
To calculate the matrix elements (\ref{eq.323}), the same
procedure can be used as it was done previously for the SIAM
(\ref{eq.239}). As a result, we find the following explicit
expressions for the self-energy matrix elements (
cf.(\ref{eq.241})
\begin{eqnarray}
\label{eq.324} M^{\uparrow}_{22}(\omega) = U^2
\int_{-\infty}^{+\infty} d\omega_{1}d{\omega}_{2}\frac{1 +
N(\omega_{1}) - n(\omega_{2})}
{\omega - \omega_{1} - \omega_{2}}\nonumber\\
(-{1 \over \pi}Im <<S^{-}_{1} \vert S^{+}_{1}>>_{\omega_{1}})\nonumber\\
(-{1 \over \pi}Im<<f_{1\downarrow}\vert
f^{\dagger}_{1\downarrow}>>_{\omega_{2}})  \\
\label{eq.325} M^{\downarrow}_{22}(\omega) = U^2
\int_{-\infty}^{+\infty} d\omega_{1}d{\omega}_{2}\frac{1 +
N(\omega_{1}) - n(\omega_{2})}
{\omega - \omega_{1} - \omega_{2}}\nonumber\\
(-{1 \over \pi}Im <<S^{+}_{1} \vert S^{-}_{1}>>_{\omega_{1}})\nonumber\\
(-{1 \over \pi}Im<<f_{1\uparrow}\vert
f^{\dagger}_{1\uparrow}>>_{\omega_{2}})  \\
\label{eq.326} M^{\uparrow}_{23}(\omega) = U^2
\int_{-\infty}^{+\infty} d\omega_{1}d{\omega}_{2}\frac{1 +
N(\omega_{1}) - n(\omega_{2})}
{\omega - \omega_{1} - \omega_{2}}\nonumber\\
(-{1 \over \pi}Im <<S^{-}_{1} \vert S^{+}_{2}>>_{\omega_{1}})\nonumber\\
(-{1 \over \pi}Im<<f_{1\downarrow}\vert
f^{\dagger}_{2\downarrow}>>_{\omega_{2}})  \\
\label{eq.327} M^{\downarrow}_{23}(\omega) = U^2
\int_{-\infty}^{+\infty} d\omega_{1}d{\omega}_{2}\frac{1 +
N(\omega_{1}) - n(\omega_{2})}
{\omega - \omega_{1} - \omega_{2}}\nonumber\\
(-{1 \over \pi}Im <<S^{+}_{2} \vert S^{-}_{1}>>_{\omega_{1}})\nonumber\\
(-{1 \over \pi}Im<<f_{1\uparrow}\vert
f^{\dagger}_{2\uparrow}>>_{\omega_{2}})
\end{eqnarray}
where the following notation  was used:
$$S^{+}_{i} = f^{\dagger}_{i\uparrow}f_{i\downarrow}; \quad
S^{-}_{i} = f^{\dagger}_{i\downarrow}f_{i\uparrow}; \quad
i=1,2$$  For $ M_{33}$ we obtain the same expressions as for
$M_{22}$  with the substitution of index 1 by 2.  For
$M^{\uparrow \downarrow}_{32}$ we must do the same. It is
possible to say that the diagonal elements $M_{22}$ and $M_{33}$
describe single-site inelastic scattering processes; off-diagonal
elements $M_{23}$ and $M_{32}$ describe   intersite inelastic
scattering processes. They are responsible for the specific
features of the dynamic behaviour of
  TIAM ( as well as the off-diagonal matrix elements of the GF
$G^{0}$) and, more generally, the cluster impurity Anderson model
(CIAM). The nonlocal contributions to the total spin
susceptibility of two well formed impurity magnetic moments at a
distance $R$ can be estimated as
\begin{equation}
\chi_{pair} \sim <<S^{-}_{1} \vert S^{+}_{2}>> \sim 2\chi -12\pi
E_{F}(\frac {\chi}{ g \mu_{B}})^{2} \frac {\cos
(2k_{F}R)}{(k_{F}R)^{3}}
\end{equation}
In the region of interplay of the RKKY and Kondo behaviour, the
key point is then to connect the partial Kondo screening effects
with the low temperature behaviour of the total spin
susceptibility. As it is known, it is quite difficult to describe
such a threshold behaviour analytically. However, progress is
expected due to a better understanding of the quasi-particle
many-body dynamics both from analytical and numerical
investigations.

\section{Conclusions}

In the present paper, we have formulated the theory of the
correlation effects for many-particle interacting systems  using
the ideas of   quantum field theory for   interacting electron
and spin systems on a lattice. The workable and self-consistent
IGF approach to the decoupling problem for the equation-of-motion
method for double-time temperature Green functions has been
presented. The main achievement of this formulation was the
derivation of the Dyson equation for double-time retarded Green
functions instead of causal ones. That formulation permits to
unify   convenient analytical properties of retarded and advanced
GF and   the formal solution of the Dyson equation (38), that, in
spite of the required approximations for the self-energy,
provides the correct functional structure of   single-particle
GF.  The main advantage of the mathematical formalism is brought
out by showing how elastic scattering corrections (generalized
mean fields) and inelastic scattering effects (damping and finite
lifetimes) could be self-consistently incorporated in a general
and compact manner. In this paper, we have thoroughly
considered   the idealized Anderson and Hubbard models  which
are  the simplest (in the sense of formulation, but not solution)
and most popular models of correlated lattice fermions. We have
presented here the novel method of calculation of quasi-particle
spectra for these and
  basic  spin lattice models, as the most representative
examples. Using   the IGF method, we were able to obtain a closed
self-consistent set of equations determining the electron GF and
self-energy. For the Hubbard and Anderson models, these equations
give a general microscopic description of correlation effects
both for the weak and strong Coulomb correlation, and,thus,
determine   the interpolation solutions of the models. Moreover,
this approach gives the workable scheme for the definition of
relevant generalized
mean fields written in terms of appropriate correlators. \\
We hope that these considerations have been done with sufficient
details to bring out their scope and power, since we believe that
this technique will have application to a variety of many-body
systems with complicated spectra and strong interaction. The
application of the IGF method to the investigation of nonlocal
correlations and quasi-particle interactions in Anderson
models~\cite{kuz8} has a particular interest for  studying of the
intersite correlation effects in the concentrated Kondo system . A
comparative study of real many-body dynamics of single-impurity,
two-impurity, and periodic Anderson model, especially for strong
but finite Coulomb correlation, when perturbation expansion in
$U$ does not work, is important  to characterize the true
quasi-particle excitations and the role of magnetic correlations.
It was shown that the physics of two-impurity Anderson model can
be understood in terms of competition between   itinerant motion
of carriers and magnetic correlations of the RKKY nature. This
issue is still very controversial and the additional efforts must
be applied in this field.\\ The application of the IGF method to
the theory of magnetic semiconductors was done in~\cite{kuz5},
\cite{kuz6}. As a remarkable result of our approach, let us
mention the  generalization of the Shastry-Mattis theory for the
magnetic polaron to the finite temperatures~\cite{kuz6} . The
quasi-particle many-body dynamics of ferromagnetic\cite{kuz5} and
aniferromagnetic semiconductors\cite{mak1},\cite{mak2} was
studied too. These studies clarified greatly the true nature of
  carriers in magnetic semiconductors. The application of the
IGF method to generalized spin-fermion models that was made in
papers~\cite{kuz12},\cite{kcf} allows one to consider carefully
the true nature of   carriers in oxides and rare-earth metals.
These applications illustrate some of   subtle details of the IGF
approach and exhibit their physical
significance in a representative  form.\\
As it is seen, this treatment has advantages in comparison with
the standard methods of decoupling of higher-order GFs within the
equation-of-motion approach, namely, the following:
\begin{itemize}
\item[(i)] At the mean-field level, the GF, one obtains, is richer
than that following from the standard procedures. The generalized
mean fields represent all elastic scattering renormalizations in
a compact form.
\item[(ii)] The approximations ( the decoupling ) are introduced at
a later stage with respect to other methods, {\it i.e.} only into
the rigorously obtained self-energy.
\item[(iii)] Many standard results of the many-particle
system  theory are reproduced mathematically incomparable more
simply.
\item[(iv)] The physical picture of   elastic and inelastic
scattering processes in the interacting many-particle systems is
clearly seen at every stage of calculations, which is not the
case with the standard methods of decoupling.
\item[(v)] The main advantage of the whole method is the
possibility of a {\it self-consistent} description of
quasi-particle spectra and their damping in a unified and coherent
fashion.
\item[(vi)] This new picture of interacting many-particle systems
on a lattice is far richer and gives more possibilities for the
analysis of phenomena which can actually take place. In this sense
the approach we suggest produces more advanced physical picture
of the quasi-particle many-body dynamics.\\
\end{itemize}
Despite the novelty of the IGF techniques introduced above and
some (not really big) complexity of the details in its
demonstrations, the major conclusions of the present paper can be
made intelligible to any reader. The most important conclusion to
be drawn from   the present consideration is that the GMF for the
case of strong Coulomb interaction has quite a nontrivial
structure and cannot be reduced to the mean-density functional.
This last statement resembles very much the situation with
strongly non-equilibrium systems, where only the single-particle
distribution function is insufficient to describe the essence of
the strongly non-equilibrium state Therefore a  more complicated
correlation functions are to be taken into account, in accordance
with general ideas of Bogoliubov and Mori-Zwanzig. The IGF method
is intimately related to the projection method in the sense, that
it   expresses the idea of   ``reduced description" of a system
in the most general form. This line of consideration is very
promising for developing the complete and self-contained theory
of strongly interacting many-body systems on a lattice. Our main
results reveal the fundamental importance of the adequate
definition of Generalized Mean Fields at finite temperatures, that
results in a   deeper insight into the nature of quasi-particle
states of the correlated lattice fermions and spins. We believe
that our approach offers a new way for   systematic constructions
of the approximate dynamic  solutions of the Hubbard, SIAM, TIAM,
PAM, spin-fermion, and other models of the strongly correlated
electron systems on a lattice. The work in this direction is in
progress.
\section{Acknowledgments} I would like to dedicate this article to
the memory of the late Professors S.V.Tyablikov, N.N.Bogoliubov,
and D.N.Zubarev. Their illuminating and deep remarks, advice, and
suggestions were indispensable stimulus for my studies.  I
express my gratitude to them.\\

\appendix
\section[{\it APPENDIX A.}]{
{\it \appendixname} {\it {\Large . The Gram-Schmidt
Orthogonalization Procedure}}}

In this appendix we briefly recall   the Gram-Schmidt
Orthogonalization Procedure. The Gram-Schmidt orthogonalization
procedure is an inductive technique to generate a mutually
orthogonal set from any linearly independent set of vectors.\\
Suppose we have an arbitrary n-dimensional Euclidean space, which
means that scalar multiplication has been introduced in some
fashion into an n-dimensional linear space. The vectors $f$ and
$g$ are orthogonal if their scalar product is zero
$$
( f,g ) = 0 \leqno (A.1)
$$
We now describe the orthogonalization process, which is a means of
passing from any linearly independent system of $k$ vectors $
f_{1}, f_{2},...f_{k}$ to an orthogonal system, also consisting
of $k$ nonzero vectors. We denote these vectors by $g_{1},
g_{2},...g_{k}$.\\ Let us put $g_{1} = f_{1}$, which is to say
that the first vector of our system will enter into the
orthogonal system we are building. After that, put
$$
g_{2} = f_{2} + \alpha g_{1} \leqno (A.2)
$$
Since $g_{1} = f_{1}$ and the vectors $f_{1}$ and $f_{2}$ are linearly
independent, it follows that the vector $g_{2}$ is different from zero for any
scalar $\alpha$. We choose this scalar from the constraint
$$
0 = (g_{1}, g_{2} ) = \alpha ( g_{1}, g_{1} ) + ( g_{1}, f_{2} )
\leqno (A.3)
$$
whence
$$
\alpha = - \frac {( g_{1}, f_{2} )}{( g_{1}, g_{1})} \leqno (A.4)
$$
In other words, we get $g_{2}$ by subtracting from $f_{2}$ the projection
of $f_{2}$ onto $g_{1}$. Proceeding inductively, we find
$$
g_{n} = f_{n} - \sum^{n-1}_{j=1} \frac{ ( g_{j}, f_{n} )}{(
g_{j}, g_{j} )} g_{j} \leqno (A.5)
$$
We are left with mutually orthogonal vectors which have the same span as the
original set.\\
Let us consider an important example of a basis $ f_{1}, f_{2},
f_{3}, f_{4}$ in a 4-dimensional space and then construct the
orthonormal basis  of the same space. Next, in the equality
$g_{3} = f_{3} + \beta_{1} g_{1} + \beta_{2} g_{2}$, chose
$\beta_{1}$ and $\beta_{2}$ such that the conditions $g_{3} \bot
g_{1},
g_{3} \bot g_{2} $ are fulfilled.\\
From the equalities
$$
( g_{1}, g_{3} ) = (g_{1}, f_{3} ) + \beta_{1} ( g_{1}, g_{1} ) + \beta_{2}( g_{1}, g_{2} ) \\
\leqno (A.6)
$$\\
$$
( g_{2}, g_{3} ) = (g_{2}, f_{3} ) + \beta_{1} ( g_{1}, g_{2} ) +
\beta_{2}( g_{2}, g_{2} ) \leqno (A.7)
$$
we obtain
$$
\beta_{1} = - \frac {( g_{1}, f_{3} )}{( g_{1}, g_{1})}~; \quad
\beta_{2} = - \frac {( g_{2}, f_{3} )}{( g_{2}, g_{2})} \leqno
(A.8)
$$
Finally, from the equality $g_{4} = f_{4} + \gamma_{1} g_{1} + \gamma_{2} g_{2}
+ \gamma_{3} g_{3}$ we find
$$
\gamma_{1} = - \frac {( g_{1}, f_{4} )}{( g_{1}, g_{1})}~; \quad
\gamma_{2} = - \frac {( g_{2}, f_{4} )}{( g_{2}, g_{2})}~; \quad
\gamma_{3} = - \frac {( g_{3}, f_{4} )}{( g_{3}, g_{3})} \leqno
(A.9)
$$
Thus, we see that with the choice of $\alpha, \beta_{1},
\beta_{2}, \gamma_{1}, \gamma_{2}, \gamma_{3}$   made, the
vectors $g_{1}, g_{2}, g_{3}, g_{4}$ are pairwise
orthogonal.\\

\section[{\it APPENDIX B.}]{{\it \appendixname} \Large {\it . Moments and Green Functions} }
It is known that the method of moments~\cite{martin2} of spectral
density is considered sometimes as an alternative approach for
describing   the many-body quasi-particle dynamics of interacting
many-particle systems. The moments technique appears naturally
when studying the particle dynamics in many-particle systems in
the context of time-dependent correlation functions ( magnetic
resonance, liquids, etc.). Qualitatively, a correlation function
describes how long a given property of a system persists until it
is averaged out by the microscopic motion of particles in the
macroscopic system. The time dependence of a particle correlation
function sometimes is approximated (at small times) via a power
series expansion about the initial time 0.
$$
<A(0)A(t)> = \sum_{n=0}^{\infty} \frac {t^{n}}{n!} \frac
{d^{n}}{dt^{n}} <A(0)A(t)>|_{t=0} = \leqno (B.1)
$$
$$
\sum_{n=0}^{\infty} \frac {(it)^{n}}{n!} <A(0)[H,[H ...
[H,A(0)]...]]]>
$$
The spectral theorem (26), (27) connects   $A(\omega)$ and the
correlation functions. From the above expression we obtain the
moments $M_{n}$ of the spectral density function
$$
M_{n} =  {1 \over 2\pi} \int_{-\infty}^{\infty}d\omega \omega^{n}
A(\omega) = (-1)^{n} < [[H,[H ... [H,A]...]], B]_{\eta} > \leqno
(B.2)
$$
So, by definition, the moments are   time-independent correlation
functions of a combination of the operators. In principle, it is
possible to calculate them in a regular way; however, in practice,
it is possible to do this only for a first few  moments. If the
moments $M_{n}$ of a given spectral density form a positive
sequence, then   GF of appropriate operators is a limit of the
sequence
$$
G(E) = \lim_{n \rightarrow \infty} G_{n}(E, \gamma) \leqno (B.3)
$$
Here the parameter $-\infty < \gamma < +\infty$ and is real. The
approximation procedure  for   GF consists in replacing    the
$G(E)$ by $G_{n}(E,\gamma)$, that depends also on the appropriate
choice of the parameter $\gamma$. The $G_{n}(E, \gamma)$ have the
properties
$$
G_{n}(E,\infty) = G_{n-1} (E, 0) \leqno (B.4)
$$
and are represented by the fraction
$$
G_{n}(E, \gamma) = M_{0} \frac {Q_{n+1}( E ) - \gamma Q_{n} ( E )}
{ P_{n+1} ( E ) - \gamma P_{n} ( E )} \leqno (B.5)
$$
The polynomials $P_{n}$ are given by the determinant
$$
P_{n \geq 1} (E) = \frac {\sqrt { M_{0}}}{ \sqrt {D_{n-1}D_{n}}}
\left|\matrix{ M_{0}&M_{1}&\ldots&M_{n}\cr
M_{1}&M_{2}&\ldots&M_{n+1}\cr \vdots&\vdots&\ddots&\vdots\cr
M_{n-1}&M_{n}&\ldots&M_{2n-1}\cr 1&E&\ldots&E^{n} \cr } \right|
\leqno (B.6)
$$
$$
P_{0} = 1
$$
where
$$
D_{n \geq 1} = \left|\matrix{ M_{0}&M_{1}&\ldots&M_{n}\cr
M_{1}&M_{2}&\ldots&M_{n+1}\cr \vdots&\vdots&\ddots&\vdots\cr
M_{n}&M_{n+1}&\ldots&M_{2n}\cr } \right|  \leqno (B.7) $$
$$
D_{0}
= D_{-1} = M_{0}
$$
The polynomial $Q_{n}(E)$ (which is of (n-1)-th order in E) is
related to the polynomial $P_{n}(E)$ (which is of n-th order in E)
via the following relation
$$
Q_{n}(E) = {1 \over 2\pi M_{0}} \int^{\infty}_{-\infty} \frac {
P_{n}(E) - P_{n}(\omega)}{ E - \omega} A(\omega) d\omega \leqno
(B.8)
$$
It is possible to find a few lowest-order terms
$$
P_{0}(E) = 1; \quad P_{1}(E) = \frac {E - {M_{1} \over
M_{0}}}{M_{2} - M_{0}^{-1}}  \leqno (B.9)
$$
$$
Q_{0}(E) = 0; \quad Q_{1} = \frac { 1}{M_{2} - M_{0}^{-1}} \leqno
(B.10)
$$
The expression (B.5) can be represented in the following form
$$
G_{n}(E, \gamma) = M_{0} \sum^{n+1}_{i=1} \frac {m_{i}(\gamma)}{
E - E _{i}(\gamma)} \leqno (B.11)
$$
Here the numbers $E_{i}(\gamma)$ are roots of the equation
$$
P_{n+1}(E) - \gamma P_{n}(E) = 0 \leqno (B.12)
$$
These relations lead to the possibility of   practical
applications of the moment expansion method. If we know the first
$(2n + 2)$ moments, then the equation (B.12) determines
 $(n+1)$ different roots $E_{i}(\gamma)$. Thus, the spectral density function
can be represented by
$$
A(\omega) = 2\pi M_{0} \sum^{n+1}_{i=1} m_{i} \delta (\omega -
E_{i}) \leqno (B.13)
$$
For example, if we know the moments $M_{0}, M_{1}, M_{2}$ then we
find, from the equation (B.11), the roots of (B.12)
$$
E_{1}(\gamma) = M_{1}M_{0}^{-1}  + \gamma (M_{2} - M_{0}^{-1})
\leqno (B.14)
$$
In this approximation, the GF  and corresponding spectral density
are represented as
$$
G_{0}(\gamma) = \frac {M_{0}}{E - E_{1}(\gamma)}; \quad A(\omega)
= 2\pi M_{0} \delta (\omega - E_{1}) \leqno (B.15)
$$
It is clear that the Tyablikov decoupling approximation (43)
corresponds to this approximation within the moment method. An
improved decoupling scheme, that conserves the first several
frequency moments of the spectral weight function for the
Heisenberg and Hubbard models was developed in paper\cite{tahir1}
({\it cf.} \cite{hein},\cite{hein1} ). \\ It was shown in
ref.~\cite{kuz5} that the IGF method permits one to calculate the
spectral density for the spin-fermion model in the approximation
that preserves the first four moments. This is valid also for the
approximation
  used for the strongly correlated Hubbard model in Section 7.2. \\
It must be clear from the above consideration that the structure
of the obtained solution for single-particle GF depends strongly
on the stage at which irreducible parts were
introduced~\cite{kuz4}. To clarify this, let us consider equation
(29) again. Instead of (30), we introduce now the IGFs in the
following way
$$
\omega G(\omega) = M_{0} + <<[A, H]_{-}\mid A^{\dagger}>>_{\omega}
$$
$$
\omega  <<[A,H] \vert A^{\dagger}>> = M_{1} +  (^{(ir)}<< [[A,H]
H] \mid A^{\dagger}>>_{\omega}) +  \leqno (B.16)
$$
$$
\alpha_{1}<<A \vert A^{\dagger}>>_{\omega} + \alpha_{2}<< [A,H]
\vert A^{\dagger}>>_{\omega}
$$
The unknown constants $\alpha_{1}$ and $\alpha_{2}$ are connected
by the orthogonality condition
$$
<[[[A,H ] H]^{(ir)} , A^{\dagger} ]> = 0 \leqno (B.17)
$$
For   illustration, we consider the simplest possibility and
write down the following equation
$$
\omega (^{(ir)}<< [[A, H] H] \vert A^{\dagger}>> ) = (^{(ir)}<<
[[A, H] H] \vert [ H, A^{\dagger} ]>>) \leqno (B.18)
$$
Then by introducing the irreducible parts for the ${\it right}$
operators, we obtain
$$
(^{(ir)}<< [[A, H] H] \vert A^{\dagger}>> )(\omega -
\alpha^{\dagger}_{1}) = (^{(ir)}<< [[A, H] H] \vert [ H,
A^{\dagger} ]>>^{(ir)}) \leqno (B.19)
$$
It is clear enough that, as a result, we arrive at the following
set of equations
$$
\omega <<A \vert A^{\dagger}>>_{\omega}  - <<[A, H]_{-}\mid
A^{\dagger}>>_{\omega}  = M_{0}
$$
$$
\alpha_{1} << A \vert A^{\dagger}>>_{\omega} + ( \omega -
\alpha_{2}) <<[A,H] \vert A^{\dagger}>>_{\omega} = M_{1} -\Phi
\leqno (B.20)
$$
where
$$
\Phi =   (^{(ir)}<< [[A,H] H] \mid [ A, H
]^{\dagger}>>^{(ir)}_{\omega}) \leqno (B.21)
$$
The solutions of the equations (B.20) are given by
$$
<< A \vert A^{\dagger} >>_{\omega} = \frac { M_{0} (\omega -
\alpha_{2}) - ( M_{1} - \Phi)}{ \omega ( \omega - \alpha_{2}) +
\alpha_{1}} \leqno (B.22)
$$
$$
<< [ A, H] \vert A^{\dagger} >>_{\omega} = \frac { \omega ( M_{1}
- \Phi) + \alpha_{1} M_{0}}{ \omega ( \omega - \alpha_{2}) +
\alpha_{1}} \leqno (B.23)
$$
$$
\alpha_{1}M_{0} + \alpha_{2} M_{1} = M_{2}
$$
It is evident that there is similarity between the obtained
solutions and the moment expansion method. The structure of
equation (B.22) corresponds to the moment expansion (B.11) except
for the factor $\Phi$ that should be calculated by considering
high-order equations of motion or by some relevant approximation.

\section[{\it APPENDIX C.}]{{\it \appendixname} \Large {\it . Projection methods and IGFs} }

The IGFs method is intimately related to the projection operator
method \cite{forst}, \cite{tser2}, that incorporates the idea of
"reduced description" of a system in the most suitable form. The
projection operation~\cite{ichi},~\cite{tser2} makes it possible
to reduce the infinite hierarchy of coupled equations to a few
relatively simple equations  that "effectively" take into account
the essential information about the system that determines the
specific nature of the given problem. Projection techniques become
standard in the study of certain dynamic processes. Projection
operator techniques of Mori-Zwanzig and similar ones~\cite{lee}
are useful for the derivation of relaxation equations and
formulas
for transport coefficients in terms of microscopic properties.\\
This approach was applied  to a large variety of phenomena
concerning the line-shape problem. It was shown that there is a
close relationship between the Mori procedure and the "classical
moment problem" of mathematical analysis.\\ Let us briefly
consider
 the projection formalism for double-time retarded
GFs~\cite{ichi},~\cite{tser2}. Ichiyanagi~\cite{ichi} constructed
the following set of equations for   GF (28):
$$
(\frac{d}{dt} - i\omega_{k}) <<A_{k}(t),A^{\dagger}_{k}(t')>> =
-i\delta(t-t') <[ A_{k},A_{k}^{\dagger}]> +
\\F(k,t-t') \leqno (C.1)
$$

$$
(\frac{d}{dt} + i\omega_{k}) F (k,t-t') = +i \delta(t-t') <[ K
(k),A_{k}^{\dagger}]> + \Pi (k, t-t') \leqno (C.2)
$$
where $F(k, t-t') = <<K(k,t), A_{k}^{\dagger}(t')>>$ and $\Pi (k,
t-t') = << K(k,t), K^{\dagger}(k,t')>>$ Here, the definitions were
introduced:
$$
i \omega_{k} = \frac{<[ \frac{d}{dt} A_{k},
A^{\dagger}_{k}]>}{<[A_{k}, A^{\dagger}_{k}]>}, \quad K(k,t) = (
1 - P ) A_{k}(t) \leqno (C.3)
$$

$$
PG = <[G, A_{k}^{\dagger}]> <[ A_{k}, A^{\dagger}_{k} ]>^{- 1}
A_{k} \leqno (C.4)
$$
The projection operator $P$ defined in (C.4) is different from
the one introduced by Mori. The main result of paper~\cite{ichi}
is that, using the projection operator, a Dyson equation that
determines an {\it irreducible} quantity, proper self-energy
part, was obtained in the following form:
$$
\bigl (\omega - \omega_{k} - \frac {2 \pi}{ <[A_{k},
A_{k}^{\dagger}]>} M (k, \omega) \bigr ) << A_{k} |
A^{\dagger}_{k}>>_{\omega} =   - \frac {< [A_{k},
A^{\dagger}_{k}]>}{ 2 \pi } \leqno (C.5)
$$
Here $M ( k, \omega )$ is the self-energy  , that, in the
diagrammatic language,   consists of irreducible diagrams.\\ Our
point of view is closely related to that of ref.~\cite{ichi} and
to the development of the this paper  by Tserkovnikov in a
systematic way~\cite{tser2}. However, our strategy is slightly
different in the time evolution aspect. We consider our IGF
technique as  more convenient from the practical computational
point of view.

\section[{\it APPENDIX D.}]{{\it \appendixname} \Large {\it . Effective
Perturbation Expansion for the Mass Operator} }

Let us consider a useful example how to iterate the initial
"trial" solution and to get an expansion for the mass
operator\cite{kuzem1},\cite{kuz2}. To be concrete, let us consider
the calculation of the mass operator for the Hubbard model in
Section 8.1. The first iteration for the equation (\ref{eq.147})
with the trial function (\ref{eq.148}) have lead us to the
expression (\ref{eq.149}), which we rewrite here in the following
form
$$
\label{eq.D1} M_{k\sigma}( \omega) = \frac{U^2}{N^2} \sum_{pq}
\frac{N_{kpq}} {\omega  -  \Omega_{kpq}} \leqno(D.1)
$$
where
$$ N_{kpq} = n_{p+q-\sigma}(1 - n_{k+p\sigma} - n_{q-\sigma}) +
n_{k+p\sigma} n_{q-\sigma}$$

$$\Omega_{kpq} = - \epsilon(p+q\sigma) +
\epsilon(k+p\sigma) + \epsilon(q\sigma)$$ Now we are able to
calculate the spectral weight function $g_{k\sigma}(\omega)$
(\ref{eq.77})  $$g_{k\sigma}(\omega) = {1 \over \pi } \frac
{\Gamma_{k\sigma}(\omega)}{ [\omega - E_{k\sigma}]^{2} +
\Gamma^{2}_{k\sigma}(\omega)} \leqno(D.2)
$$ We approximate
this expression by the following way
$$g_{k\sigma}(\omega) \approx  (1 - \alpha_{k\sigma} ) \delta (\omega
- E_{k\sigma}) + {1 \over \pi } \frac {\Gamma_{k\sigma}(\omega)}{
[\omega - E_{k\sigma}]^{2} } \leqno(D.3)$$ Here
$$\Gamma_{k\sigma}(\omega) = \pi \frac{U^2}{N^2} \sum_{pq}
N_{kpq} \delta (\omega - \Omega_{kpq})$$
$$ E_{k\sigma} = \epsilon(k\sigma) + \Delta_{k\sigma}$$
$$\Delta_{k\sigma} = Re M_{k\sigma}( \omega + i\epsilon)$$
The unknown factor $(1 - \alpha_{k\sigma} )$ is determined by the
normalization condition
$$ \int^{\infty}_{-\infty} d\omega g_{k\sigma}(\omega) = 1 $$
whence
$$ \alpha_{k\sigma} =  \frac{U^2}{N^2} \sum_{pq}
\frac{N_{kpq}}{\Omega_{kpq} - E_{k\sigma} }$$ Then, using
(\ref{eq.22}), we find for the mean occupation numbers
$$ n_{\sigma} =    \frac{1}{N}\sum_{k}n(E_{k\sigma}) + \frac{U^2}{N^3} \sum_{kpq}
\frac{N_{kpq}}{(\Omega_{kpq} - E_{k\sigma})^{2}} [n(\Omega_{kpq})
- n(E_{k\sigma})] \leqno(D.4)
$$
Now we can use the spectral weight function ( D.2) to iterate the
equation (\ref{eq.147}) and to get a perturbation expansion for
the self-energy  $M_{k\sigma}$ in the pair approximation. Instead
of the initial trial solution in the form of   delta-function
(\ref{eq.148}), we   take the expression (D.3). It is easy to
check that we get an expansion up to 6th order in U.


\end{document}